\documentclass[aps,pre,preprint,floatfix,showpacs]{revtex4-2} 
\usepackage{colordvi,graphicx,color,amsbsy}
\usepackage{amsmath,amssymb}
\usepackage{wrapfig}
\usepackage{mwe}
\usepackage[export]{adjustbox}

\usepackage{graphicx}
\usepackage{dcolumn}
\usepackage{bm}
\usepackage{enumerate}
\usepackage[latin1]{inputenc}
\newcommand{\comment}[1]{}
\newcommand\etal{\mbox{\textit{et al.}}}

\usepackage{xcolor}

\usepackage[normalem]{ulem}

\usepackage{etoolbox}

\makeatletter
\def\@email#1#2{%
 \endgroup
 \patchcmd{\titleblock@produce}
  {\frontmatter@RRAPformat}
  {\frontmatter@RRAPformat{\produce@RRAP{*#1\href{mailto:#2}{#2}}}\frontmatter@RRAPformat}
  {}{}
}%

\makeatother

\begin{document}


\title[Transport, flow topology and Lagrangian statistics]{Transport, flow topology and Lagrangian conditional statistics \\ in edge plasma turbulence}

\author{Benjamin Kadoch}
\affiliation{Aix-Marseille Universit\'e, CNRS, IUSTI, Marseille, France}

\author{Diego del-Castillo-Negrete}
\affiliation{Oak Ridge National Laboratory, Oak Ridge, Tennessee, USA}

\author{Wouter J.T. Bos}
\affiliation{LMFA-CNRS - \'Ecole Centrale de Lyon, France}

\author{Kai Schneider}
\affiliation{Aix-Marseille Universit\'e, CNRS, I2M, Marseille, France}
\email{kai.schneider@univ-amu.fr}

\date{\today}


\begin{abstract}

Lagrangian statistics and particle transport in edge plasma turbulence are investigated using the Hasegawa--Wakatani model and its modified version. The latter shows the emergence of pronounced zonal flows. Different values of the adiabaticity parameter are considered. The main goal is to characterize the role of coherent structures, {\it i.e.}, vortices and zonal flows, and their impact on the Lagrangian statistics of particles. Computationally intensive long time simulations following ensembles of test particles over hundreds of eddy turnover times are considered in statistically stationary turbulent flows. The flow topology is characterized using the Lagrangian Okubo--Weiss criterion, 
and the flow can thus be split into topologically different domains. In elliptic and hyperbolic regions, the probability density functions (pdfs) of the residence time have self-similar algebraic decaying tails. However, in the intermediate regions the pdfs do exhibit exponentially decaying tails. Topologically conditioned pdfs of the Lagrangian velocity,  and acceleration and density fluctuations are likewise computed. The differences between the classical Hasegawa--Wakatani system and its modified version are assessed and the role of zonal flows is highlighted. The density flux spectrum which characterizes the contributions of different length scales is studied and its inertial scaling is found to be in agreement with predictions based on dimensional arguments. Analyzing the angular change of particle tracers at different time scales, corresponding to coarse grained curvature, completes the study and the multiscale geometric statistics quantify the directional properties of the
particle motion in the different flow regimes.

\end{abstract}

\maketitle

\vfill

\section{Introduction}

Drift-wave turbulence and zonal flows in the tokamak edge are crucial for the dynamics of magnetically confined plasma flow, its confinement properties and the non-diffusive transport.
In recent years there has been a significant interest in
understanding this problem from a Lagrangian perspective. In this approach, the transport properties of the system are studied by tracking the trajectories of large ensembles of tracers. In numerical simulations, this is accomplished by solving the equations for  the trajectories of test particles in a given velocity field, {\it e.g.}, the ${\bm E} \times {\bm B}$ velocity field in the guiding center description of a plasma. 
One of the main lessons learned from the Lagrangian approach is that coherent structures have a profound influence on transport. In particular, the combined effect of the  trapping by eddies and the long displacements induced by zonal shear flows typically gives rise to non-diffusive transport \cite{Krasheninnikov_2008,del_castillo_etal_2004,raul_etal_2004}. From the Lagrangian point of view, non-diffusive transport is characterized by the anomalous scaling of the statistical moments of the particle's displacements $\langle \delta x^2 \rangle \sim t^\gamma$, where $\delta x$ denotes the particle's displacements, $t$ denotes the time, $0<\gamma <2$, and $\langle \, \rangle$ denotes ensemble average. In the case of diffusive transport, $\gamma=1$, whereas for non-diffusive transport $\gamma \neq 1$. 

It has been experimentally and numerically observed that edge turbulence typically develops coherent structures, {\it e.g.}, blobs, streamers and shear flows, see for example \cite{Krasheninnikov_2008} and references therein. The understanding of the role of these coherent structures on the particle and heat transport is a critical step to guarantee the confinement needed in a fusion reactor. However, establishing a quantitative connection between coherent structures (\textit{e.g.}, vortices and zonal flow) and transport, and in particular anomalous diffusion (\textit{e.g.}, sub-diffusion and super-diffusion), can be a highly nontrivial task. Studies based on simple transport models (\textit{e.g.}, chaotic advection models, see for example \cite{del_castillo_1998}) and on generalized random walk models (\textit{e.g.}, continuous time random walks) have shown that typically, when the trapping effects of eddys dominate, $\gamma <1$, and transport is sub-diffusive ({\it i.e.}, slower than diffusion). But, when zonal flows dominate, transport {becomes highly anisotropic and is observed to be} super-diffusive ({\it i.e.}, faster than diffusion) and $\gamma >1$. However, establishing this connection at a quantitative level has proved elusive in the case of flows with complex spatio-temporal structures ({\it e.g.} turbulent plasmas and fluids), because in these systems the coherent structures typically exhibit complicate dynamics. For example, in $2$-D turbulence, the size and location of trapping regions changes  unpredictably in space and time due to vortex merging and decay. In the case of bounded flows, boundary layers create intense vorticity dominated region  near the walls that might temporally trap particles. Moreover, a particle that is ``trapped" in an eddy will not necessarily be confined to a region in space since the eddy might be moving throughout the whole domain. Further complications appear in systems that exhibit rapid transport phenomena that give rise to intermittent large particle displacements. This type of phenomena, sometimes referred to as ``avalanche-like" transport, is routinely observed in plasma turbulence models near marginal stability. In this case, tracers are transported large distances by short-lived coherent structures. 

The key question then is, given a Lagrangian trajectory of a tracer, ${\bm x}(t)$, in a flow with a complex spatio-temporal dynamics, how can we characterize the signatures of trapping due to eddies and long-displacements (flights), from the information contained in the trajectory?  The answer to this question is important because it is the first step to characterize the statistics of trapping and flights in turbulent transport. This statistics is the cornerstone in the construction of effective models for nondiffusive transport in the context of the continuous time random walk model and in fractional diffusion, see for example \cite{del_castillo_etal_2004,raul_etal_2004}.

In Ref.~\cite{Umansky_POF_2011}, the authors showed that turbulence fluctuations have non-Gaussian characteristics, while the radial flux of plasma density is similar to Bohm diffusion. 
Garland {\it et al.}~\cite{Garland_POF_2017} presented a study investigating the influence of collisionality on intermittency in drift-wave turbulence using both numerical and experimental approaches, the latter for the TJ-K stellerator; they showed an increase in intermittency with increasing collisionality for density fluctuations. The study in \cite{Garland_POF_2020} revealed the importance of considering local curvature properties as a factor in the decoupling of density and potential fluctuations leading to intermittency in drift-wave turbulence. 

As mentioned before, when the coherent structures are time independent and fixed in space, the answer to this question is simple: trapping (flight) events correspond to sections of the trajectory for which ${\bm x}(t)$ is contained in an eddy (shear flow). For general flows,  Ref.~\cite{metzler_klafter_2000} proposed a simple characterization based on the intuition that the radial component of the Lagrangian velocity would stay small (and varying sign rapidly) while trapped, and become large (with a well-defined sign) during a flight event. This idea was implemented in Ref.~\cite{mier_etal_2008} to study the non-Gaussian statistics of trapping and flight events in near-critical dissipative-trapped-electron-mode turbulence. Ref.~\cite{carreras_etal_2001} proposed a definition of trapping based on the time that nearby particles stay close, to study anomalous diffusion and exit time statistics of tracers in a fluid model of resistive pressure-gradient-driven plasma turbulence. In the study of mesoscale transport in near critical resistive pressure-gradient-driven turbulence in toroidal geometry presented in Ref.~\cite{carreras_garcia_2006} a flight was defined as the portion of the trajectory that keeps the same sign in the radial velocity. Despite the fact that these studies provided valuable insights on the nature of nondiffusive transport in plasma turbulence, there is a need to provide a more conclusive, systematic, and quantitative measure of trapping and flights events from a Lagrangian perspective. As a first step to address this important challenging problem, in this paper we propose a method to characterize the Lagrangian statistics of tracers based on the topological properties of the turbulence. Based on a previous previous work in Ref.~\cite{kadoch_etal_2011}, we characterize the topology using the Okubo--Weiss criterion, which provides a conceptually simple tool to partition the flow into topologically different regions: elliptic (vortex dominated), hyperbolic (deformation dominated), and intermediate (turbulent background). However, different from Ref.~\cite{kadoch_etal_2011}, which was limited to Navier--Stokes fluid turbulence, we consider the Hasegawa--Wakatani system that provides one of the simplest models to study cross-field transport by electrostatic drift waves in magnetically confined plasmas in general and in the plasma edge in particular.
In addition to the classical Hasegawa--Wakatani model we consider a modified version proposed in \cite{nbd2007} and used in \cite{Pushkarev_POF_2013}. The later exhibits pronounced zonal flows for large adiabaticity values and the flow characteristics differ significantly from those obtained in the classical case.

As a second step, geometric statistics of the particle trajectories are performed by analyzing the angular change of particle tracers at different time scales. The thus obtained angular curvature angle corresponding to coarse grained curvature of the trajectories, quantifies the directional properties of the complex particle motion in the different flow regimes from a Lagrangian perspective.
%


The remainder of the manuscript describes the Hasegawa--Wakatani turbulence model in Sect.~\ref{sec:drift-wave} and recalls the Okubo--Weiss criterion used to partition the flow into distinct regions.
Tools for performing directional Lagrangian statistics, including the coarse-grained curvature angle are likewise introduced in Sect.~\ref{sec:drift-wave}. 
Results on Eulerian statistics and Lagrangian conditional statistics for different flow regimes are presented and discussed in Sect.~\ref{sec:results}. Finally, we conclude and give some perspectives for future work in Sect.~\ref{sec:conclusions}.

\section{Edge turbulence model}
\label{sec:drift-wave}
\begin{figure}[!htb]
 \begin{center}
 \begin{tabular}{cc}
   \includegraphics[width=0.8\linewidth]{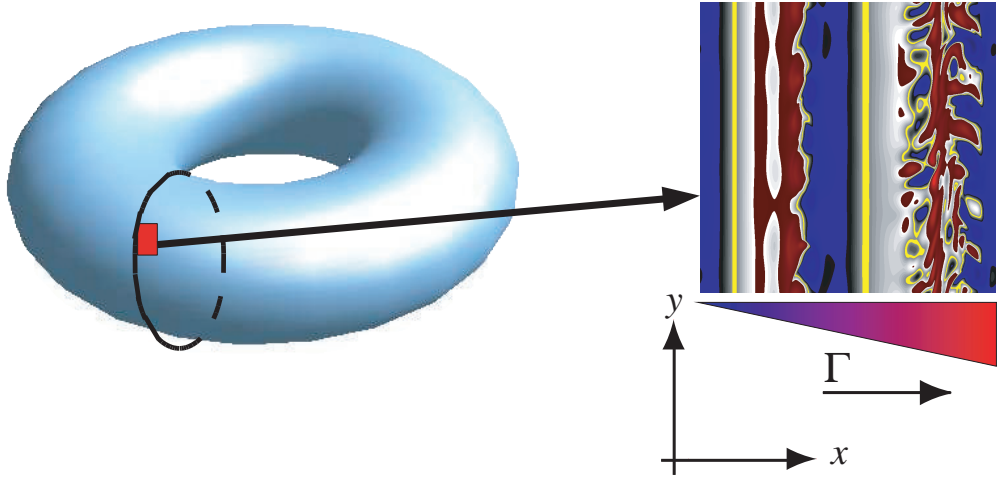}
  \end{tabular}
\caption{Sketch of the flow configuration in the 2-D  slab geometry in the tokamak edge modeled by the Hasegawa--Wakatani system, where $x$ denotes the radial direction and $y$ the poloidal direction.
A mean plasma density gradient in the radial direction
$\Gamma$
is imposed and the 2-D flow in the domain of size 
64 $\rho_s$ is computed, where $\rho_s$ is the 
Larmor radius. The computational domain is illustrated by the red square. A snapshot of the vorticity field for modified Hasegawa-Wakatani ({\it mHW}) with zonal flows is shown on the top, right. 
\label{sketch}}
 \end{center}
\end{figure}

As a numerically tractable model for edge turbulence we consider the Hasegawa-Wakatani turbulence model describing the drift wave-zonal flow interaction in a two-field coupled system in a shearless slab geometry \cite{Hasegawa_PRL_1983}.
A sketch illustrating the flow configuration, considering a two-dimensional slab in the radial-poloidal plane, is given in Fig.~\ref{sketch}.

\subsection{Eulerian description}

The closed set of equations,
describing the evolution of vorticity $\omega=\nabla^2 \phi$ of the $E\times B$ motion (with $\phi$ being the electrostatic potential) and of the plasma density fluctuations $n$, read:
\begin{eqnarray}
\left(\frac{\partial}{\partial{t}}- \nu\nabla^2\right)\nabla^2 \phi=\left[\nabla^2 \phi,\phi\right]+c(\phi-n),\label{hw1}\\
\left(\frac{\partial}{\partial{t}}- D\nabla^2\right)n=\left[n,\phi\right]
 - \bm u \cdot \nabla \ln(\left<n\right>) 
+c(\phi-n),\label{hw2}
\end{eqnarray}
All quantities are dimensionless and have been suitably normalized, as described in 
\cite{Bos2010-1,futatani2011coherent}. 
The coordinates $x$ and $y$ denote the radial and the poloidal direction, respectively. 
The constant parameters $D$ and $\nu$ are the respectively the cross-field diffusion of the plasma density fluctuations $n$ and the kinematic viscosity. 
The adiabaticity $c$ is given by
\begin{equation}
c=\frac{T_e k_{z}^2}{e^2n_0\eta\omega_{ci}},
\end{equation}
with $T_e$ the electron-temperature, $k_{z}$ the effective parallel wavenumber ({\it i.e.}, parallel to the magnetic field), $e$ the electron charge,   $n_0$ the reference plasma density, $\eta$ the electron resistivity and $\omega_{ci}$ the ion-gyro-frequency.
The Poisson brackets are defined as {$[a,b]=\frac{\partial a}{\partial x}\frac{\partial b}{\partial y}-\frac{\partial a}{\partial y}\frac{\partial b}{\partial x}$}.
The mean plasma density $\left<n\right>$ acts as a source driving the density fluctuation dynamics. In particular, for a profile of the form $\left<n\right> =n_0(x) = N_0 \exp(-x/L_n)$ where $N_0$ is a constant number density and $L_n$ is a characteristic length scale for the density gradient, the second term on the right-hand side of eq.~(\ref{hw2}) reduces to $-\Gamma u_x$, where $\Gamma=(N_0/L_n)$ is the mean density gradient.
The electrostatic potential $\phi$ plays for the $E\times B$ velocity the role of a stream-function, $\bm u=\nabla^\perp \phi$, where $\nabla^\perp = (-\partial_y, \partial_x)$, \emph{i.e.}, we have $u_x=-\partial \phi/\partial y$ 
and $u_y=\partial \phi/\partial x$. 
The coupling term $c (n - \phi)$, present in both equations, is related to the parallel current density and triggers the electrostatic plasma instability. The adiabaticity value $c$ which quantifies the collisionality of ions and electrons determines the dynamics. 
Large collisionality corresponding to small $c$ values coincides with the {\it hydrodynamic 2D limit} characterized by the presence of long living coherent vortices in the ${\bm E} \times {\bm B}$ flow and almost passive advection of the density fluctuations. 
Intermediate values of $c$ of order unity yield dynamics which are supposed to be close to tokamak edge-turbulence, {\it the so-called quasi-adiabatic regime}. 
In the limit $c \rightarrow \infty$ we obtain the Hasegawa-Mima model \cite{hasegawa1977stationary} which corresponds to the Charney equation for Rossby waves used for modeling geophysical flows, {\it also known as geostrophic regime.}
Note that geometrical variations of the magnetic field can be neglected as the domain is chosen sufficiently small, {\it i.e.}, of size 64 $\rho_s$, where $\rho_s$ is the  
Larmor radius. 
The magnetic field lines of the unperturbed constant magnetic field are straight and they are assumed to be {perpendicular to} the slab.
In addition to the classical Hasegawa--Wakatani model {\it(cHW)}, described above,  we also consider a revised version, named modified Hasegawa--Wakatani model {\it (mHW)}, which was introduced in \cite{nbd2007}. To obtain zonal flows for large adiabaticity values, $k_y=0$ modes of the coupling term ($c(\phi-n)$) are set to zero in {\it mHW}, similarly to what has been done in \cite{Pushkarev_POF_2013}. 

The Lagrangian acceleration of tracer particles $\bm a_L$, advected by the $E\times B$ velocity can be defined in the Eulerian reference frame using the momentum evolution equation,
\begin{equation}\label{eqaL}
\bm a_L=\frac{\partial \nabla^\perp \phi}{\partial t}+\left[\phi,\nabla^\perp\phi\right] =-\nabla p+\nu\nabla^2\bm u- \frac{\nabla^\perp}{\nabla^2} \left[c(n-\phi)\right],
\end{equation}
where $p$ denotes pressure and $\nabla^\perp / \nabla^2$ denotes the Biot-Savart operator.

The above equations (\ref{hw1},\ref{hw2}) are solved numerically in a double-periodic domain of size $64^2$.
To this end a fully dealiased pseudo-spectral method with a resolution of $1024^2$ grid points is used.
The time step is $5 \cdot 10^{-4}$ and for the diffusivity of the plasma density and the kinematic viscosity we use  $D= 5 \cdot 10^{-3}$ and $\nu= 5 \cdot 10^{-3}$, respectively. This results in a Schmidt number of unity ($Sc=\nu/D=1$). The value of $\Gamma$ in the mean density gradient of plasma density fluctuations is equal to one, {\it i.e.}, $\Gamma =1$.

The simulations are initialized with Gaussian random 
initial conditions and run until a saturated, fully developed turbulent flow is obtained. 
This transition phase is typically quite time consuming and may take millions of time steps.
Subsequently $10^4$ particles were uniformly injected and their velocity and acceleration were monitored during a large number of large-scale turn-over times (table \ref{tab: Parameters}). 
The eddy turn-over time $t_k$ is defined as $1/\sqrt{2 Z_{rms}}$ where $Z_{rms}$ is the RMS vorticity.
It is of the same order of magnitude in the different regimes, $\sim 0.5$ for $c\leq 2$ for classical Hasegawa--Wakatani {\it(cHW)} and $\sim 1$ for modified Hasegawa--Wakatani {\it (mHW)}.
The adiabaticity is varied between $c=0.01$ and $c=4$, to obtain different flow regimes.  
Series of long time simulations have been performed for both {\it cHW} and {\it mHW}.
Compared to the simulations in \cite{Pushkarev_POF_2013} we use instead of hyperviscosity classical Newtonian viscosity. Qualitatively the results are similar, which shows that the dynamics does not critically depend on the choice of the small scale damping.

\subsection{Lagrangian description: trajectories, curvature and coarse-grained curvature} 
\label{subsec:lagrange}

Considering passive tracer particles in drift-wave turbulence is motivated by analyzing for instance the transport of impurities in the plasma edge.
In the Lagrangian setting the time evolution of a tracer particle 
position $\bm x (t)$ obeys to the classical differential equation:
\begin{equation}\label{eq: part evol}
\frac{d \bm x}{dt}=\bm u \left( \bm x(t),t \right)
\end{equation}
completed with the initial position of the tracer particle ${\bm x}(t=0) = {\bm x}_0$.
The velocity $\bm u$ is obtained at the particle position $\bm x(t)$ using bicubic interpolation from the velocity field, and the time advancement of particles is done using a second-order Runge--Kutta scheme with time step $5 \cdot 10^{-4}$. The curvature $\kappa$ of the particle trajectory ${\bm x}(t)$ is also considered in this study, in particular for analyzing Lagrangian intermittency \cite{kadoch_etal_2011}, and defined by
\begin{equation}\label{eq: curvature}
\kappa=\frac{a_n}{||\bm{u}_L||^2}
\end{equation}
where $a_n$ is the normal component of Lagrangian acceleration, defined in eq.~(\ref{eqaL}). The curvature vanishes when the particle velocity and the acceleration vectors are parallel and we can note that $\kappa \ge 0$. All the Lagrangian statistics are computed using ensemble and time averages (with the exception of the residence time where only ensemble averaging is applied) and the index $\cdot_L$ is used to denote Lagrangian quantities. 
Further details on the simulations using equations (\ref{hw1},\ref{hw2}) can be found in [\onlinecite{Bos2010-1}].
For the Lagrangian part of the study we refer to \cite{Kadoch2008} where a similar investigation was performed for Navier--Stokes turbulence. 

\begin{figure}[!htb]
 \begin{center}
 \begin{tabular}{cc}
   \includegraphics[scale=0.6]{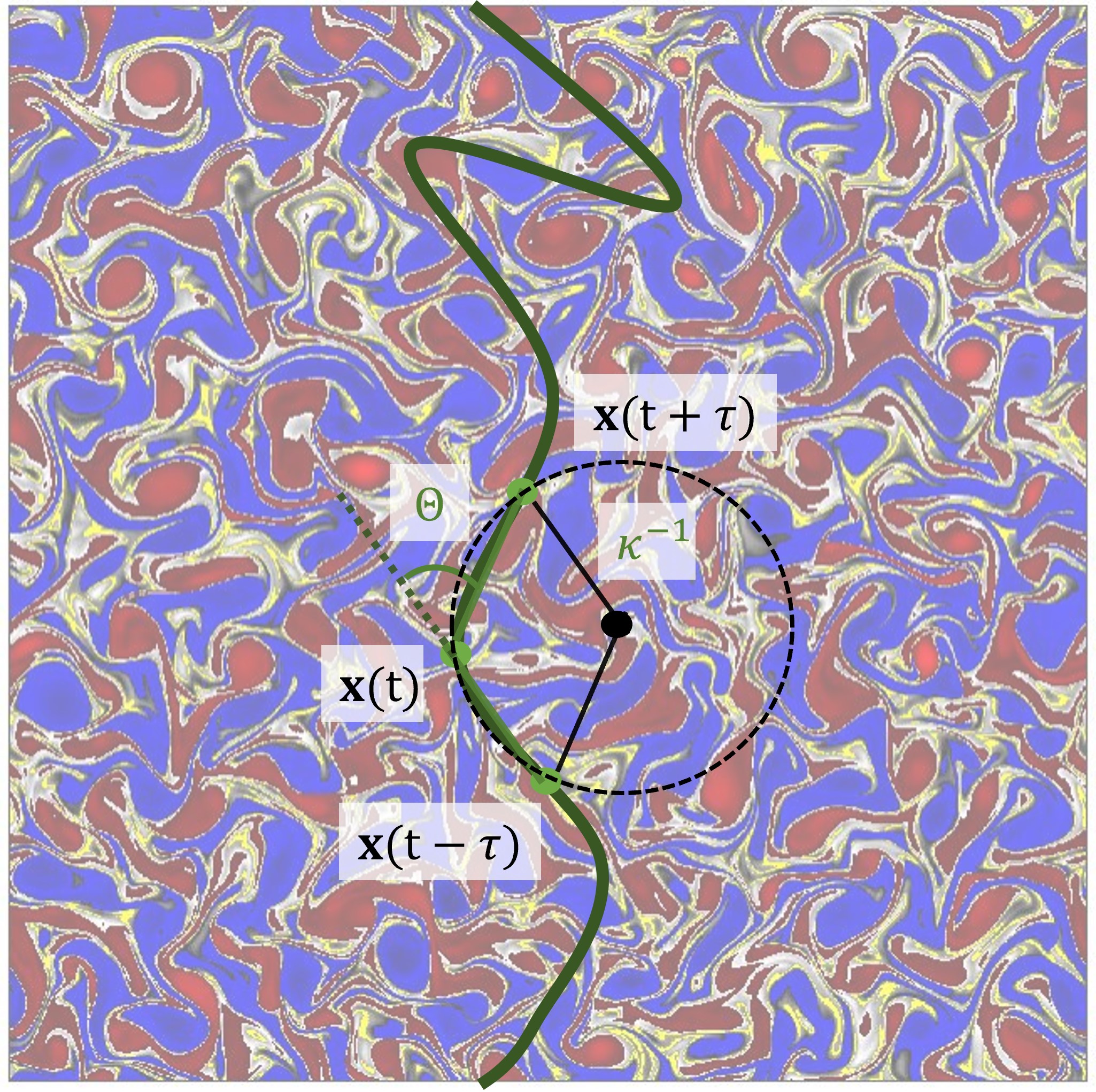}
  \end{tabular}
\caption{Sketch of the coarse-grained curvature angle $\Theta$ and the curvature $\kappa$. A sample trajectory is superimposed to a vorticity field. Three trajectory positions ${\bm x}(t)$ at three time instants, $ t- \tau, t$ and $t + \tau$ are marked by green dots and used to determine the curvature angle $\Theta$.  \label{sketch_ang}}
 \end{center}
\end{figure}

To get insight into the complex multiscale dynamics of drift-wave turbulence and zonal flows from a Lagrangian perspective we quantify directional motion in stochastic trajectories statistically at different time scales using the curvature angle~\cite{Burov2013,Bos_PRL_2015}.
To determine this directional change the angle between subsequent particle displacement increments is evaluated as a function of the timelag, and thus multi-scale geometric statistics can be performed. We define the Lagrangian spatial increment as $\delta \bm X(\bm x_0,t,\tau)=\bm x(t)-\bm x(t-\tau)$ where $\bm x(t)$ is the position of a fluid particle at time $t$, passing through point $\bm x_0$ at the reference time $t=t_0$ and advected by a velocity field $\bm u$, eq.~(\ref{eq: part evol}), as illustrated in Fig.\ref{sketch_ang}. The cosine of the angle $\Theta(t,\tau)$ between subsequent particle increments, introduced in \cite{Burov2013} and analyzed in three-dimensional homogeneous isotropic turbulence \cite{Bos_PRL_2015} and in two-dimensional homogeneous isotropic and confined turbulence \cite{Kadoch_PRF_2017}, is 
\begin{equation}\label{eq:cos}
 \cos(\Theta(t,\tau))=\frac{\delta \bm X(\bm x_0,t,\tau)\cdot\delta \bm X(\bm x_0,t+\tau,\tau)}{|\delta \bm X(\bm x_0,t,\tau)|~|\delta \bm X(\bm x_0,t+\tau,\tau)|}.
\end{equation}
The curvature angle $\Theta$ can be related to
the curvature $\kappa$ introduced in eq.~(\ref{eq: curvature}).
In the limit when $\tau$ goes to zero the curvature angle yields the curvature
$\kappa$, as discussed in \cite{kadoch2016ctr, Bos_PRL_2015}. 
For finite values of $\tau$ we obtain a finite time curvature measure $K(t,\tau)$, corresponding to a coarse grained curvature and the influence of $\tau$ can be analyzed.
To this end the scale-dependent curvature angle must be properly normalized, {\it i.e.}, dividing $\Theta(t,\tau)$ by $ 2 \tau || {\bm u} ||$ and thus we obtain $K(t,\tau) = \Theta(\tau) / (2 \tau || {\bm u} ||)$. 

\subsection{Lagrangian flow topology}
\label{sec:weiss}

The Okubo--Weiss criterion \cite{Okubo_1970,Weiss_PysD_1991} is a well established tool to characterize the topology of turbulent flow fields.
Hereby the flow field can be  partitioned into topologically distinct regions, {\it i.e.}, into vorticity dominated regions, which correspond to elliptic zones, strong deformation regions which correspond to hyperbolic zones and likewise into intermediate regions. 
At a given time instant the flow parameter $Q$, also called Weiss value, is determined,
\begin{equation}
\label{weiss}
Q=s^2-\omega^2 \, ,
\end{equation}
where $\omega= \partial_x u_y-\partial_y u_x$ is the vorticity and $s^2=s_1^2+s_2^2$ is the deformation with
  \begin{equation}
  s_1=\frac{\partial u_x}{\partial x}-\frac{\partial u_y}{\partial y} \, , \qquad  s_2=\frac{\partial u_y}{\partial x}+\frac{\partial  u_x}{\partial y} \, .
\end{equation}
Then using the local value of $Q$, the flow domain can be partitioned in three disjoint regions, as suggested for instance in \cite{kadoch_etal_2011}:
\begin{itemize}
\item[i)]  strongly elliptic for which $Q \leq -Q_0$ (vorticity dominated)
\item[ii)] strongly hyperbolic for which $Q \geq Q_0$(deformation dominated)
\item[iii)] intermediate regions for which $-Q_0<Q<Q_0$
\end{itemize}

The threshold value $Q_0$ can be chosen for instance as the standard deviation of  $Q$, {\it i.e.}, $Q_0=\sqrt{\langle Q^2 \rangle}$ where $\langle \cdot \rangle$ is the
ensemble average. 
It is important to mention that we use here the Lagrangian Weiss field, {\it i.e.}, we compute $Q$ along particle trajectories, a technique we introduced in \cite{kadoch_etal_2011}. Classically the Okubo--Weiss criterion is applied in an Eulerian framework only and comes with a theoretical justification.
The Okubo--Weiss criterion is based on the linearization of the Navier--Stokes equation and the assumption that the strain field remains frozen. Some of its limitations are discussed in~\cite{Basdevant_PysD_1994}.

Based on this partitioning of the flow the residence time can be considered for the different regions. The time during which a particle remains in the same zone (strong elliptic, strong hyperbolic or intermediate regions) can be determined.

\section{Results}
\label{sec:results}
\subsection{Flow topology and Eulerian statistics}

A series of six simulations has been performed for the classical Hasegawa--Wakatani system and two simulations have been conducted 
for the modified system.
Table~\ref{tab: Parameters} summarizes the physical parameters of the eight configurations. 
Different values of the adiabaticity are considered ranging from 0.01, corresponding to a hydrodynamic regime, via 0.7 relevant for edge turbulence in fusion plasmas and called quasi-adiabatic regime, up to values of 4.0 which yields a geostrophic regime and is similar to Hasegawa--Mima flows. For the two modified cases {\it mHW}, we consider $c =2$ and $c=4$, with the motivation to obtain zonal flows.

After some transition time, exhibiting drift-wave instabilities, all configurations reach a statistically steady regime, 
see appendix~\ref{sec:appendix_flowdynamics}.
The present study focuses on the statistically steady state obtained for different cases.

\begin{table}[!htb]
\begin{center}
\begin{tabular}{|l| c c c c |l|}
\hline
Configurations &
$\lambda=\frac{\sqrt{E_{rms}}}{{Z_{rms}}}$ & 
$R_\lambda=\lambda \frac{\sqrt{E_{rms}}}{\nu}$ &
$t_k=\frac{1}{\sqrt{2 Z_{rms}}}$ &
$\frac{t_d}{t_k}$ \\
& & & & \\
\hline
$c   =0.01~cHW$ & $1.53$ & $679$    & $0.49$ & $307$ \\
$c   =0.05~cHW$ & $1.09$ & $432$    & $0.39$ & $388$ \\
$c   =0.10~cHW$ & $0.95$ & $341$    & $0.37$ & $404$ \\
$c   =0.70~cHW$ & $0.74$ & $218$    & $0.35$ & $427$ \\
$c   =2.00~cHW$ & $0.82$ & $225$    & $0.42$ & $354$ \\
$c   =4.00~cHW$ & $1.13$ & $283$    & $0.64$ & $234$ \\
$c   =2.00~mHW$ & $3.27$ & $380$    & $3.96$ & $ 38$ \\
$c   =4.00~mHW$ & $3.14$ & $345$    & $4.03$ & $ 37$ \\
\hline
\end{tabular}
\caption{\label{tab: Parameters} Physical parameters of the {\it cHW} and {\it mHW} simulations, where $c$ denotes the adiabaticity. Root mean squares of the total energy and enstrophy are denoted respectively by $E_{rms}$ and $Z_{rms}$. The Reynolds number $R_{\lambda}$ is based on the Taylor microscale $\lambda$ and $\nu$ denotes the kinematic viscosity. The mean eddy turn over time is denoted by $t_k$ and the total duration time of the simulations by $t_d$. }
\end{center}
\end{table}

\begin{figure}[!htb]
  \begin{center}
  \begin{tabular}{cc}
    $E_{kin}(k)$ &
    $E_{n}(k)$\\
   \includegraphics[scale=0.7]{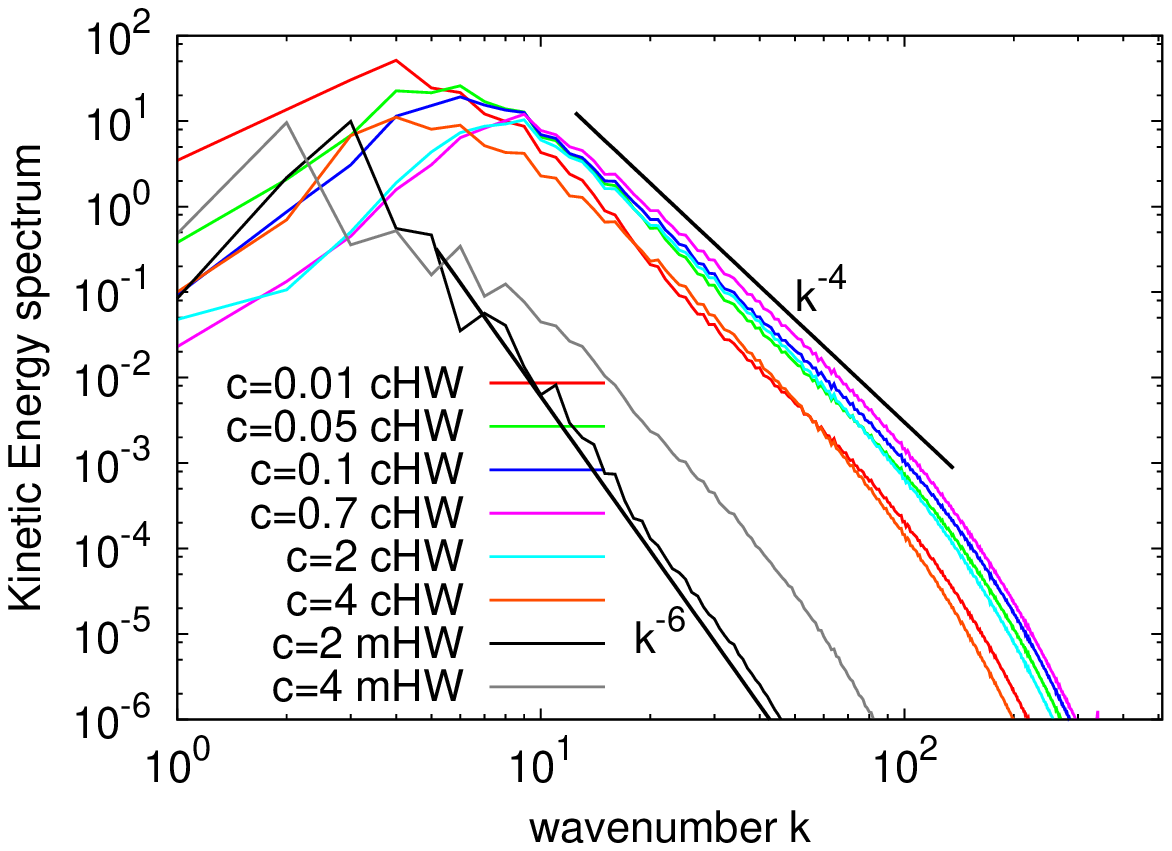} &
   \includegraphics[scale=0.7]{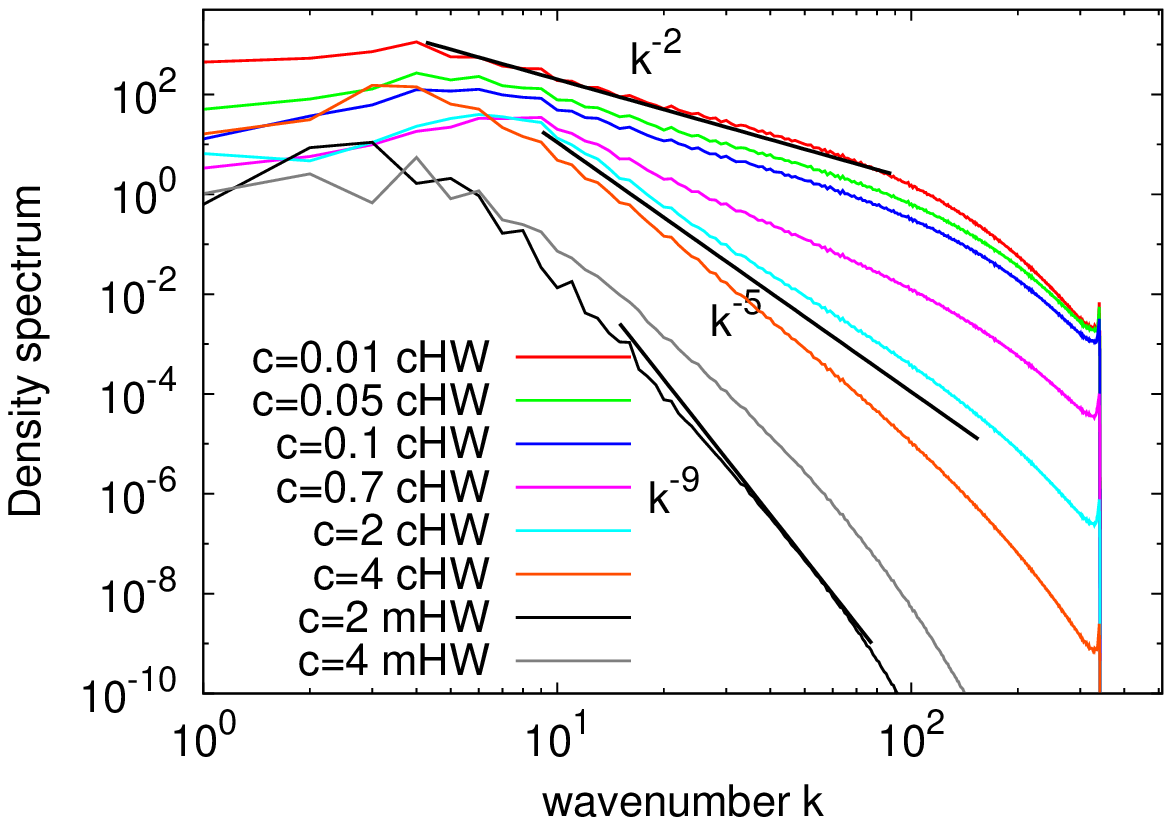}\\
    $F(k)$ &
    $\tau(k)$\\   
   \includegraphics[scale=0.7]{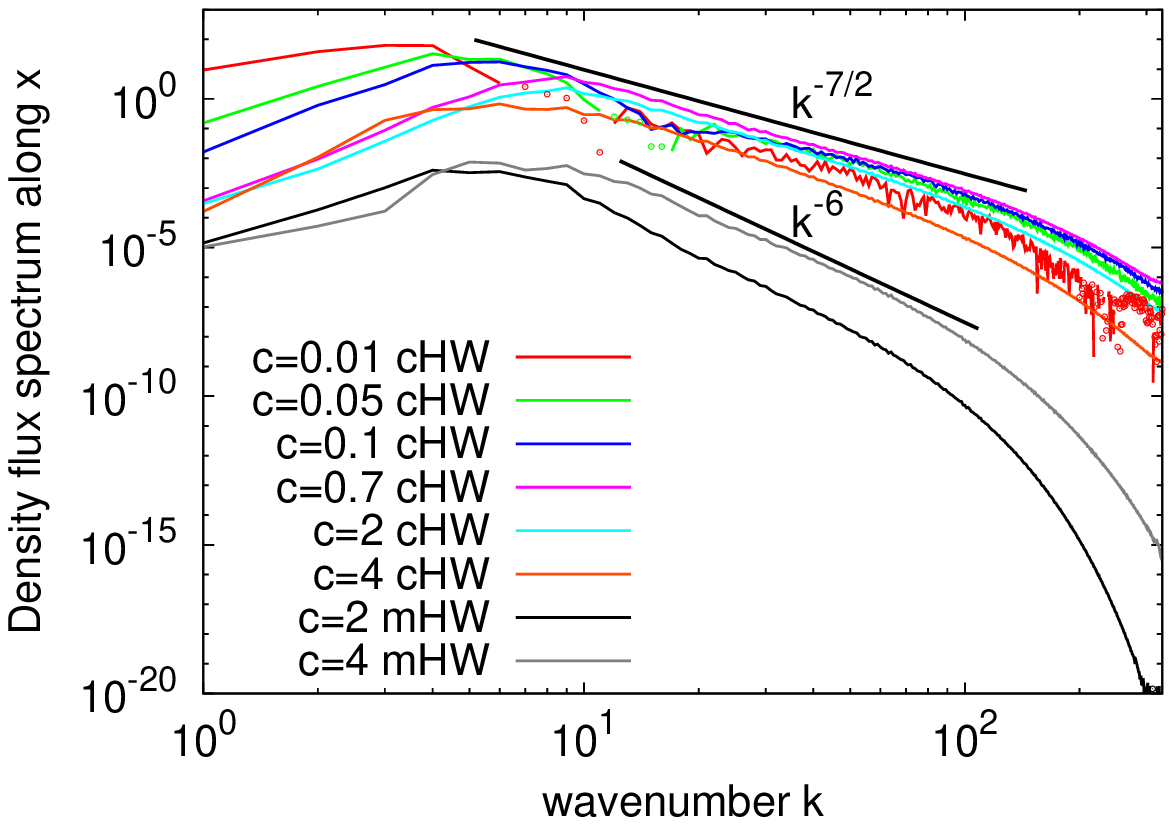} &
   \includegraphics[scale=0.7]{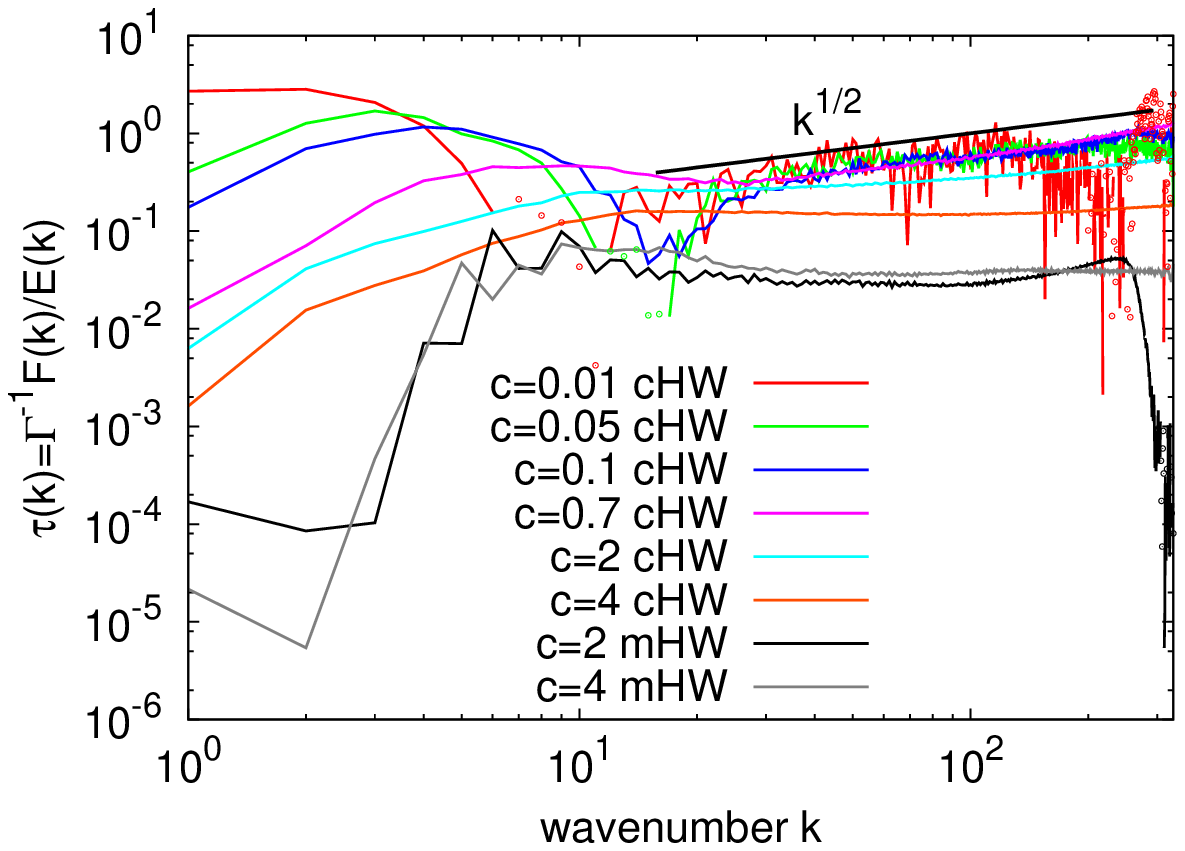}   
   \end{tabular}
\caption{Top: Eulerian kinetic spectra $E_{kin}(k)$ (left) and Eulerian spectra $E_{n}(k)$ of density fluctuations (right). 
Bottom : Density flux spectra $F(k)=F_{u_{x} n}(k)$ (left) and spectral timescale $\tau(k)=\Gamma^{-1}F_{u_x n}(k)/E_{kin}(k)$ (right). All the spectra are obtained using time-averaging with 75 snapshots taken at regular time intervals in the statistically stationary regimes. Dots in the spectra indicate negative values, plotted by taking the absolute value.}
\label{Fig: SPEC}
  \end{center}
\end{figure}

The Eulerian spectrum of kinetic energy is defined as 
\begin{equation}
E_{kin}(k,t) = \frac{1}{2} \int_{\Sigma (k)} \, {\cal F}\vert_{\bm x -  \bm x'}
[\overline{{\bm u}({\bm x},t) \cdot {\bm u}({\bm x}',t)}] \, d\Sigma(k)
\label{eq:kin_energy_spectrum}    
\end{equation}
where ${\cal F}$ denotes the 2D Fourier transform with respect to the separation vector ${\bm x -  \bm x'}$, and $\Sigma(k)$ is the circular wavenumber shell of radius $k$. The spectrum of density fluctuations $E_{n}(k,t)$ is defined correspondingly, replacing $\bm u$ by $n$.
The kinetic energy spectrum in Fig.~\ref{Fig: SPEC} (top, left) quantifies the contribution of the different length scales (or wavenumbers) and exhibits a $k^{-4}$ power law behavior for all considered {\it cHW} cases. 
This is consistent with previous work, see {\it e.g.} \cite{bcs1994}.
The two {\it mHW} cases yield likewise clear power laws, however the slope is steeper and found to be close to $-6$.
For the spectra of the density fluctuations also power laws can be observed (Fig.~\ref{Fig: SPEC} top, right), but their slopes differ and vary from $-2$ in the hydrodynamic case (which is different from the classical Batchelor scaling corresponding to $-1$) to $-5$ and even $-9$ for the largest adiabaticity value $c=4$ for classical and modified HW, respectively. For {\it cHW} this is consistent with what is found in \cite{Umansky_POF_2011} for large azimuthal wavenumbers, {\it i.e.}, a power law close to $\propto k^{-6}$ using the BOUT code \cite{XuCo1998}.   \\


Similar to what can be done in the context of passive scalar with stationary mean scalar gradient in incompressible Navier--Stokes turbulence \cite{bos2009inertial}, we now consider the mean density flux $\overline{u_x n}$ where $u_x$ and $n$ are the radial velocity fluctuations and the density fluctuations, respectively. 
Note that for large collisionality, {\it i.e.} for small $c$ values the density equation (\ref{hw2}) corresponds indeed to a passive scalar equation with mean scalar gradient $\Gamma u_x$.
The mean flux contribution at the different length scales can be quantified considering the density flux spectrum, likewise called co-spectrum in the literature, see {\it e.g.} \cite{bos2009inertial} and references therein. It is defined similar as the energy spectrum (\ref{eq:kin_energy_spectrum}) by computing the Fourier transform of the correlation between $u_x$ and $n$,
\begin{equation}
F_{u_x n}(k,t) = \frac{1}{2} \int_{\Sigma (k)} \,  {\cal F}\vert_{\bm x -  \bm x'}
[\overline{{u_x}({\bm x},t) \, {n}({\bm x}',t)}] \, d\Sigma(k)
\label{eq:flux_spectrum}    
\end{equation}
The spectrum $F_{u_x n}$ is real valued and by construction we have
\begin{equation}
     \overline{u_x \, n} = \int_{0}^{\infty} \, F_{u_x n}(k) \, dk
    \label{eq:flux}
\end{equation}
which nicely illustrates that the flux density $dF(k) = F_{u_x n}(k) \, dk$ quantifies the contributions of the mean density flux across different length scales (or wavenumbers). 

Interestingly the co-spectrum naturally connects the Eulerian and the Lagrangian flow description, as the density field yields to some extend insight into the Lagrangian dynamics of the velocity field. 
The Lagrangian time scale relates the energy spectrum with the scalar flux spectrum \cite{bos2006single} and thus provides a link between the scalar field and the Lagrangian dynamics of the turbulent velocity field.
The Lagrangian spectral timescale $\tau(k)$ can be defined by
\begin{equation}
    \tau(k) = \frac{\Gamma^{-1} F_{u_x n}(k)}{E_{kin}(k)}
\label{eq:scalarfluxspectrum}
\end{equation}
and dimensional analysis yields a spectrally-local estimate, 
\begin{equation}
\tau(k) \propto [k^3 E_{kin}(k)]^{-1/2}
\label{eq:scalarfluxspectrum_estimation}
\end{equation}
which was proposed in \cite{bos2009inertial}. 
In our case the scalar field corresponds to the density fluctuations and the above results
shall be valid at least in the limit of small $c$ values in the {\it cHW} case.
We thus obtain an estimation for the scaling of the density flux in the inertial range,
\begin{equation}
    F_{u_x n}(k) \propto  \Gamma \; \sqrt{\frac{E_{kin}(k)}{k^3}}.
    \label{eq:fluxscaling}
\end{equation}
{From a physical point of view the timescale $\tau(k)$ corresponds to a correlation time. The decorrelation of turbulent scales in an inertial range is generally due to straining or shearing. If we consider a {\it steep} kinetic energy distribution, this straining at a scale $k$ is not only due to scales around $k$ but is also due to the interaction with all scales with wavenumbers smaller than $k$ \cite{kraichnan1971inertial}. This is taken into account by using instead of a spectrally local estimate, a `straining' time 
\begin{equation}
\tau(k) \propto \left[ \int_{k_0}^{k} \, p^2 E_{kin}(p) \, dp \right]^{-1/2}.   
\label{eq:straining}
\end{equation}
In particular this time behaves qualitatively differently from the local timescale for  kinetic energy spectra $E_{kin}(k) \propto k^{-n}$ for $n>3$. Note that $k_0$ is a cut-off wavenumber to remove the infrared divergence. Using the dimensional analysis in eq.~(\ref{eq:scalarfluxspectrum_estimation}) the time scale would satisfy $\tau(k) \sim k^{(n-3)/2}$, while the `straining' time estimate in eq.~(\ref{eq:straining}) yields $\tau(k) \sim k^{0}$, valid for steep kinetic energy spectra. The latter implies also that the scalar flux spectrum and the kinetic energy spectrum satisfy the same scaling, $F_{u_x n}(k) \propto E_{kin}(k)$. }

Inspecting Fig.~\ref{Fig: SPEC} (bottom, left) shows that for all {\it cHW} cases we obtain a clear scaling behavior of $k^{-7/2}$ for the density flux spectrum as predicted by eq.~(\ref{eq:fluxscaling}) with the observed $k^{-4}$ scaling of the kinetic energy spectrum (Fig.~\ref{Fig: SPEC} (top, left)). For $c=0.7$, the relevant case for edge turbulence in fusion plasmas, this scaling is particularly well pronounced.
The spectral time scale $\tau(k)$ shown in Fig.~\ref{Fig: SPEC} (bottom, right) exhibits the corresponding $k^{1/2}$ scaling, according to eq.~(\ref{eq:scalarfluxspectrum_estimation}).
Note that the rough estimation, based on Cauchy--Schwarz inequality and proposed by Smith \etal \, \cite{smith2002turbulent} yields $F(k) \propto E_{kin}(k)^{1/2} E_n(k)^{1/2}$. It would thus predict slopes ranging from $-3$ to $-9/2$ for the {\it cHW} cases, which are clearly not found, as it was already the case for 2D Navier--Stokes turbulence \cite{bos2009inertial}.
For {\it mHW} we find a scaling of $F_{u_x n}(k) \propto k^{-6}$ which is steeper than the expected $-9/2$ slope and thus cannot be justified by eq.~(\ref{eq:fluxscaling}).
The corresponding spectral time scale $\tau(k)$ in Fig.~\ref{Fig: SPEC} (bottom, right) shows a constant behavior, {\it i.e.} a horizontal line for more than one decade, and is hence independent of the wavenumber. This is in agreement with the `straining' time scale estimate for steep kinetic energy spectra. 
The observed $k^{-6}$ scaling of the density flux spectra, {\it i.e.}, we have $F_{u_x n}(k) \propto E_{kin}(k)$ can therefore be justified in retrospect.
\\

\begin{figure}[!htb]
  \begin{center}
  \begin{tabular}{cc}
   \includegraphics[scale=0.7]{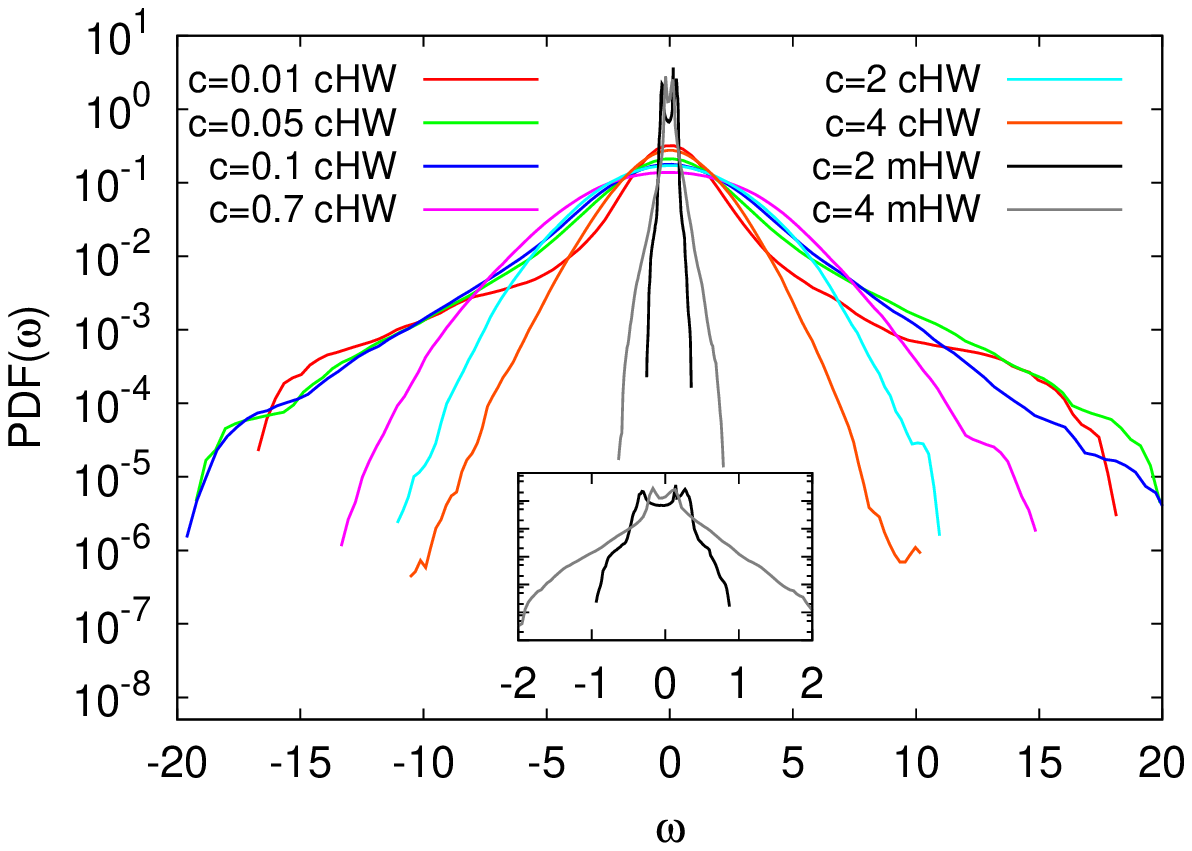}&
   \includegraphics[scale=0.7]{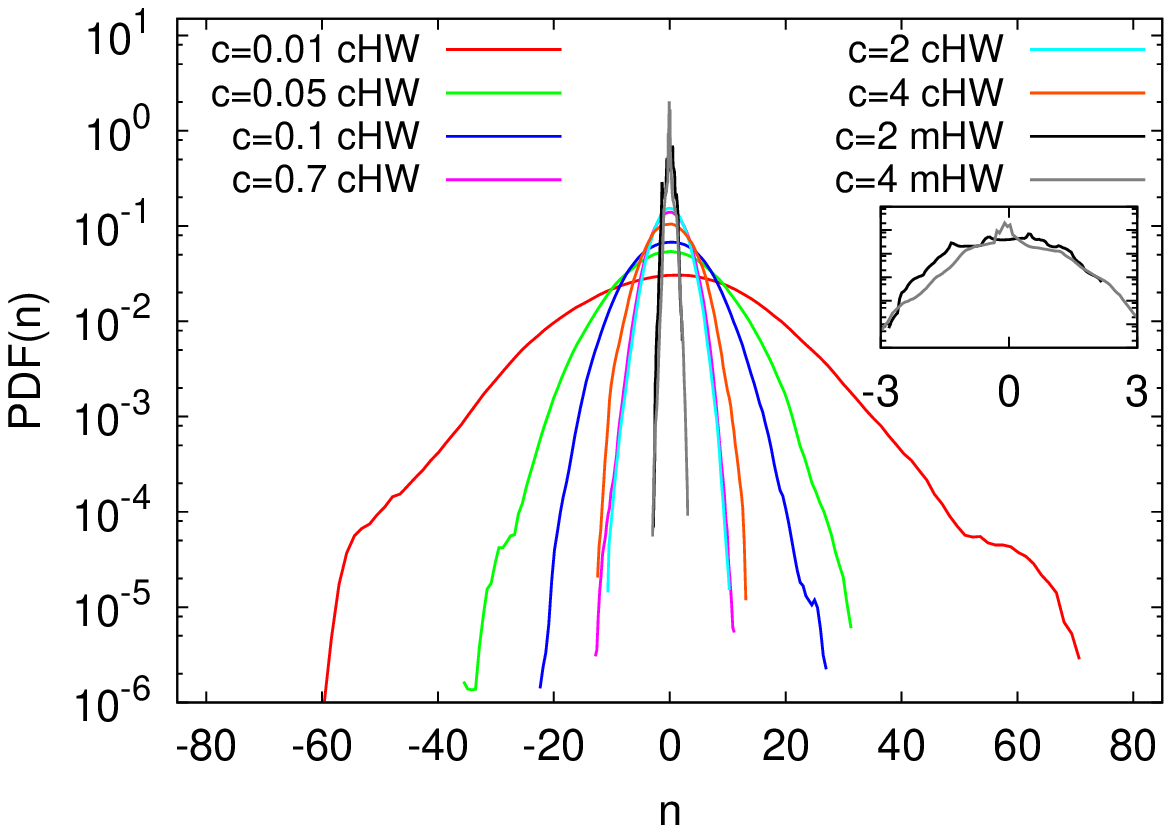}\\
   \includegraphics[scale=0.7]{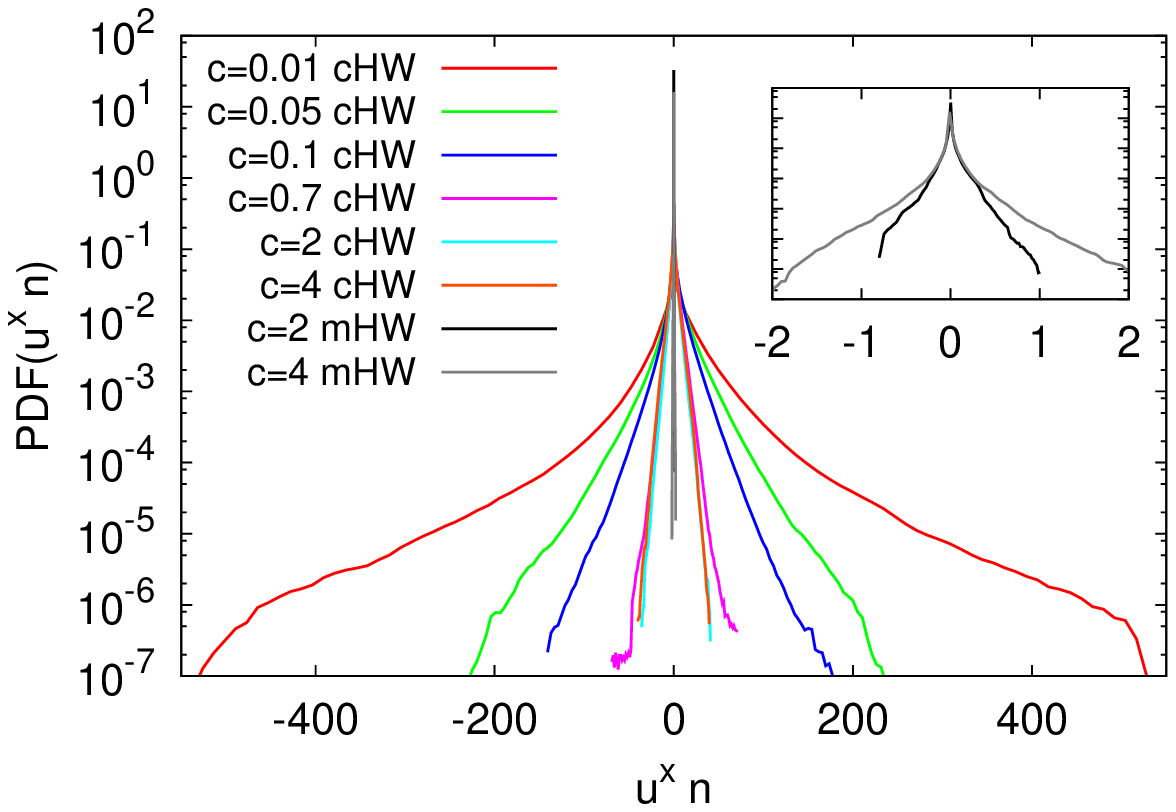}&
   \includegraphics[scale=0.7]{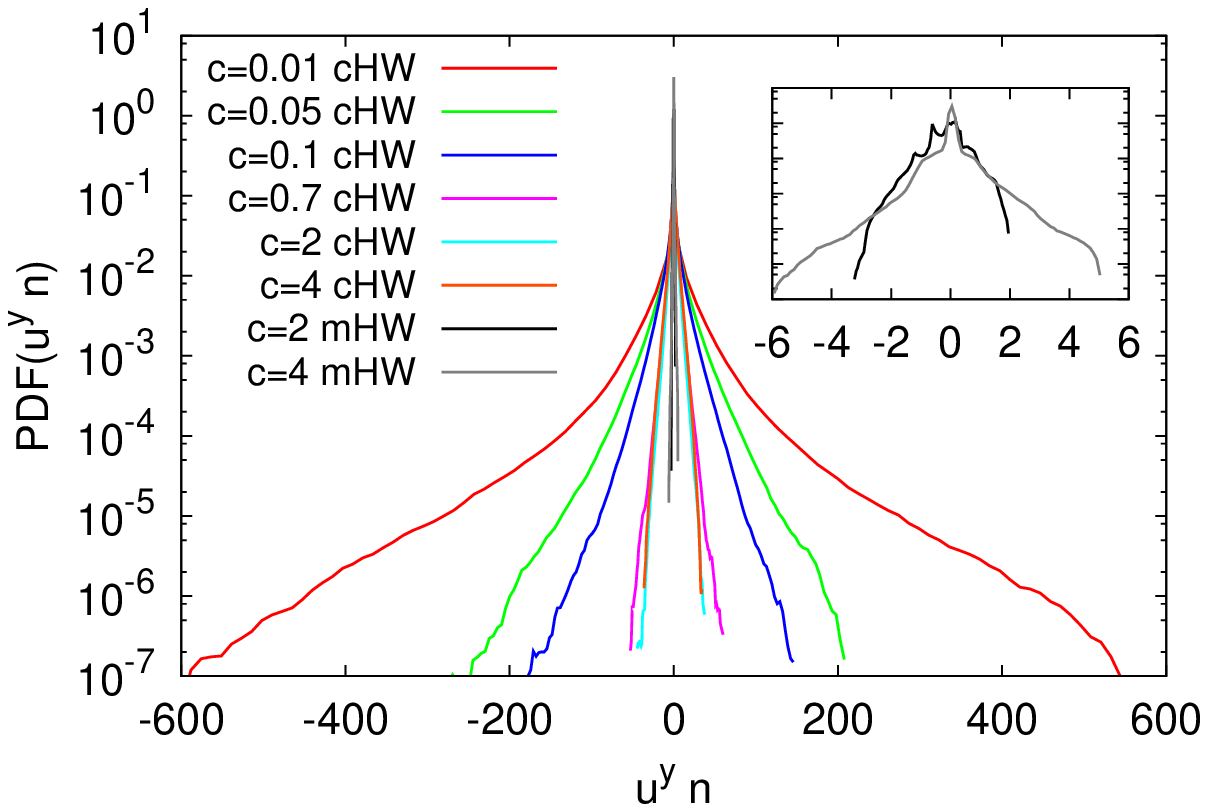}
   \end{tabular}
\caption{Top: PDFs of Eulerian vorticity (left) and density fluctuations (right). Bottom: PDFs of Eulerian density flux in $x$ and $y$-directions. The insets show zooms for the two {\it mHW} cases with $c=2$ and $4$.}
\label{Fig: PDF EUL}
  \end{center}
\end{figure}

Now we consider higher order statistics and analyze the PDFs of the different Eulerian quantities, which are computed using ensemble averaging.
To this end histograms of snapshots of the considered fields are computed using 100 bins. Subsequently time-averaging with 75 snapshots taken at regular time intervals in the statistically stationary regime is applied.
The PDFs of Eulerian vorticity, shown in Fig.~\ref{Fig: PDF EUL} (left), exhibit a change of behavior with $c$.  
For the adiabatic regime ($c=0.7$) and the case $c=2~cHW$ the PDFs have an almost Gaussian shape. The PDFs of Eulerian density in Fig.~\ref{Fig: PDF EUL} (right) appear to be Gaussian, except for the hydrodynamic regime where we can observe a small deviation from the parabola shape for negative values. Concerning the modified case (cf. the insets in Fig.~\ref{Fig: PDF EUL}), the PDFs of density and vorticity show strongly reduced fluctuations around $0$, which can be probably attributed to the presence of the shear flows.

The PDFs of Eulerian density flux along $x$ and $y$-directions are plotted in Fig.~\ref{Fig: PDF EUL} (bottom). The PDFs are almost symmetric and exhibit heavy tails for $c=0.01$. For larger values of $c$ an exponential behavior can be observed and the variance decreases with increasing $c$ and thus the range of values likewise shrinks. 
The difference is more significant for small adiabaticity values and almost insignificant for $c>0.7 $ ({\it cHW}). As a conclusion, the density fluxes have larger values and are more important for hydrodynamic regimes with extreme values much larger for $c=0.01$. For the two {\it mHW} cases (cf. the insets in Fig.~\ref{Fig: PDF EUL}) we observe strongly reduced density flux values, which again might be due to the presence of zonal flows. 

\begin{figure}[!htb]   
  \begin{center}
  \begin{tabular}{cccccc}
   \hspace{-2.5cm}
   &   
   $c   =0.01,~cHW$  &
   $c   =0.70,~cHW$  &
   $c   =4.00,~cHW$ &
   $c   =4.00,~mHW$ \\
     
   \hspace{-2.5cm} 
   $\omega$ &     
   \includegraphics[valign=c,scale=0.36]{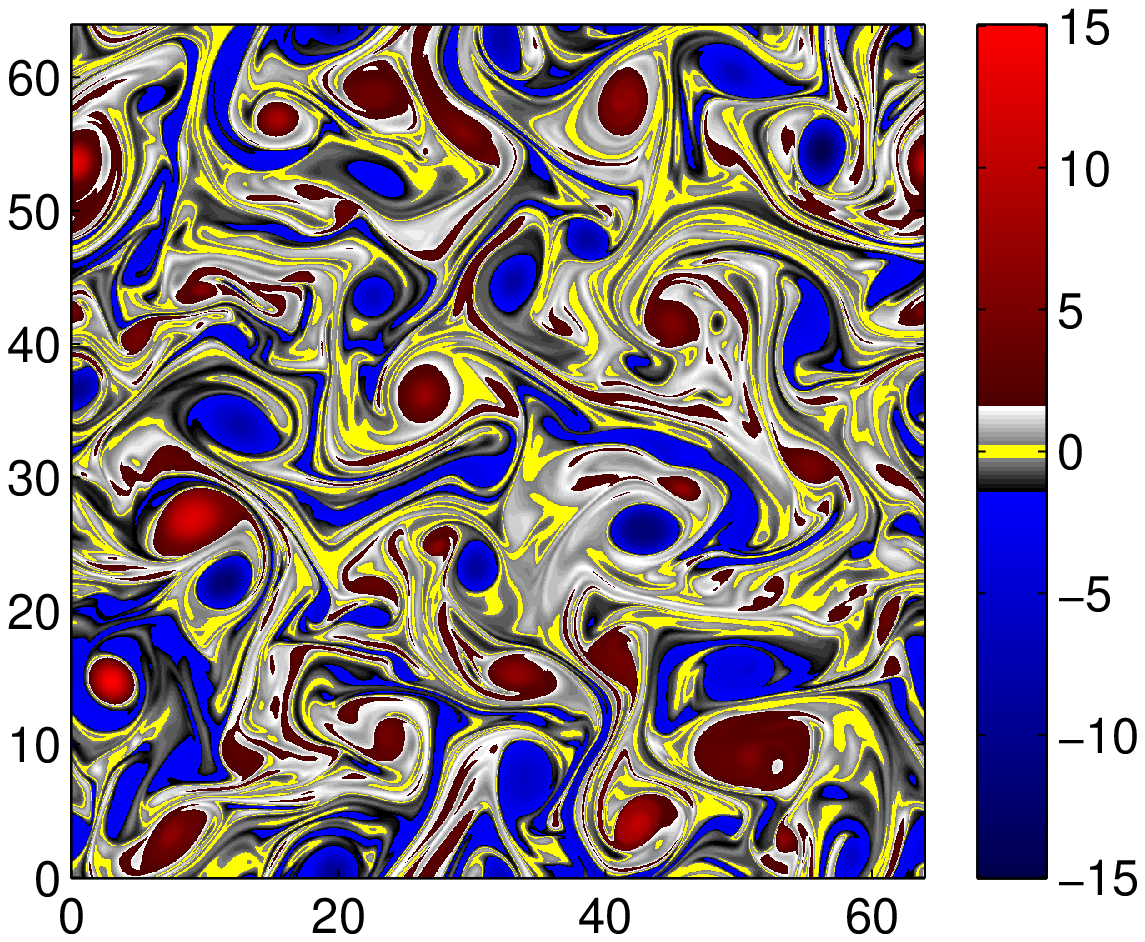}&
   \includegraphics[valign=c,scale=0.36]{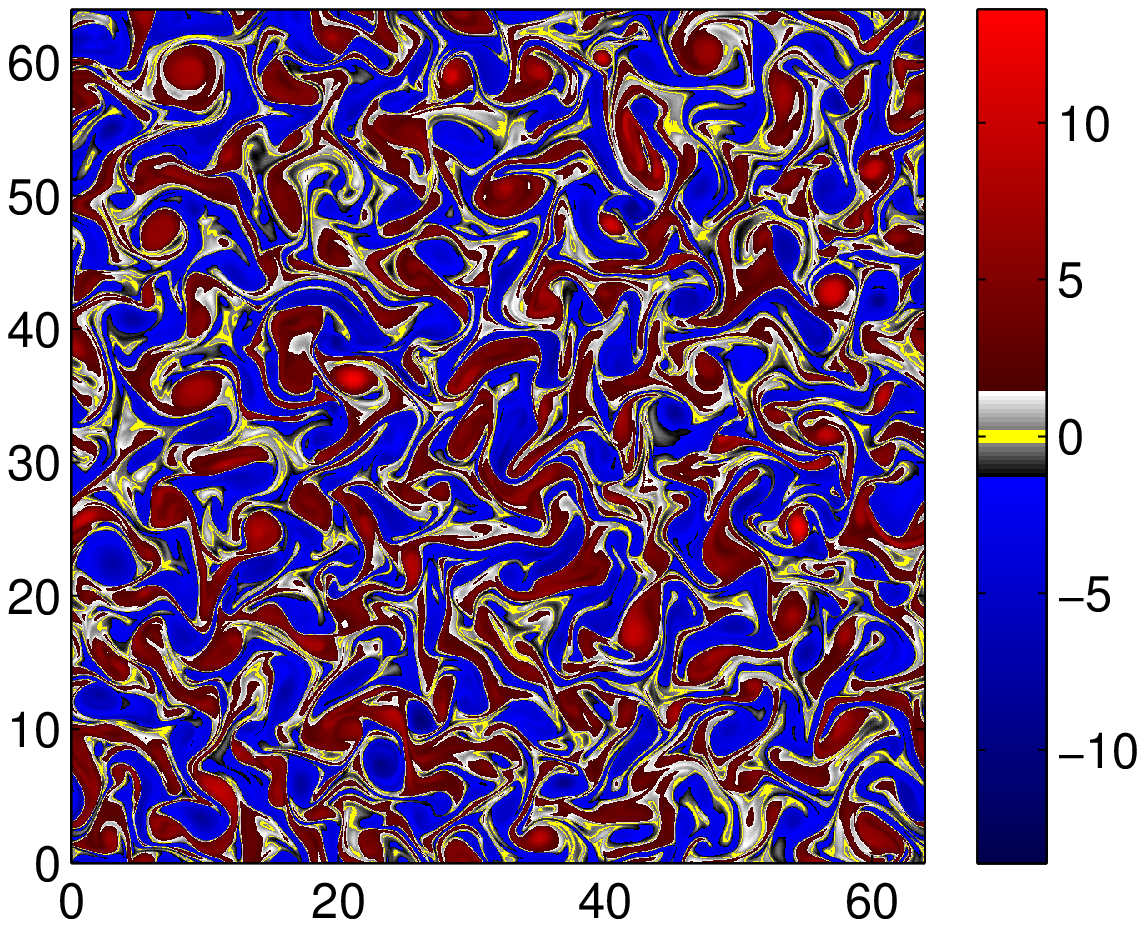}&
   \includegraphics[valign=c,scale=0.36]{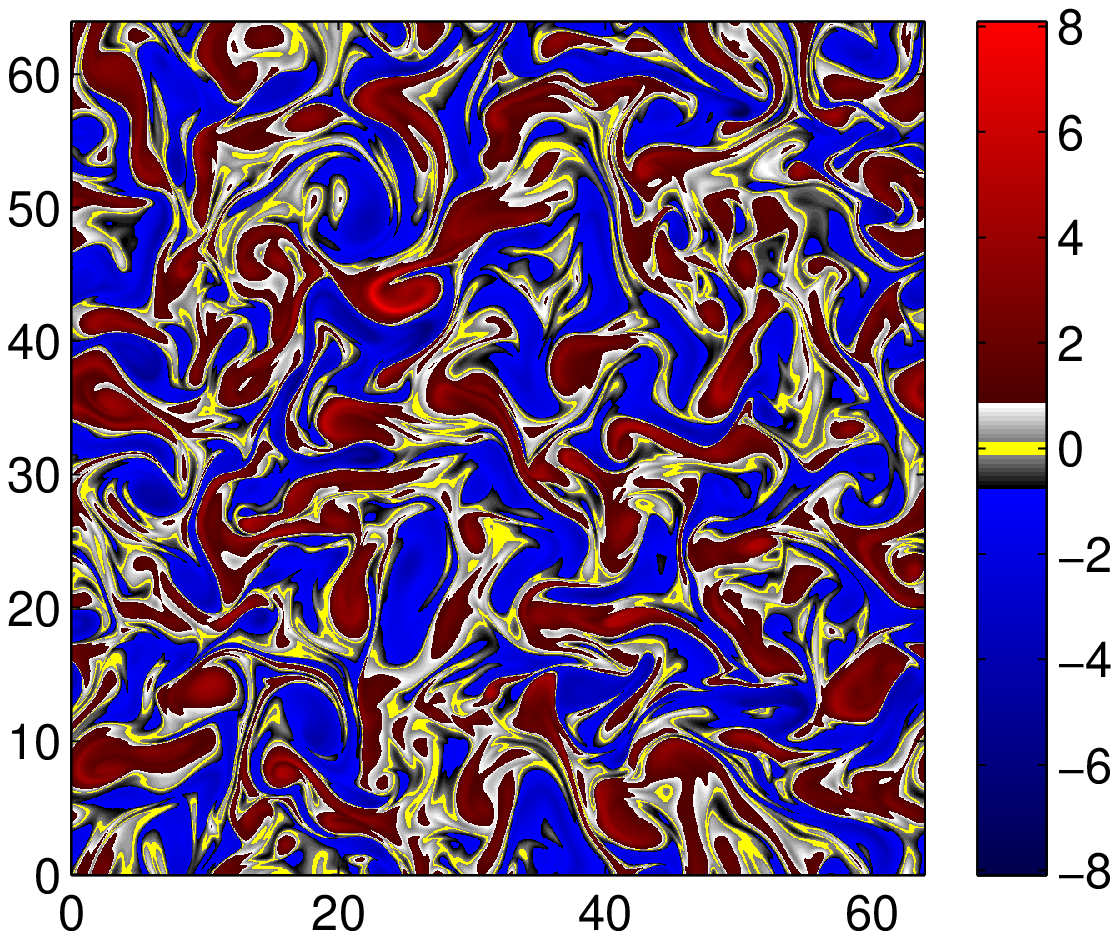}&
   \includegraphics[valign=c,scale=0.36]{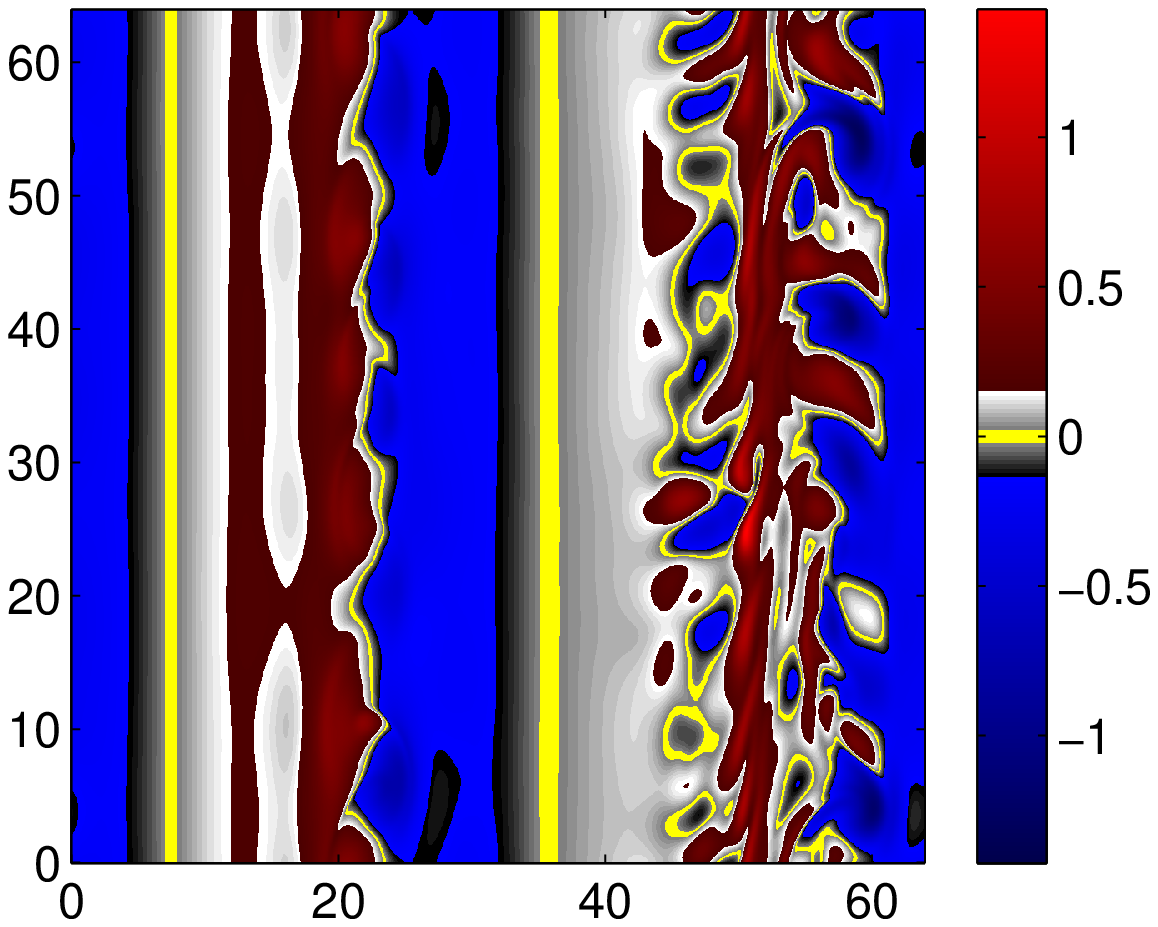}\\

   \hspace{-2.5cm}
   $n$ &     
   \includegraphics[valign=c,scale=0.36]{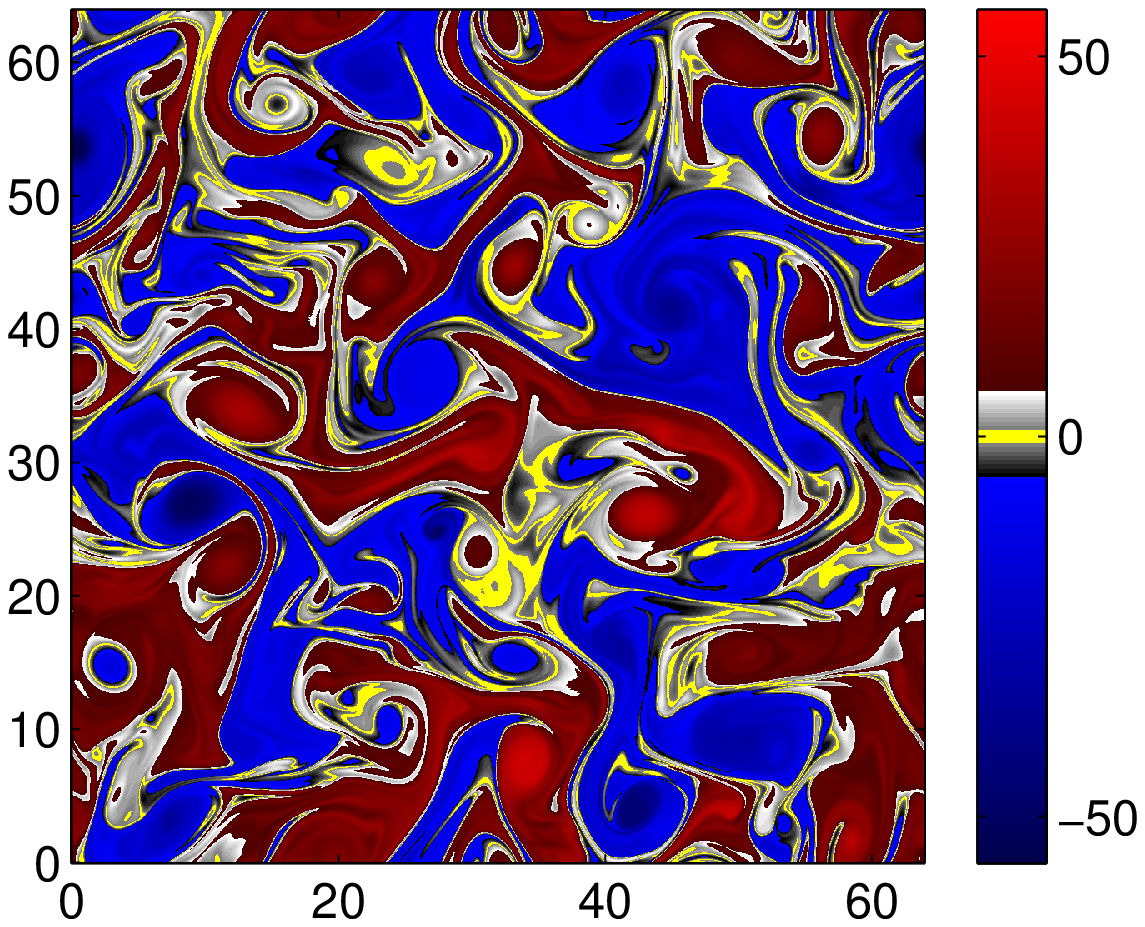}&
   \includegraphics[valign=c,scale=0.36]{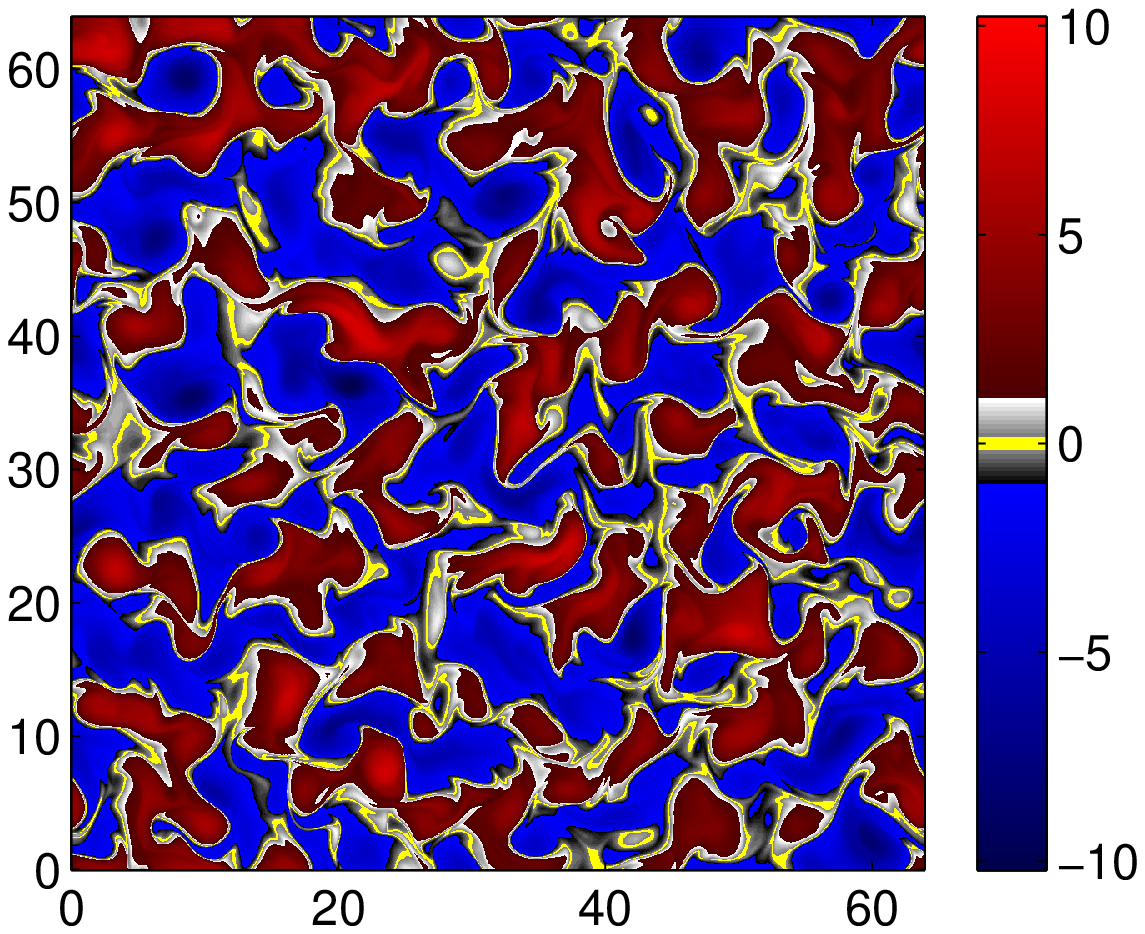}&
   \includegraphics[valign=c,scale=0.36]{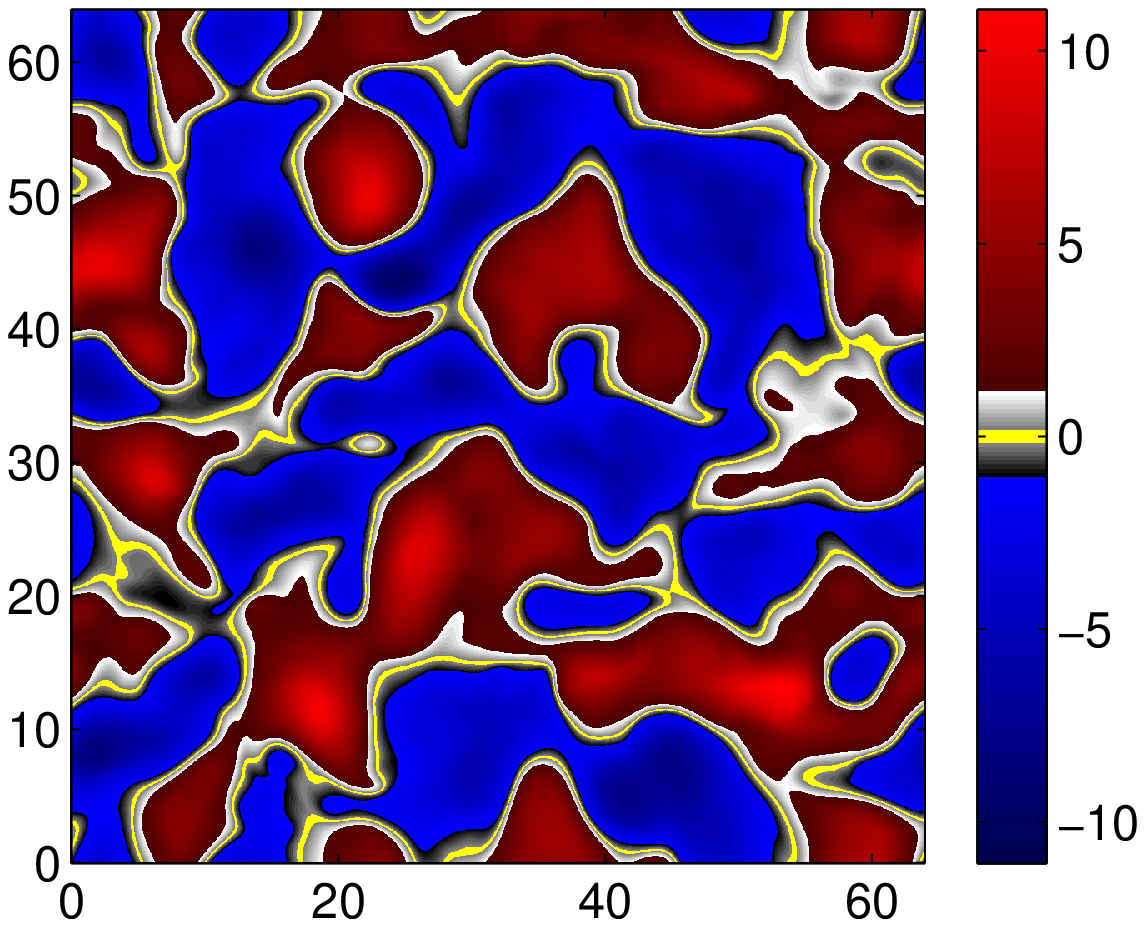}&
   \includegraphics[valign=c,scale=0.36]{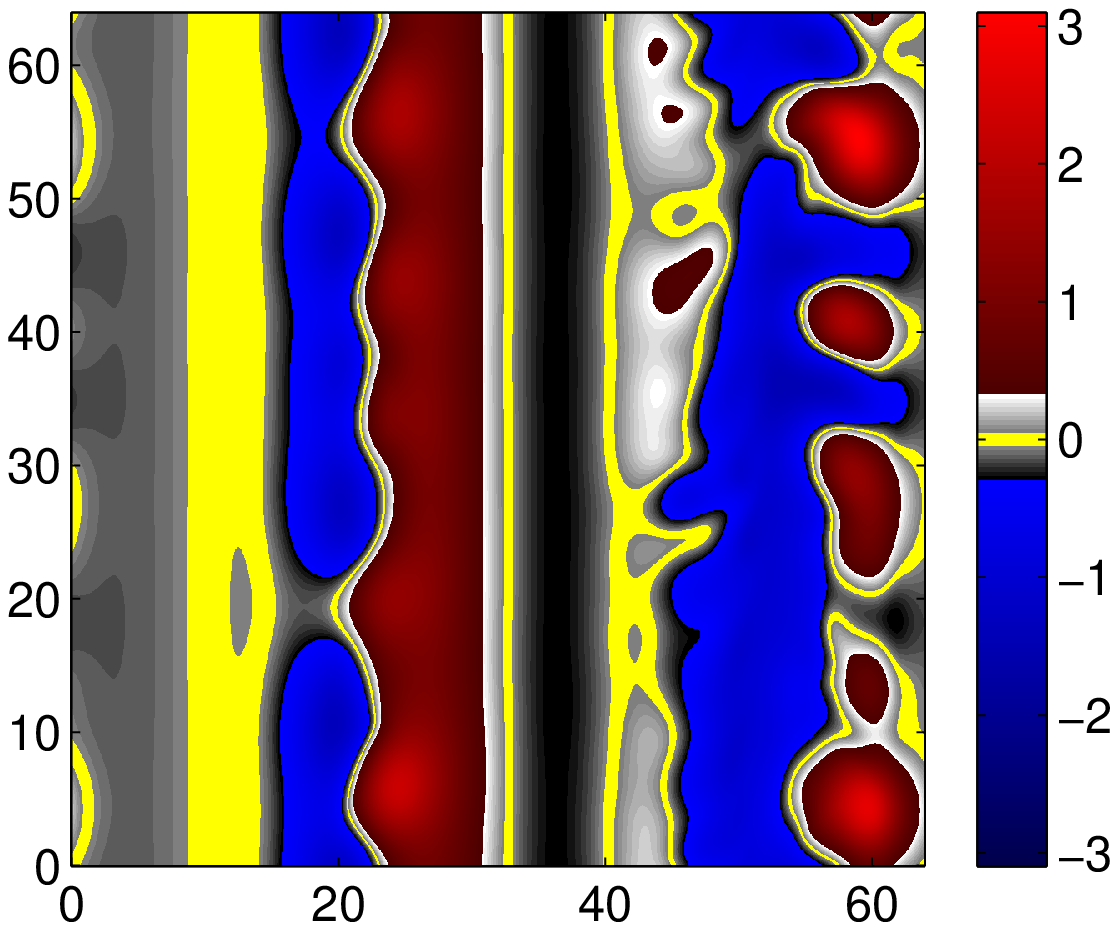}\\

   \hspace{-2.5cm}
   $Q_3$ &     
   \includegraphics[valign=c,scale=0.36]{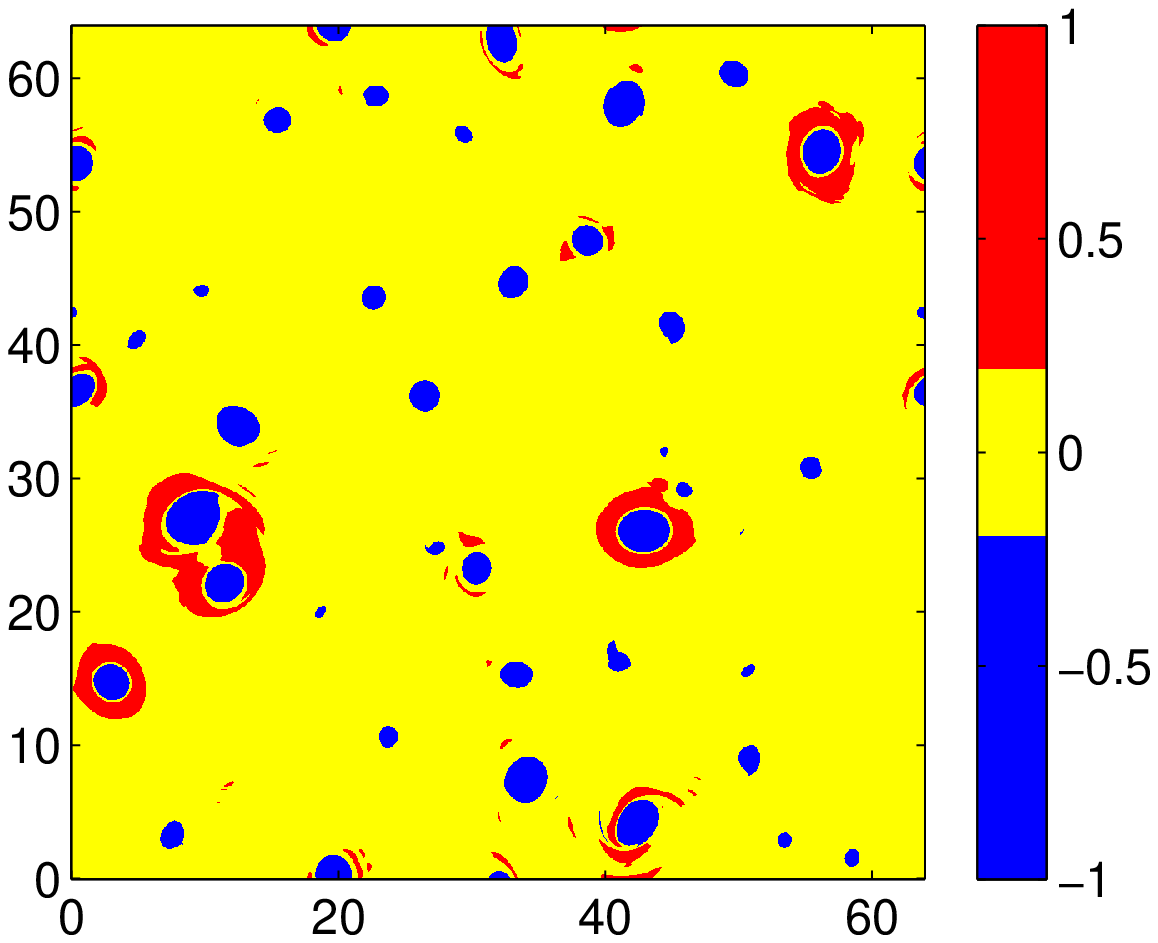}&
   \includegraphics[valign=c,scale=0.36]{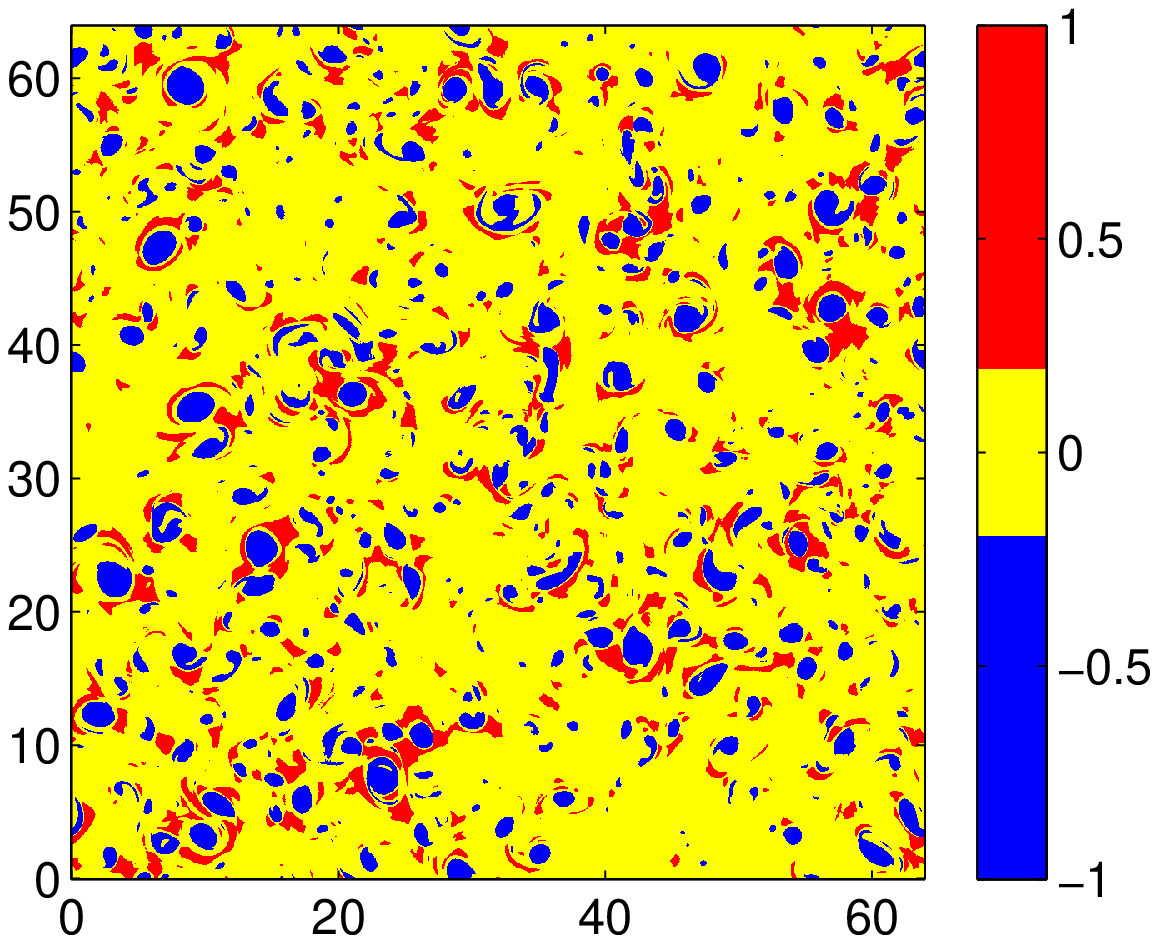}&
   \includegraphics[valign=c,scale=0.36]{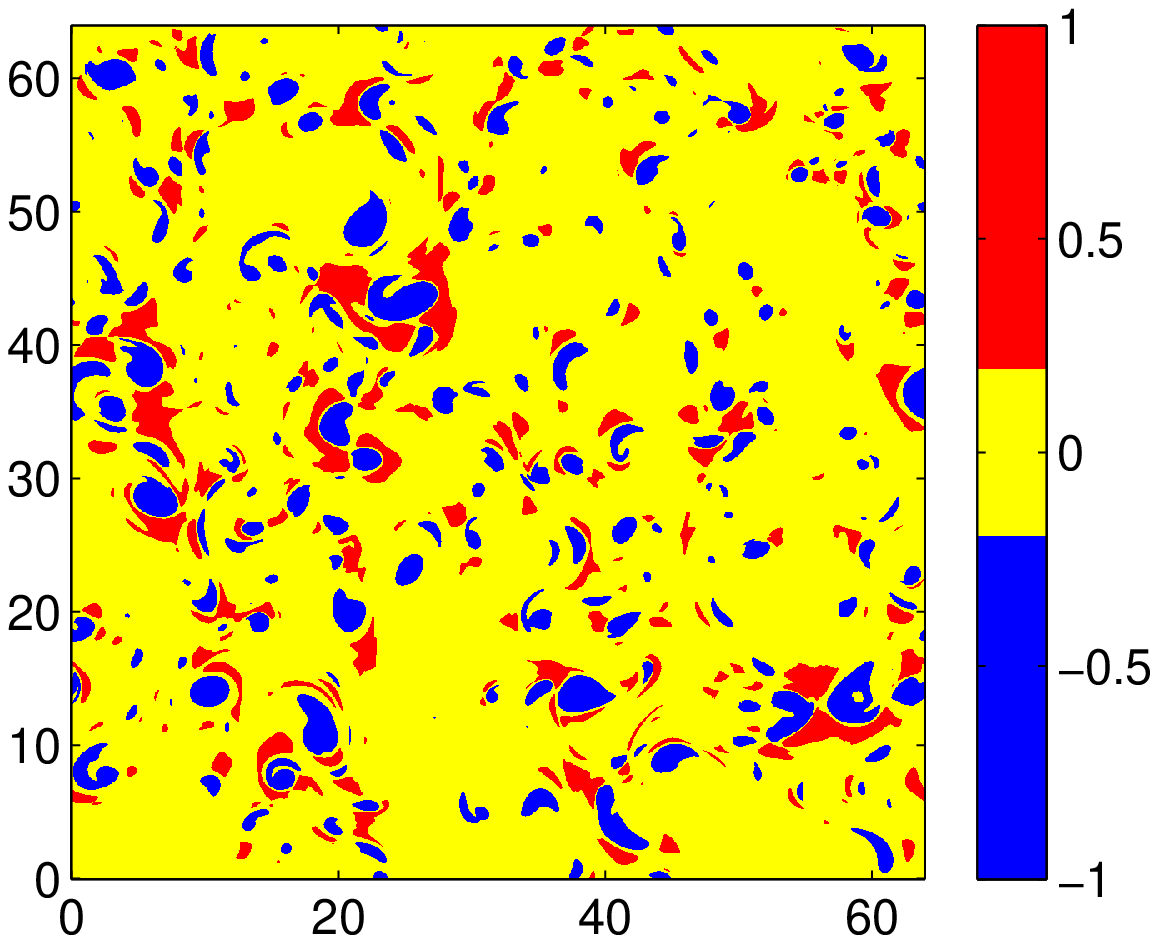}&
   \includegraphics[valign=c,scale=0.36]{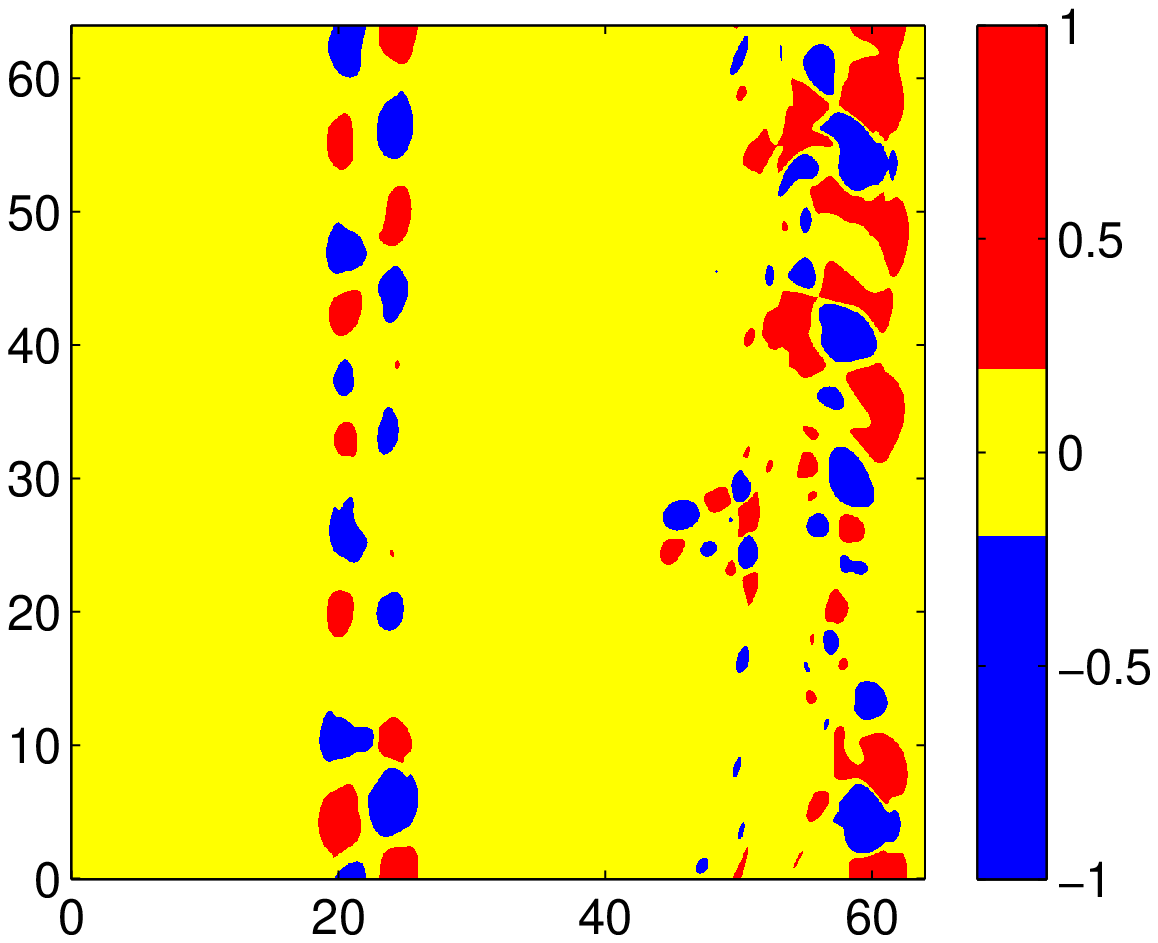}\\

   \hspace{-2.5cm}
   $u_x \, n$ &  
   \includegraphics[valign=c,scale=0.36]{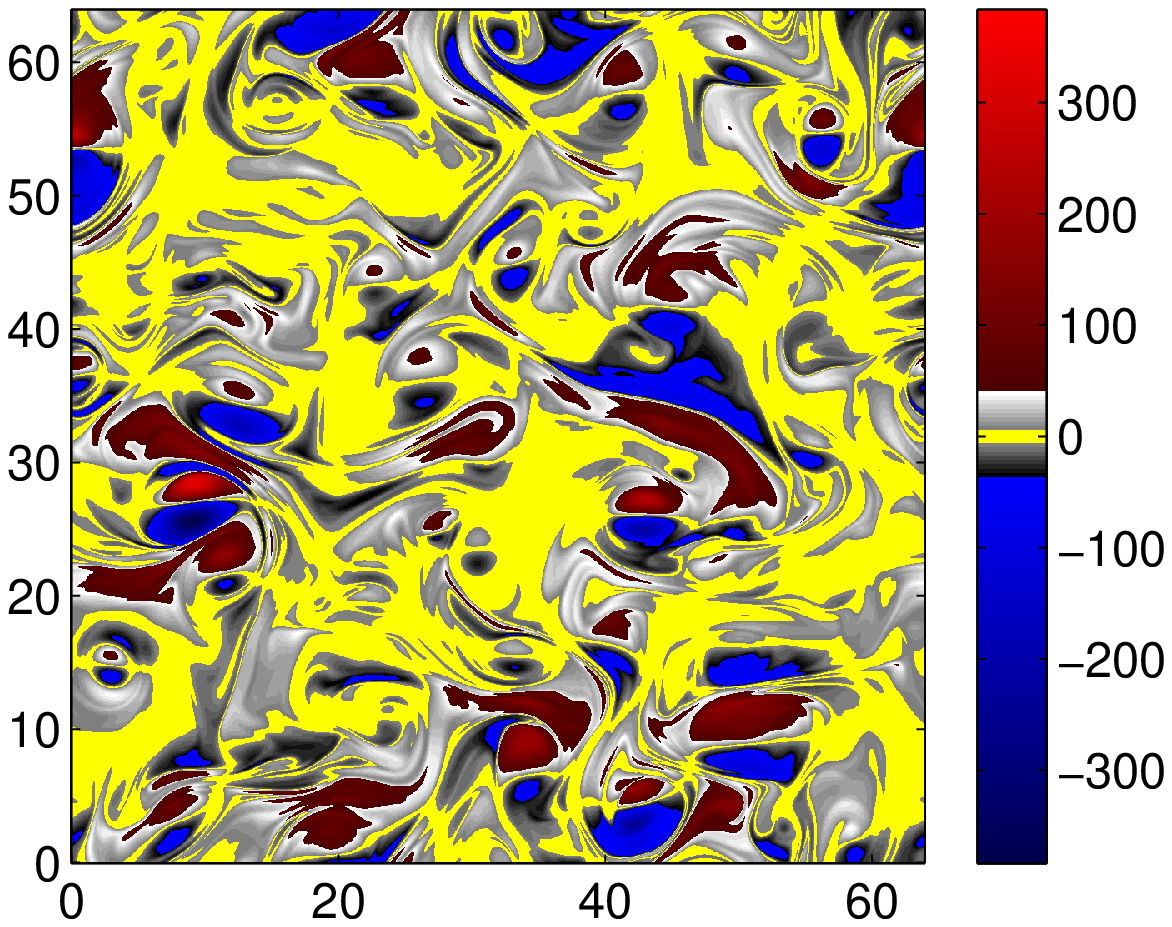}&
   \includegraphics[valign=c,scale=0.36]{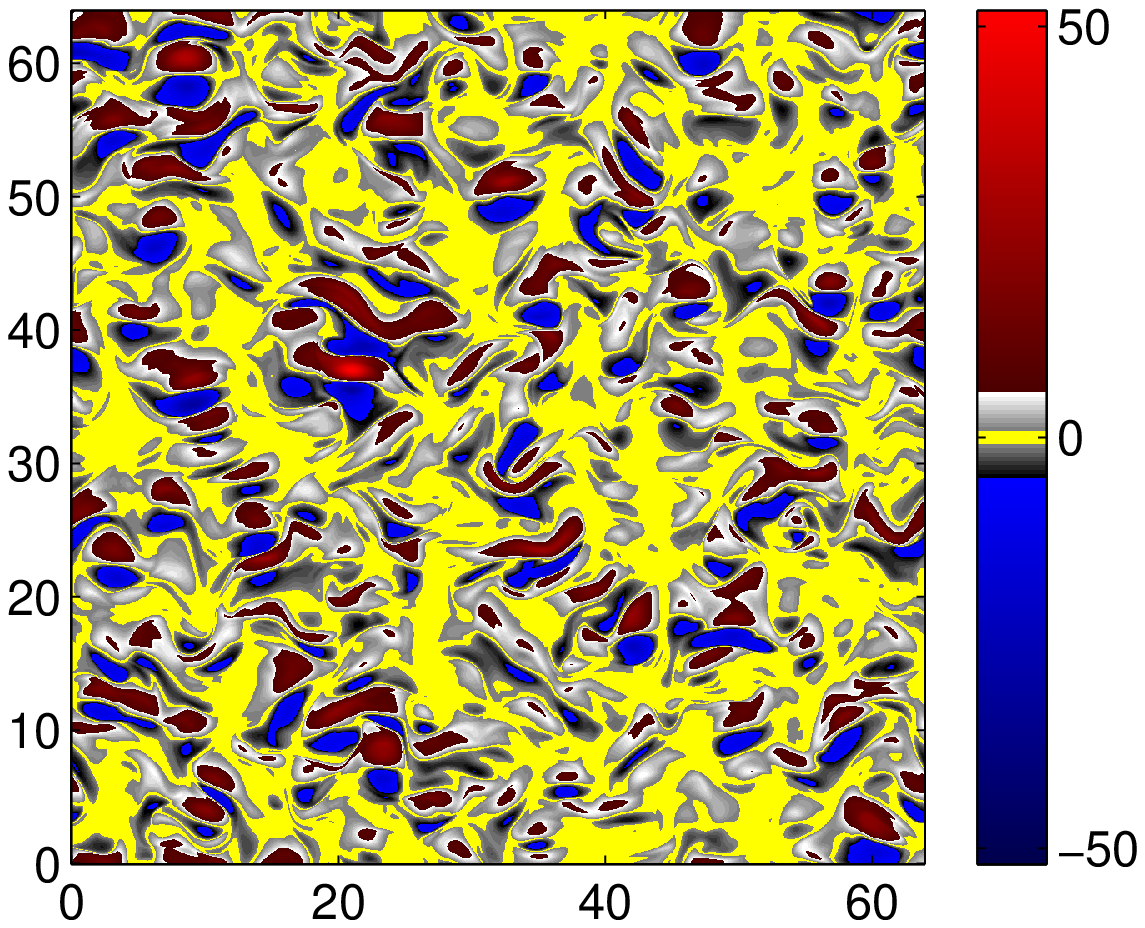}&
   \includegraphics[valign=c,scale=0.36]{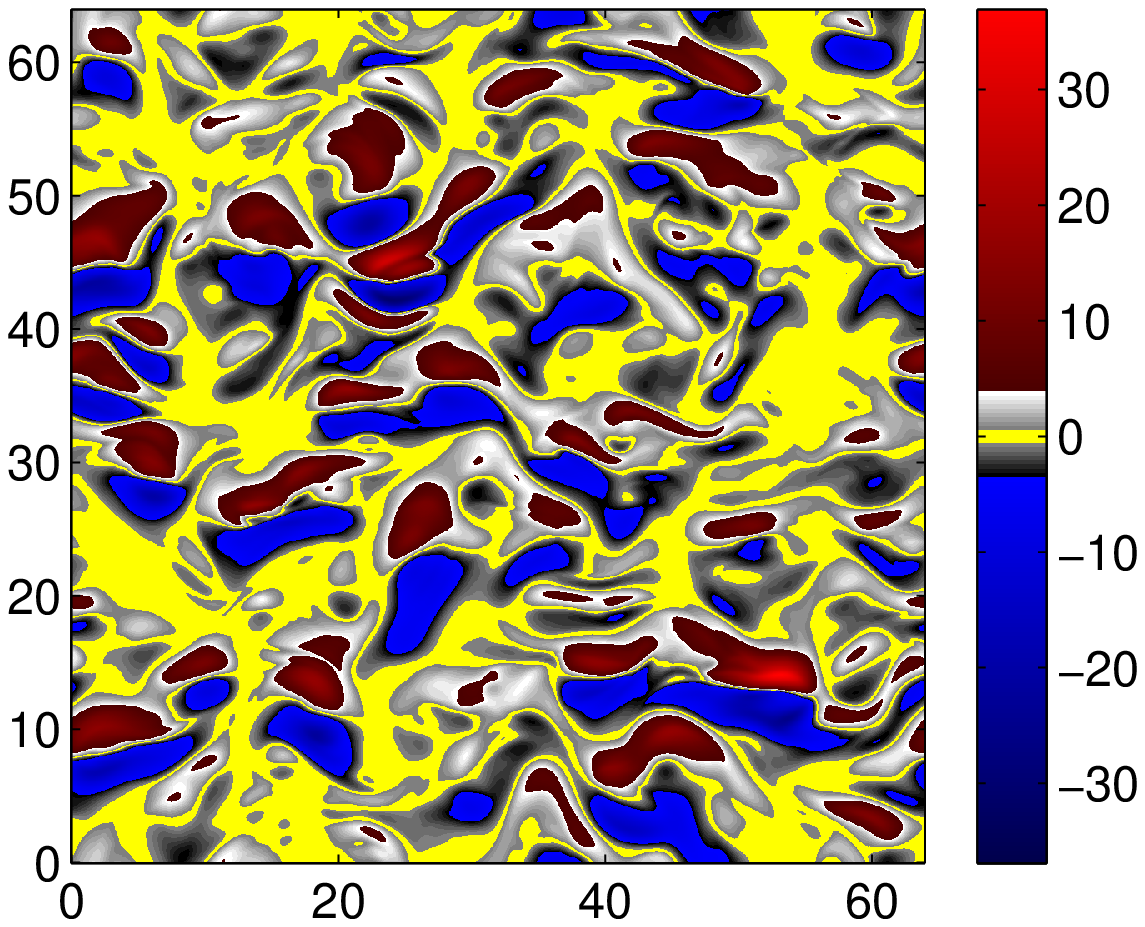}&
   \includegraphics[valign=c,scale=0.36]{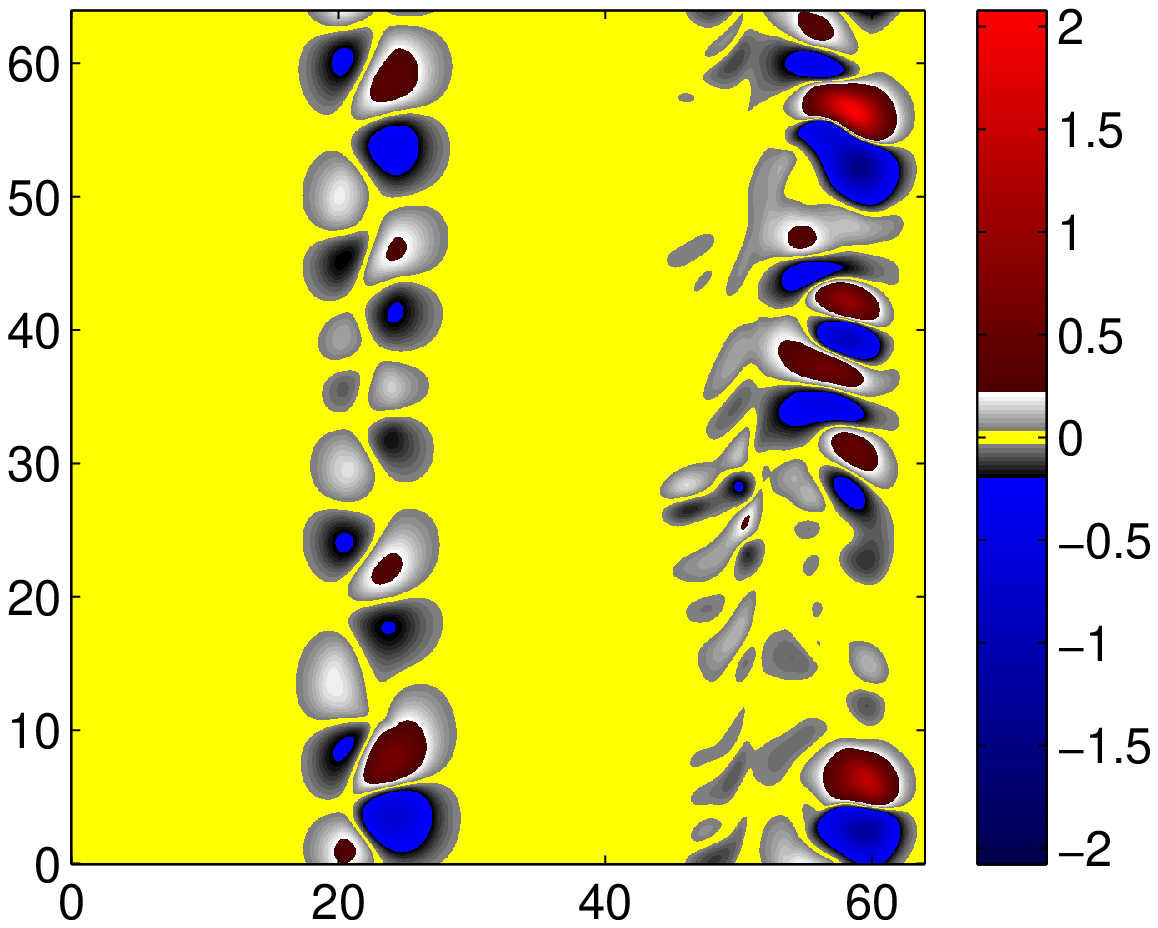}\\
   
   \hspace{-2.5cm}
   Traj. &     
   \includegraphics[valign=c,scale=0.68]{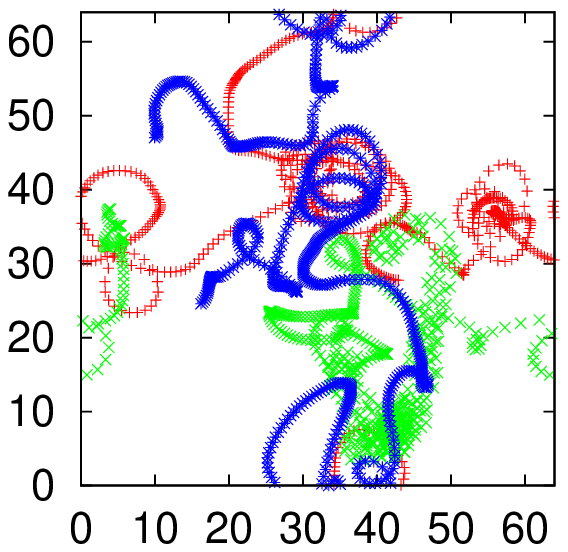}&
   \includegraphics[valign=c,scale=0.68]{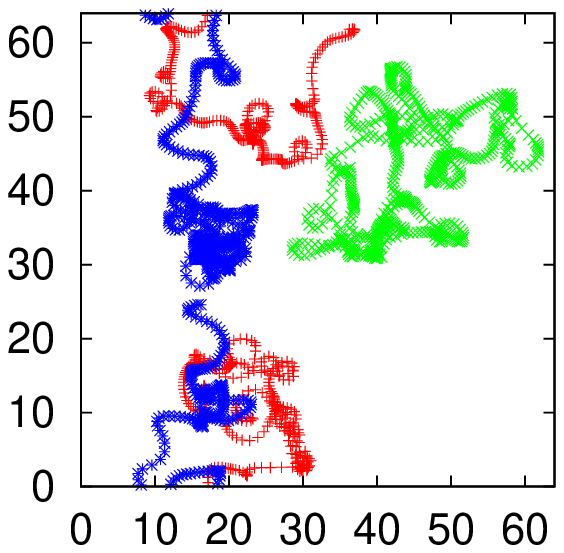}&
   \includegraphics[valign=c,scale=0.68]{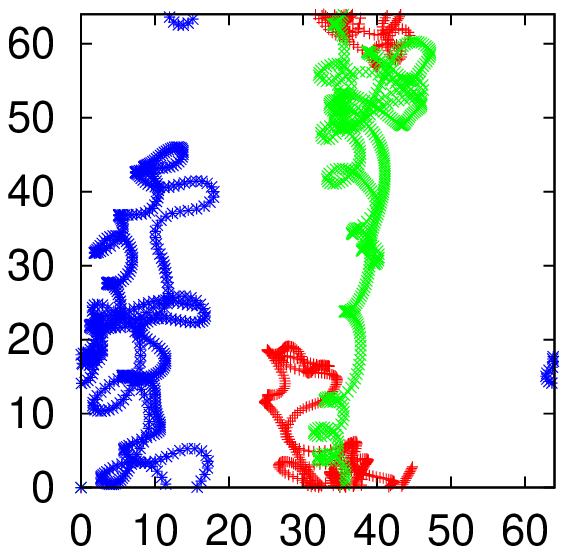}&
   \includegraphics[valign=c,scale=0.68]{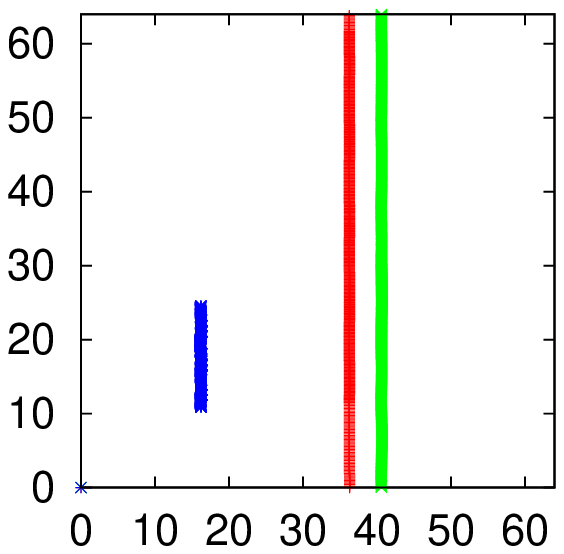}\\
   
   \end{tabular}
\caption{Flow visualizations for {\it cHW} with $c=0.01, 0.7$ and 4.0, and for {\it mHW} with $c=4.0$ (from left to right). From top to bottom: snapshots of vorticity field $\omega$, density fluctuation field $n$, three-level Weiss field $Q_3$, density flux in $x$-direction $u_x \, n$  and three typical tracer trajectories in the statistically stationary regime. Here $Q_3$ is the trinary Weiss field with values $-1, 0$ and $1$ computed from $Q$ using the threshold $Q_0$ defined in sect.~\ref{sec:weiss}.}
\label{Fig: VISU}
  \end{center}
\end{figure}

The snapshots of vorticity, density, three-level Eulerian Weiss fields and the density flux in the radial direction, in Fig.~\ref{Fig: VISU}, illustrate the different flow regimes obtained by varying the adiabaticity. The snapshots are taken at a time which is representative for a statistically stationary state. For low adiabaticity, $c = 0.01$, the flow regime is close to the ones obtained in hydrodynamic turbulence. 
The simulations for large $c$ values are computational very demanding due to a long transient phase (cf. appendix~\ref{sec:appendix_flowdynamics}). 
Zonal flows occur periodically after subsequent destabilization by shear flows.
The flow visualizations in Figure~\ref{Fig: VISU}, {\it i.e.}, vorticity, density, three-level Eulerian Weiss fields and the density flux in the radial direction also include some sample trajectories. We observe that the flow structures in the vorticity field change with $c$ and in the {\it mHW} case we observe pronounced zonal flows in the poloidal direction. This change of behavior can be likewise found in the corresponding density and density flux fields.
In the Weiss fields we can see that the elliptic regions dominate the flow fluid for low adiabaticity and become decreasingly important for increasing adiabaticity, which is in agreement with the observations in \cite{Bos2008-1}. The strong hyperbolic regions are localized in the cells which are surrounding the vortices. We can also note that the vortices with strong magnitude are circular for low adiabaticity and become thinner and more elongated for large adiabaticity. It reveals also that the strong hyperbolic regions are not concentrated anymore in cells surrounding the vortices 
for large adiabaticity values. As a consequence the strong elliptic regions are less localized. Concerning the zonal flows, strong elliptical and hyperbolic regions are located in the ascending regions and exhibit similar large shape, while only intermediate zones are present in the descending regions.
Finally, some sample trajectories of three particles for the different configurations are plotted in Fig.~\ref{Fig: VISU} (bottom). From $c=0.01~cHW$ to $c=0.1~cHW$, the trajectories are similar. For increasing $c$ values from $0.7$ on for {\it cHW}, they are more elongated. 
For modified HW, the particles exhibit almost straight line trajectories which are more pronounced for $c=4~mHW$. This is due to the presence of zonal flows which destroy the formation and presence of vortices. The particles are confined in a band with the same velocity along $y$-direction and only `few' particles are able to escape and change zones.\\

\subsection{Lagrangian statistics}

\begin{figure}[!htb]
  \begin{center}
  \begin{tabular}{cc}
\includegraphics[scale=0.7]{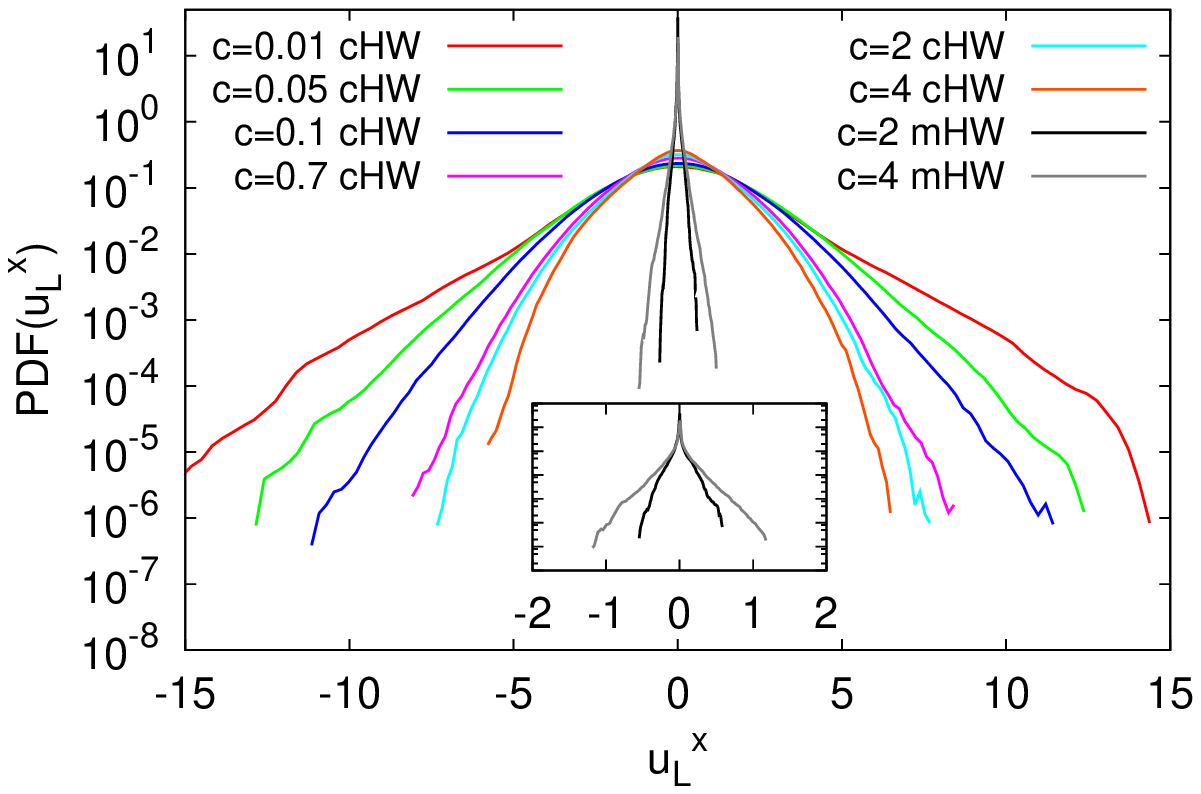}& 
\includegraphics[scale=0.7]{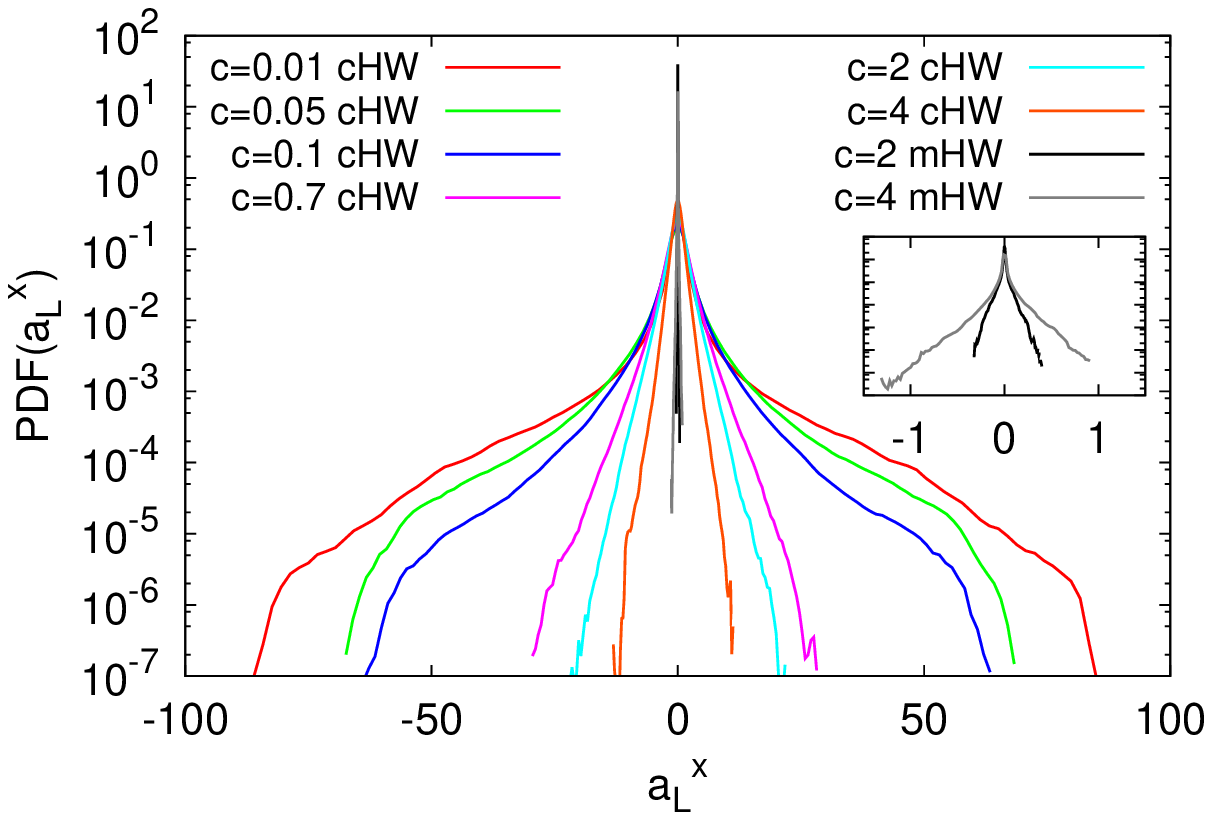}
   \end{tabular}
\caption{PDFs of the $x$-component of Lagrangian velocity (left) and Lagrangian acceleration (right). The insets show zooms for the two {\it mHW} cases with $c=2$ and $4$.}
\label{Fig: PDF VEL ACC}
  \end{center}
\end{figure}

The PDFs of Lagrangian velocity and acceleration, the latter computed using eq.~(\ref{eqaL}), shown in Fig.~\ref{Fig: PDF VEL ACC}, exhibit a change of behavior with $c$. The corresponding values of the centered second order moments and the flatness are given in table~\ref{tab: Norm Flatness STAT WEISS Stat 2}. The PDFs of Lagrangian velocity display Gaussian shape for classical HW, except for large $c$ values where small asymmetries appear for the velocity along the $y$-direction. The PDFs of Lagrangian velocity for modified HW are of a different nature with symmetric heavy tails for the $x$-direction, cf. the inset in Fig.~\ref{Fig: PDF VEL ACC} (left). 
The PDFs of Lagrangian acceleration exhibit exponential shape for large adiabaticity values (the flatness values in table~\ref{tab: Norm Flatness STAT WEISS Stat 2} are close to $6$, corresponding to a Laplace distribution) with a smaller width for increasing $c$, corresponding to reduced variance quantified in table~\ref{tab: Norm Flatness STAT WEISS Stat 2}, and heavy tails for small $c$ corresponding to large flatness values, which is in agreement with the findings in \cite{Bos2010-1}. For $c=2~mHW$, shown in the inset in Fig.~\ref{Fig: PDF VEL ACC} (right), some asymmetry does appear for the $x$-direction. \\

\begin{figure}[!htb]
  \begin{center}
  \begin{tabular}{cc}
   \includegraphics[scale=0.7]{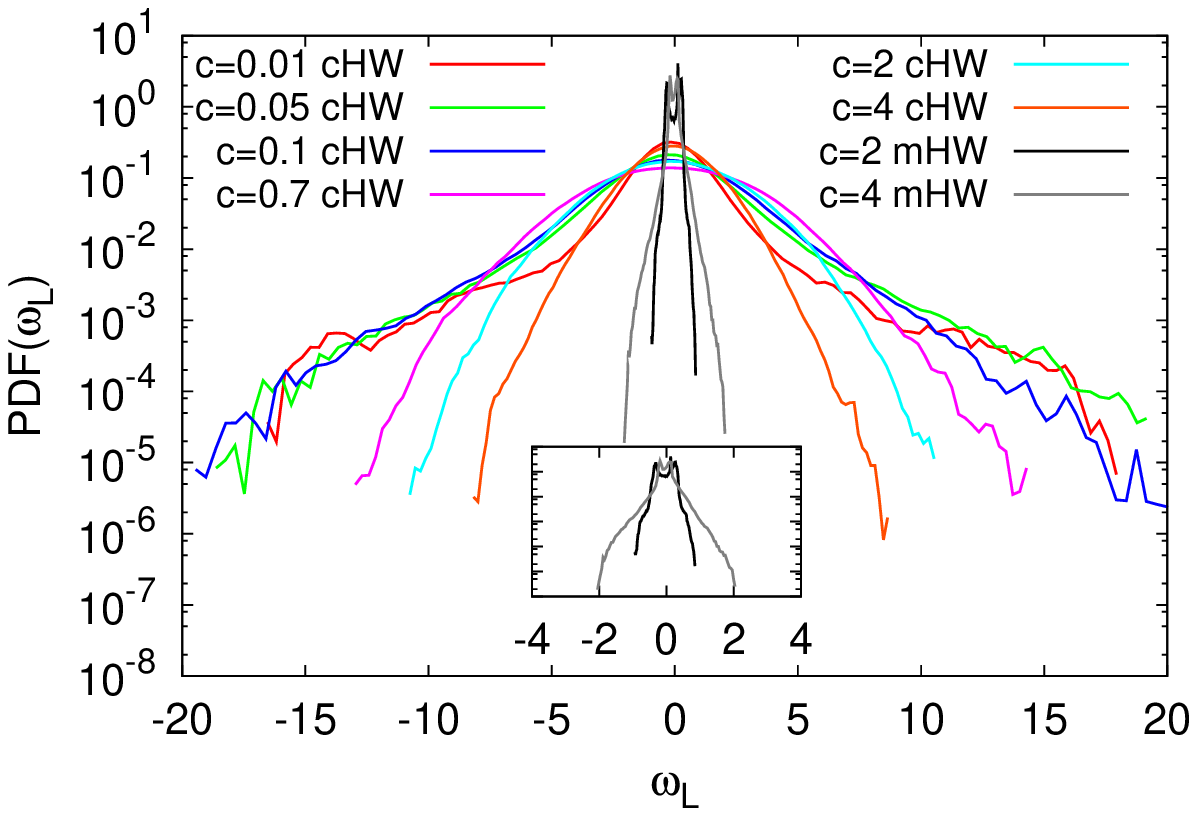}&
   \includegraphics[scale=0.7]{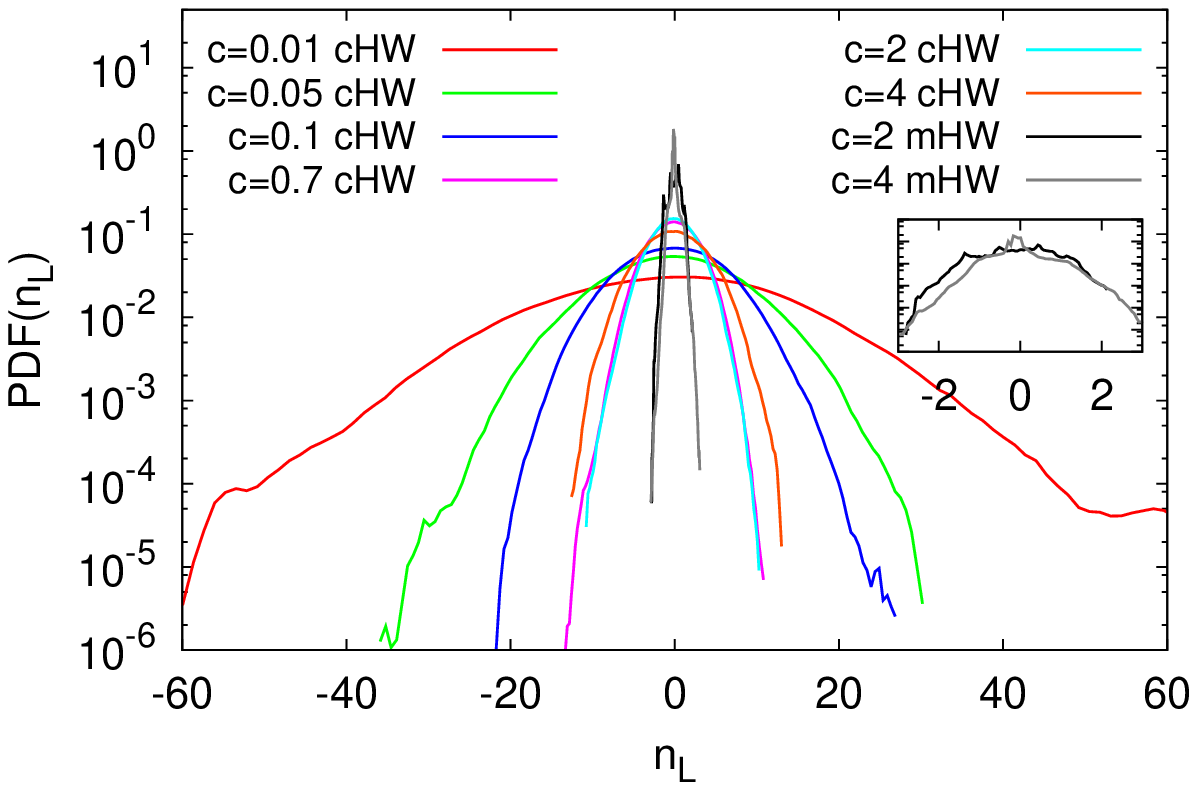}\\
   \end{tabular}
   \includegraphics[scale=0.7]{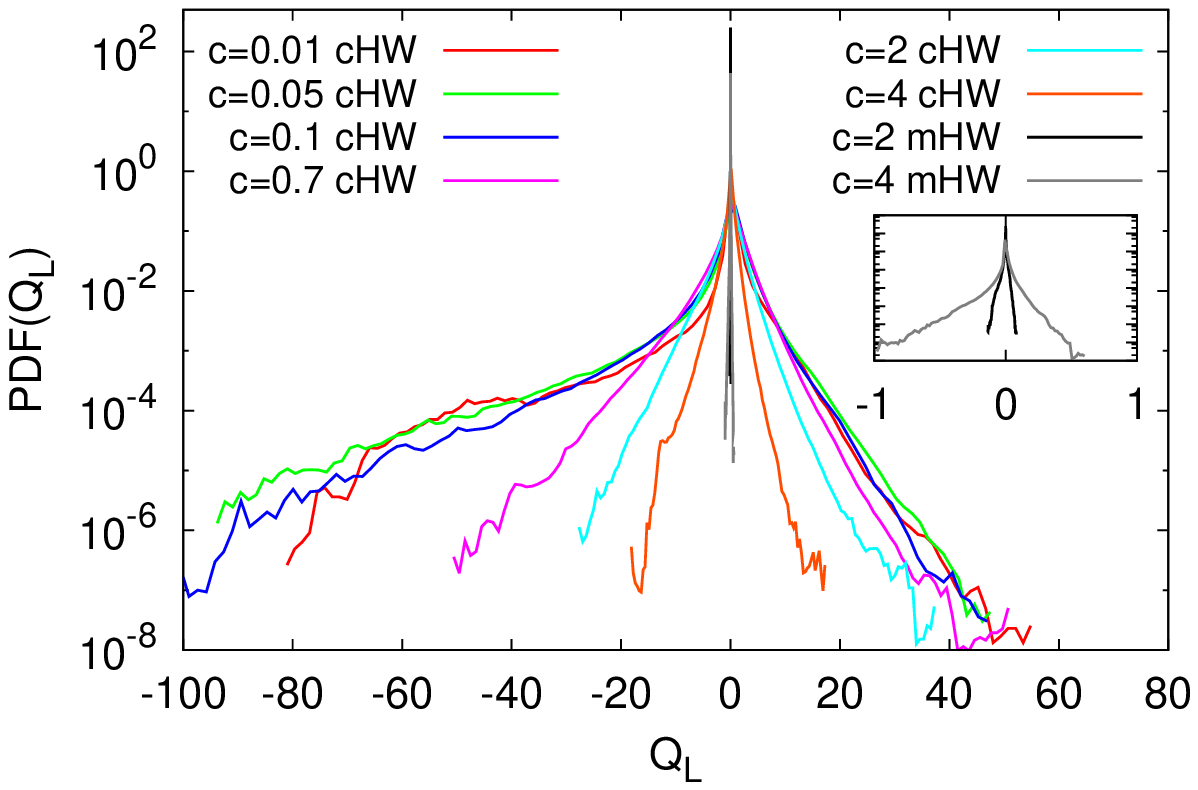}\\
\caption{PDFs of Lagrangian vorticity (top, left), Lagrangian density fluctuations (top, right) and Lagrangian Weiss value (bottom, center). The insets show zooms for the two {\it mHW} cases with $c=2$ and $4$.}
\label{Fig: PDF LAGR N}
  \end{center}
\end{figure}

In Fig.~\ref{Fig: PDF LAGR N} (top), we can observe that the PDFs of Lagrangian vorticity and density are almost identical
with the PDFs of the corresponding Eulerian quantities shown in Fig.~\ref{Fig: PDF EUL}. For a statistically stationary flow which is incompressible it is expected that for advected quantities (here vorticity and density) the Lagrangian particle samples the Eulerian flow domain uniformly due to the volume preserving property of the flow field and thus for sufficient long time averages the statistics coincide, assuming ergodicity.
The almost perfect agreement between Eulerian and Lagrangian PDFs thus confirms that the statistical sampling is sufficient.
In the PDFs of Lagrangian vorticity it is found, for low adiabaticity values, that asymmetries are present, corresponding to positively skewed PDFs. For the adiabatic regime and $c=2~(cHW)$ the PDFs have an almost Gaussian shape. Furthermore, the PDFs of Lagrangian density fluctuations are Gaussian for large adiabaticities. The mean values are almost equal to zero (as expected and values are omitted). In contrast we can observe strong differences for second order moments and flatness values, as quantified in~table~\ref{tab: Norm Flatness STAT WEISS Stat 2}. We observe that the variance of density decreases with increasing adiabaticity, except for $c=4$ for which we find a larger value. The flatness values likewise decrease with increasing $c$ and we can identify a sub-Gaussian behavior with values below $3$.

The PDFs of Lagrangian Weiss values, in Fig.~\ref{Fig: PDF LAGR N} (bottom), are negatively skewed, {\it i.e.}, they show more negative values for low adiabaticities, similar to what was observed in~\cite{Bos2008-1}. The corresponding values are given in table~\ref{tab: Norm Flatness STAT WEISS Stat 2}. This is reflected in the presence of more pronounced vortical structures in the flow, as it can be observed in Fig.~\ref{Fig: VISU}. For increasing adiabaticities, the PDFs become increasingly symmetric which corresponds to the absence of the vortical structures and thus a significant change of flow topology. However, for modified HW (inset in Fig.~\ref{Fig: PDF LAGR N}), the increase of asymmetry in the PDFs of Lagrangian Weiss value for large $c$ reveals that some vortical structures are formed due to the presence of shear flows, which tend to destabilize and generate Kelvin--Helmholtz vortices. \\

Table~\ref{tab: Norm Flatness STAT WEISS Stat 2} shows the second order moment values and the flatness of the different Lagrangian quantities. The skewness values, which quantify the asymmetry of the PDF, are only given for the Weiss value, as for the other quantities values close to zero are found.\\

\begin{table}[!htb]
\begin{center} \begin{footnotesize}
\begin{tabular}{|l| c c c c c |c c c c c| c| }
\hline
Configurations & $u_L^x$    & $a_L^x$    & $Q$   & $\omega$    &    $n$ & $u_L^x$    & $a_L^x$    & $Q$   & $\omega$    &    $n$   & $Q$ \\ 
& & & \textbf{$M_2$} & & & & & \textbf{$F$} & & & \textbf{$S$} \\
\hline
$c   =0.01~cHW$ & 5.0086 & 28.4189 & 15.6211 & 4.2059 & 182.0711 & 4.9765 & 40.6484 & 86.0252 & 14.3067 & 3.3713 & -7.1784  \\
$c   =0.05~cHW$ & 4.1107 & 20.3524 & 20.5109 & 6.8190 & 56.3207  & 3.7661 & 32.8554 & 67.1198 & 7.7039 & 3.0216 & -5.9186  \\ 
$c   =0.10~cHW$ & 3.3151 & 12.5433 & 16.4683 & 7.4005 & 31.5241 & 3.4671 & 28.6267 & 58.1253 & 5.5382 & 2.8321 & -5.0000  \\  
$c   =0.70~cHW$ & 2.2778 & 5.2765 & 9.1490 & 8.1136 & 7.6012 & 3.2973 & 8.6850 & 13.5941 & 3.0570 & 2.8147 & -1.5099  \\  
$c   =2.00~cHW$ & 1.9671 & 3.2987 & 4.2101 & 5.6229 & 6.6556  & 3.4499 & 6.9236 & 11.1652 & 3.0240 & 2.9646 & -1.0842  \\ 
$c   =4.00~cHW$ & 1.6056 & 1.1687 & 0.7853 & 2.3188 & 13.0173 & 3.4526 & 6.2109 & 18.3221 & 3.4360 & 2.7814 & -1.6767  \\  
$c   =2.00~mHW$ & 0.0046 & 0.0011 & 0.0001 & 0.0636 & 0.6399  & 12.2982 & 18.2834 & 33.5518 & 1.8105 & 2.5214 & -2.9332  \\  
$c   =4.00~mHW$ & 0.0148 & 0.0050 & 0.0009 & 0.0621 & 0.3197 & 14.3581 & 28.3987 & 92.1782 & 8.2050 & 4.9547  & -4.7475 \\   
\hline
\end{tabular} \end{footnotesize}
\caption{\label{tab: Norm Flatness STAT WEISS Stat 2} Second order centered moments, $M_2$, and flatness, $F$, of different Lagrangian quantities, {\it i.e.}, $x$-component of velocity ($u^x_L$) and acceleration ($a^x_L$), Weiss value ($Q$), vorticity ($\omega$) and density ($n$) computed from the corresponding PDFs. Only the skewness $S$ of $Q$ is shown, since the other quantities yield skewness values close to zero.}
\vspace{-0.1cm}
\end{center}
\end{table}

\subsection{Residence time}

\begin{figure}[!htb]
  \begin{center}
  \begin{tabular}{cc}
   $c=0.01~cHW$ & $c=0.7~cHW$ \\
   \includegraphics[scale=0.70]{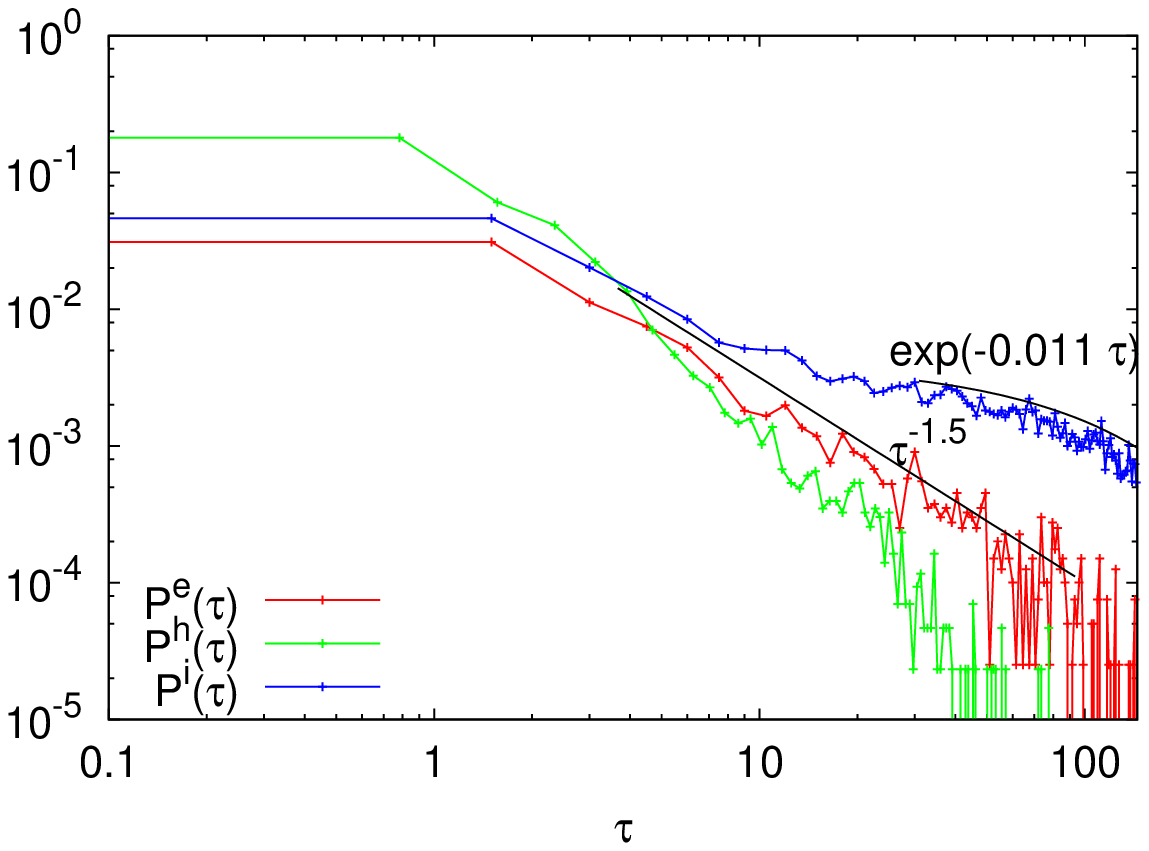}&
   \includegraphics[scale=0.70]{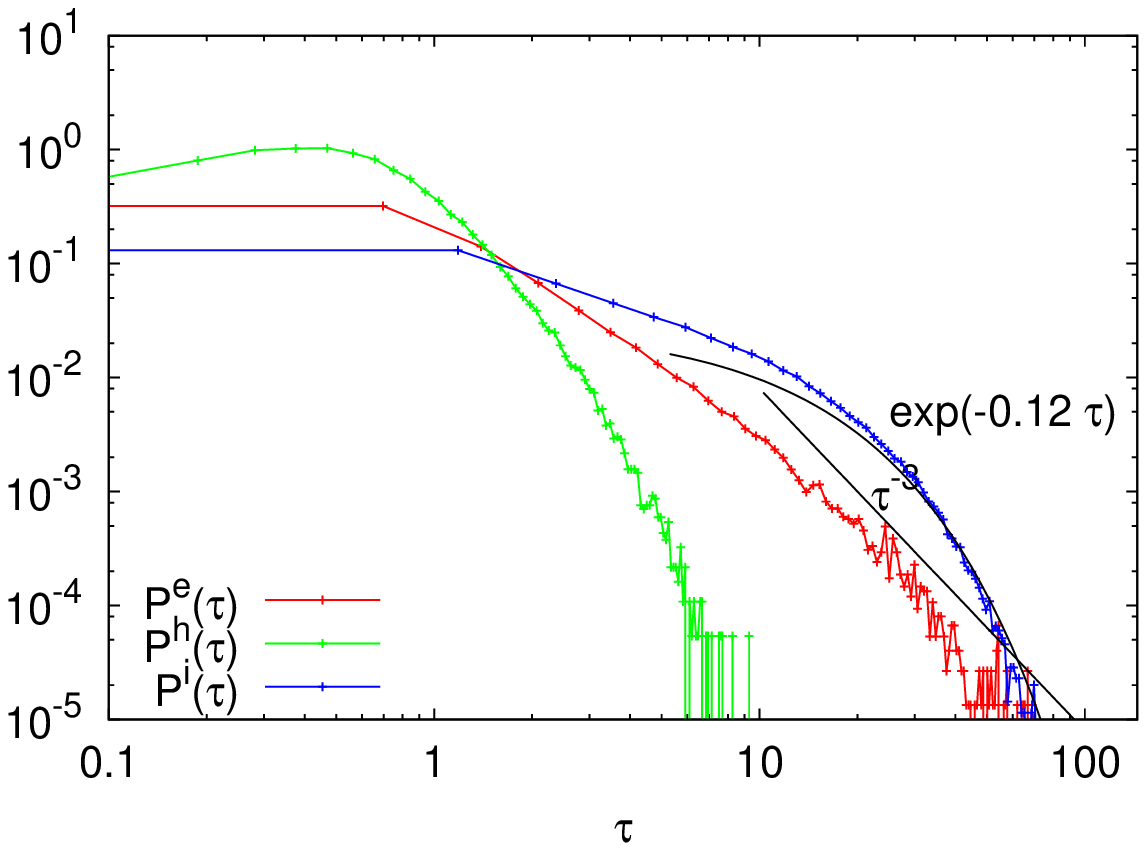}\\
   $c=4~cHW$ & $c=4~mHW$ \\   
   \includegraphics[scale=0.70]{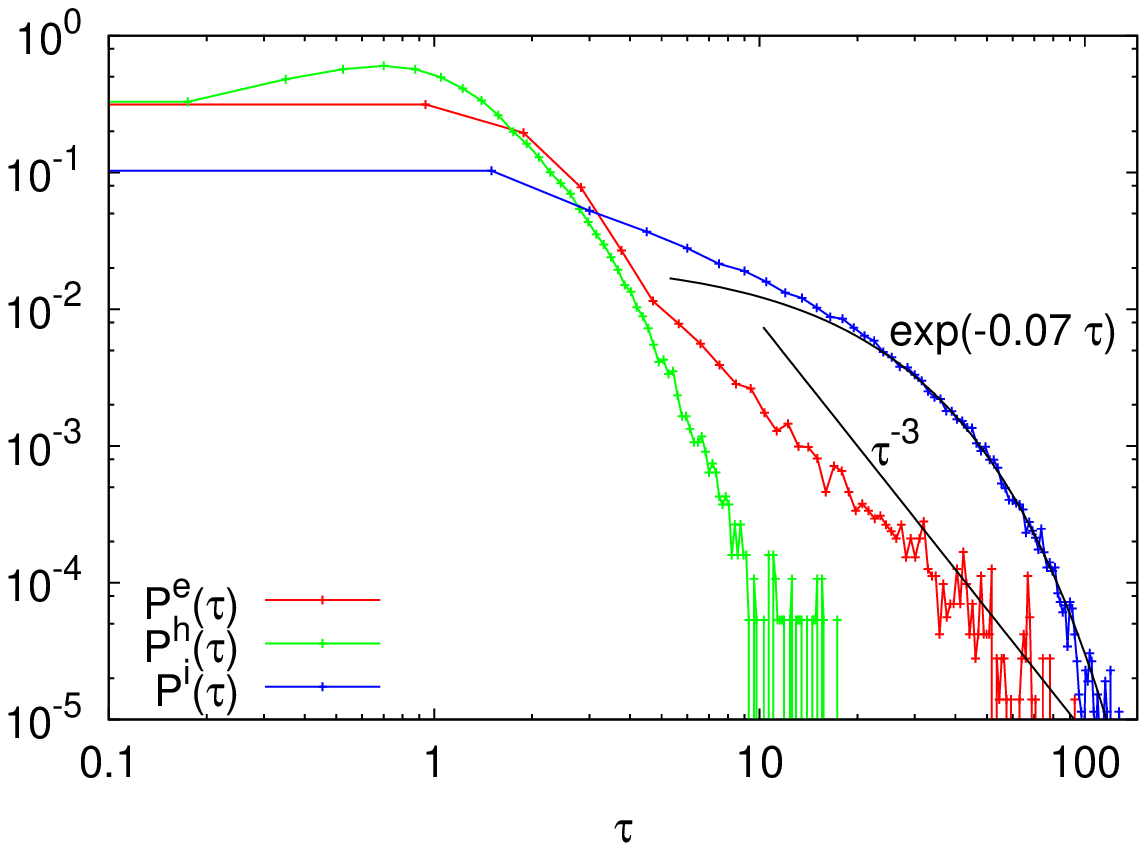}&
   \includegraphics[scale=0.70]{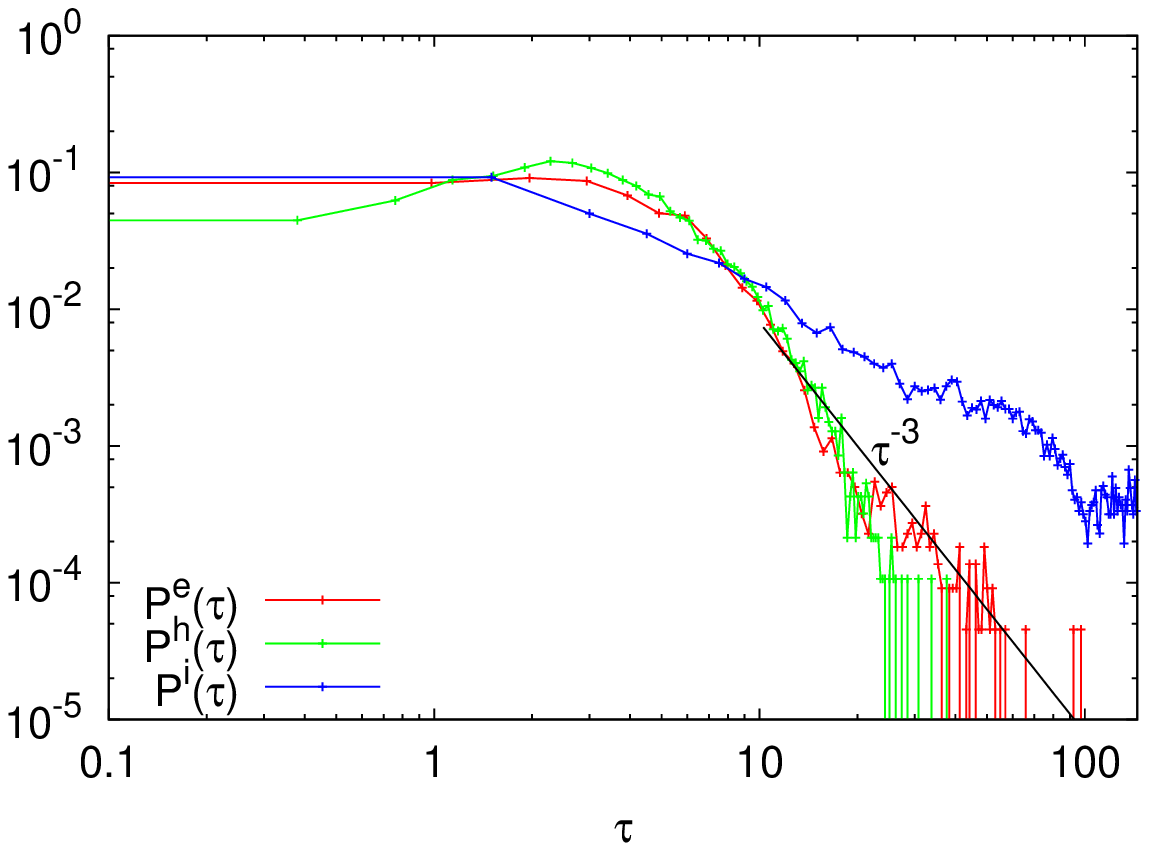}
   \end{tabular}
\caption{PDFs of residence time conditioned with respect to the three-level Lagrangian Weiss value for {\it cHW} with $c=0.01$ (top, left), $c=0.7$ (top, right), $c=4$ (bottom, left) and {\it mHW} with $c=4$ (bottom, right).}
\label{Fig: PDF RESIDENCE TIME}
  \end{center}
\end{figure}

Motivated by the work in \cite{kadoch_etal_2011} for 2D Navier--Stokes turbulence we consider the PDFs of the residence time conditioned with respect to the three-level Lagrangian Weiss value, shown in Fig.~\ref{Fig: PDF RESIDENCE TIME}. 
These PDFs determine the probability that a given Lagrangian tracer stays in a region with the same value of the three-level normalized Weiss field for a given time $\tau$. Note that for Gaussian fluctuations, characteristic  of the turbulent background, results in the exponential decay of the residence time PDF.
The main behavior, observed for 2D Navier--Stokes in \cite{kadoch_etal_2011}, is still present for different flow regimes obtained with the Hasegawa--Wakatani model, that is to say, algebraic tails for {strongly} hyperbolic and elliptic regions, and exponential decay for intermediate regions which can be modeled by a Poisson process. The longest residence times are found for intermediate regions. The times are longer for {strongly} elliptic regions than for {strongly} hyperbolic regions, which is explained by the particle trapping in elliptic zones. The difference between the elliptic and hyperbolic regions tends to disappear for the largest $c$ value. This is confirmed by the mean and centered second order moments shown in table~\ref{tab: Curvature WEISS Stat 2}. 
Moreover, we can remark that the strongest mean residence time and centered second order moment are found for intermediate regions, while the weakest values appear for {strongly} hyperbolic regions.
The influence of adiabaticity is weak for {strongly} elliptic and hyperbolic regions, {illustrated by the fact that} the exponents do not differ much. For intermediate regions, the adiabaticity has a stronger influence by changing significantly the slope between small and large adiabaticity. We can note also that the behaviors are the same for large adiabaticity $c\geq 0.7~cHW$. Concerning the modified HW case, the residence times are a little longer for {strongly} hyperbolic regions than for elliptic ones because the shapes of these two zones are of the same nature as confirmed by the visualizations in Fig.~\ref{Fig: VISU}.\\

\begin{table}[!htb]
\begin{center}
\begin{tabular}{|l|c c c |c c c |}
\hline
Configurations &  Elliptic  & Hyperbolic & Intermediate &  Elliptic  & Hyperbolic & Intermediate \\
& & $M_1$ & & & $M_2$ & \\
\hline
$c   =0.01~cHW$ & 1.6293  & 0.6383 & 18.9589 & 89.7212  & 5.1834 & 1510.6079 \\  
$c   =0.05~cHW$ & 1.3667  & 0.4343 & 10.1370 & 42.7289  & 1.3205 & 404.3433 \\   
$c   =0.10~cHW$ & 1.2164  & 0.4726 & 7.3025  & 23.7842  & 0.6241 & 196.5689 \\  
$c   =0.70~cHW$ & 1.1092  & 0.6207 & 3.5919  & 7.7833  & 0.2904 & 40.5284 \\   
$c   =2.00~cHW$ & 1.1889  & 0.7495 & 4.1236  & 9.0669  & 0.4016 & 59.8395 \\    
$c   =4.00~cHW$ & 1.4298  & 1.0490 & 6.3722  & 9.9281  & 0.8595 & 125.5260  \\  
$c   =2.00~mHW$ & 3.2255  & 3.7634 & 29.7768 & 19.1782  & 19.7126 & 2487.1401 \\    
$c   =4.00~mHW$ & 2.4195  & 2.5951 & 35.3828 & 15.7420  & 10.5522 & 2986.2272 \\      
\hline
\end{tabular}
\caption{\label{tab: Curvature WEISS Stat 2}
Mean values, $M_1$, and centered second order moments, $M_2$, of the residence time conditioned with respect to the three-level Lagrangian Weiss value in elliptic, hyperbolic and intermediate flow regions.}
\vspace{-0.1cm}
\end{center}
\end{table}

\subsection{Conditional Lagrangian statistics}

\begin{figure}[!htb]
  \begin{center}
  \begin{tabular}{cc}
   $c=0.01~cHW$ & $c=0.7~cHW$ \\
   \includegraphics[scale=0.70]{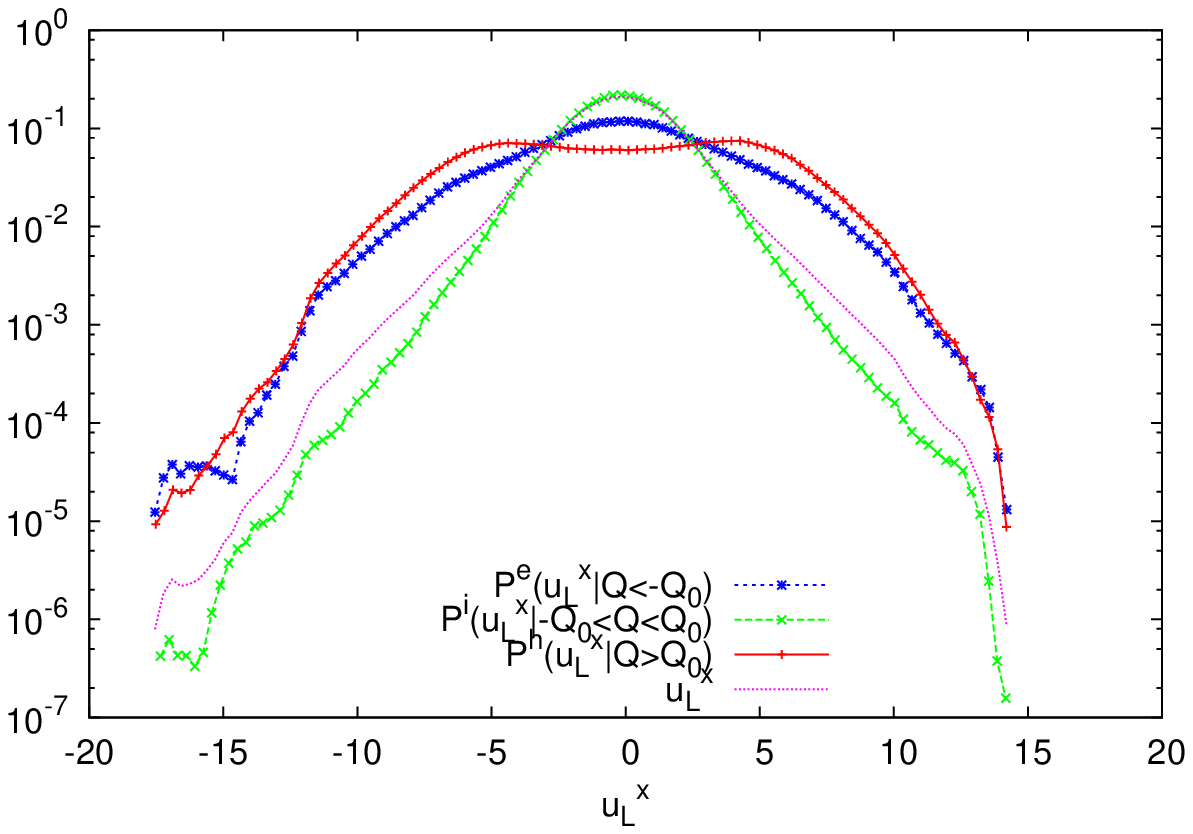}&
   \includegraphics[scale=0.70]{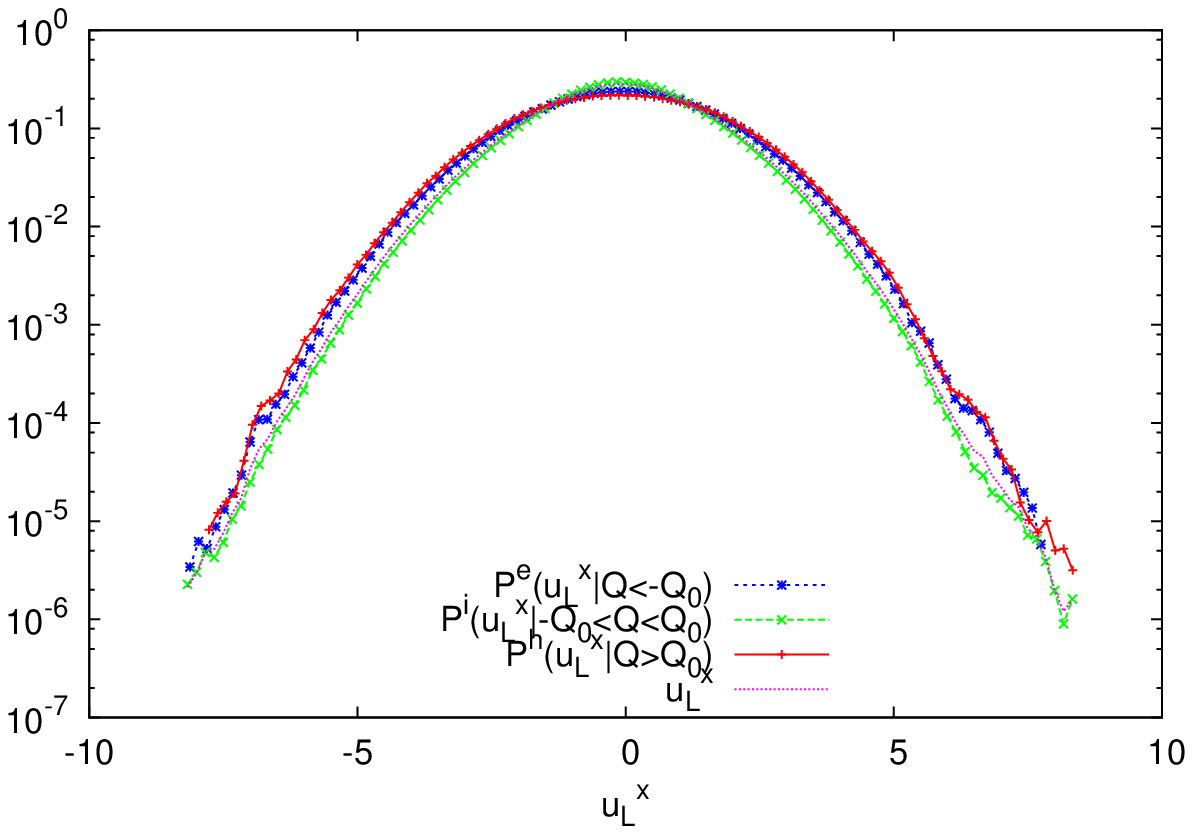}\\
   $c=4~cHW$ & $c=4~mHW$ \\   
   \includegraphics[scale=0.70]{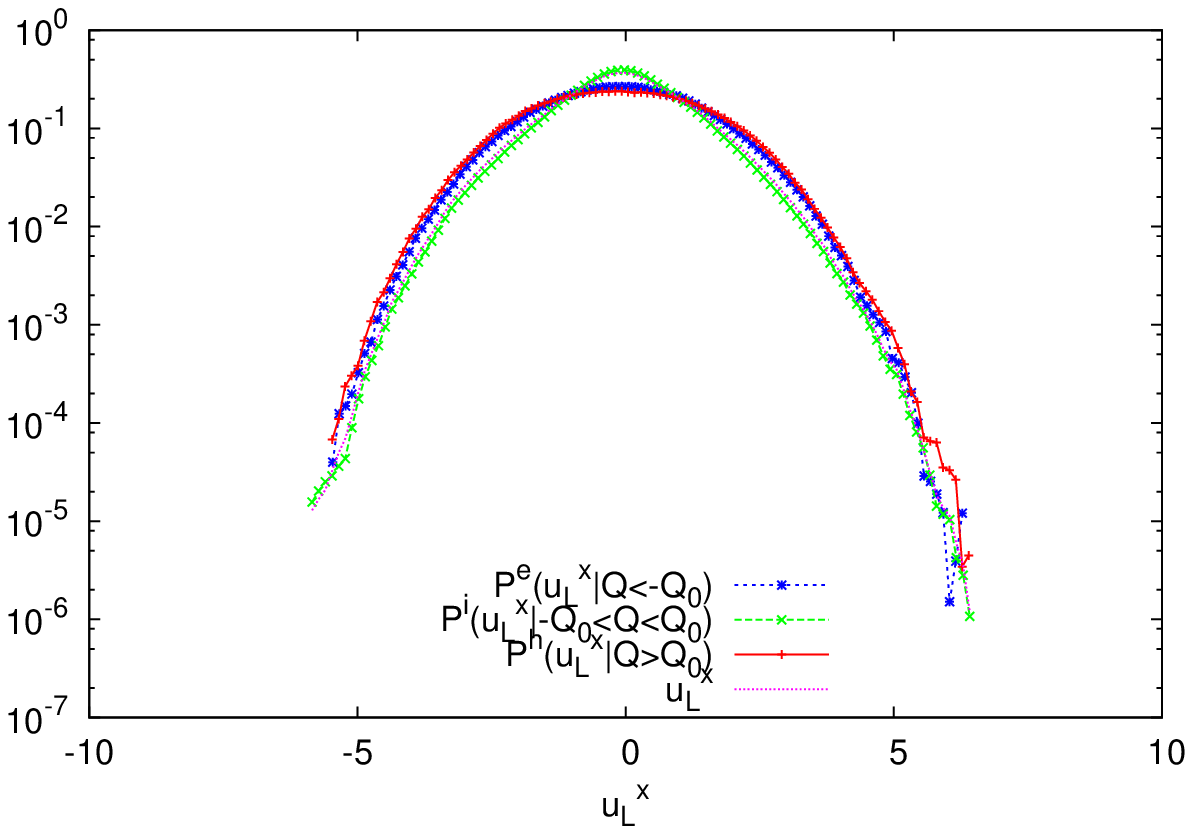}&
   \includegraphics[scale=0.70]{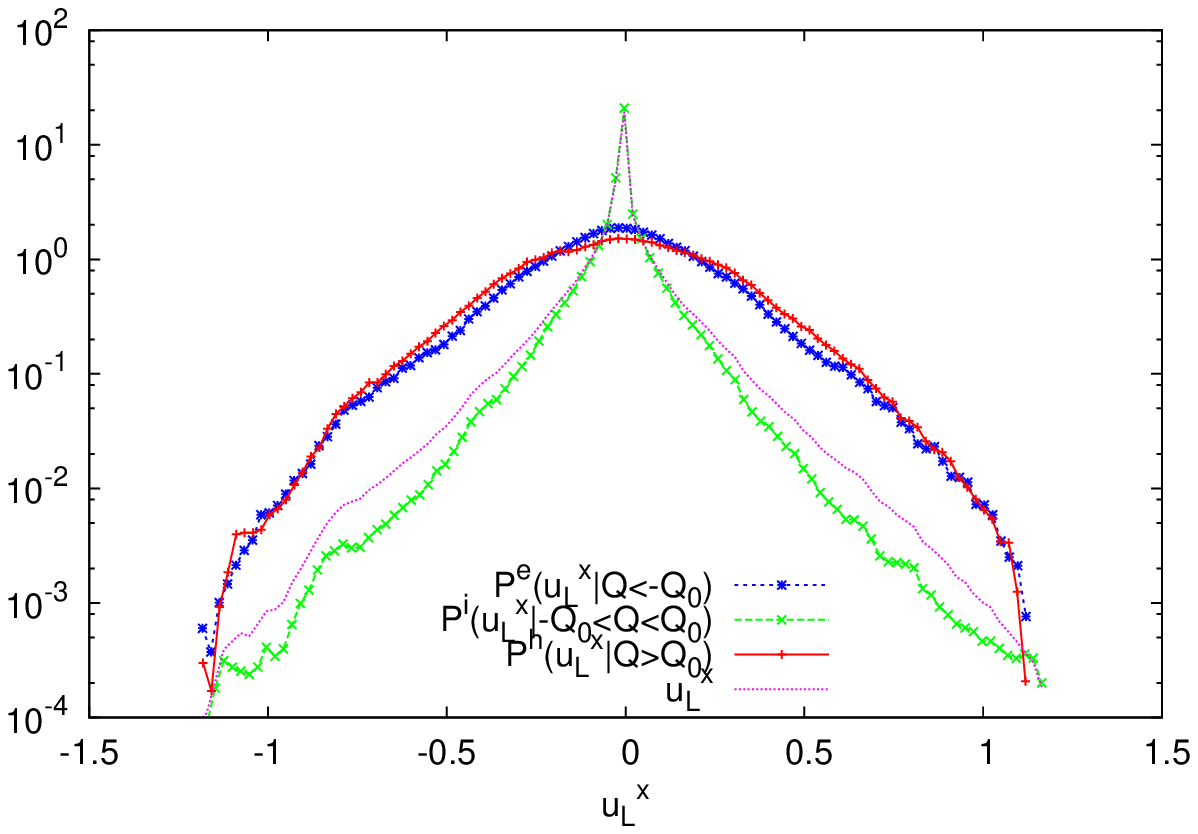}
   \end{tabular}
\caption{PDFs of Lagrangian conditional $x$-component velocity with respect to three-level Lagrangian Weiss value for {\it cHW} with $c=0.01$ (top, left), $c=0.7$ (top, right), $c=4$ (bottom, left) and {\it mHW} with $c=4$ (bottom, right).}
\label{Fig: PDF COND VELX}
  \end{center}
\end{figure}

Figure \ref{Fig: PDF COND VELX} shows the PDFs of Lagrangian velocity in $x$-direction conditioned with respect to the three-level Weiss value. First we find that the intermediate zones dominantly determine the shape of 
the total PDF. The differences between the zones tend to disappear with increasing adiabaticity for classical HW. Moreover the differences are much more pronounced for the modified HW case. The behavior of the $y$-component of the Lagrangian velocity is very similar (omitted here) with respect to the contribution of the different zones and their shape. An exception is the modified HW case due to presence of zonal flows in the $y$-direction.\\
 
\begin{figure}[!htb]
  \begin{center}  \begin{tabular}{cc}
   $c=0.01~cHW$ & $c=0.7~cHW$ \\
   \includegraphics[scale=0.70]{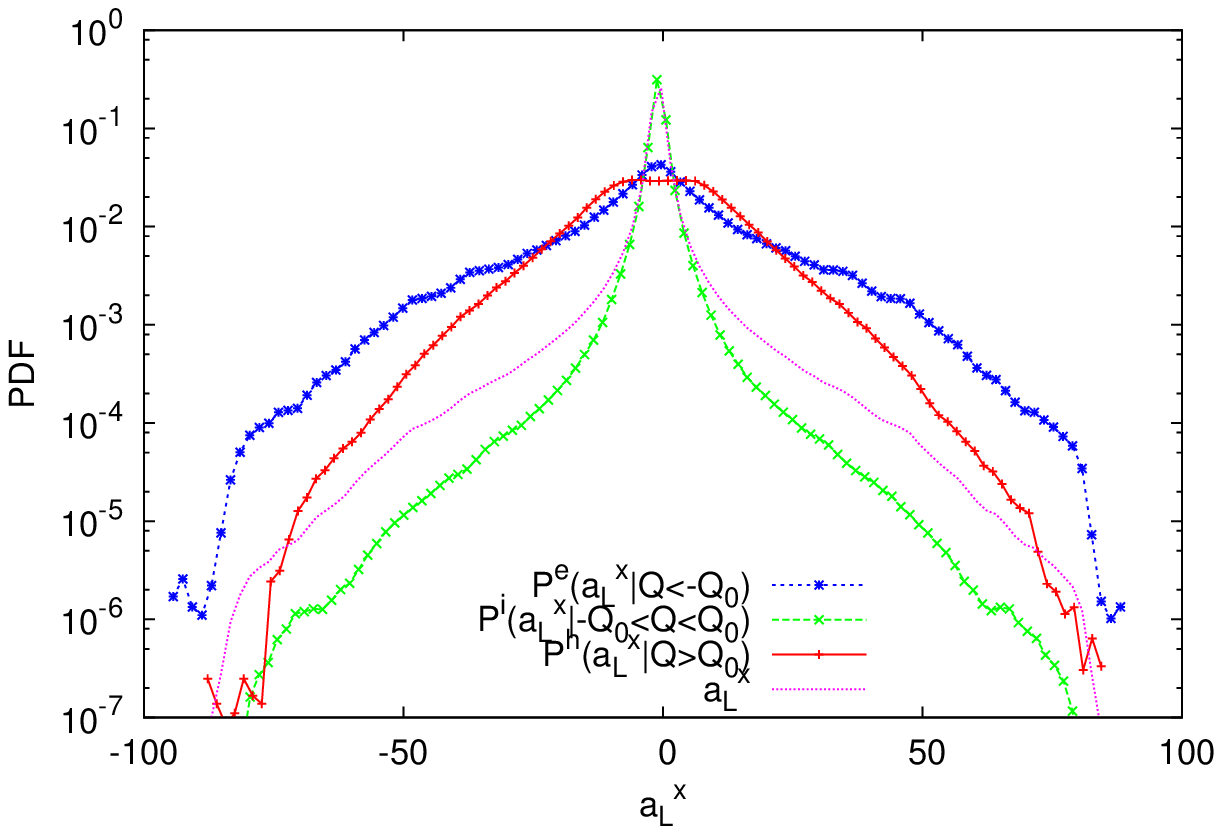}&
   \includegraphics[scale=0.70]{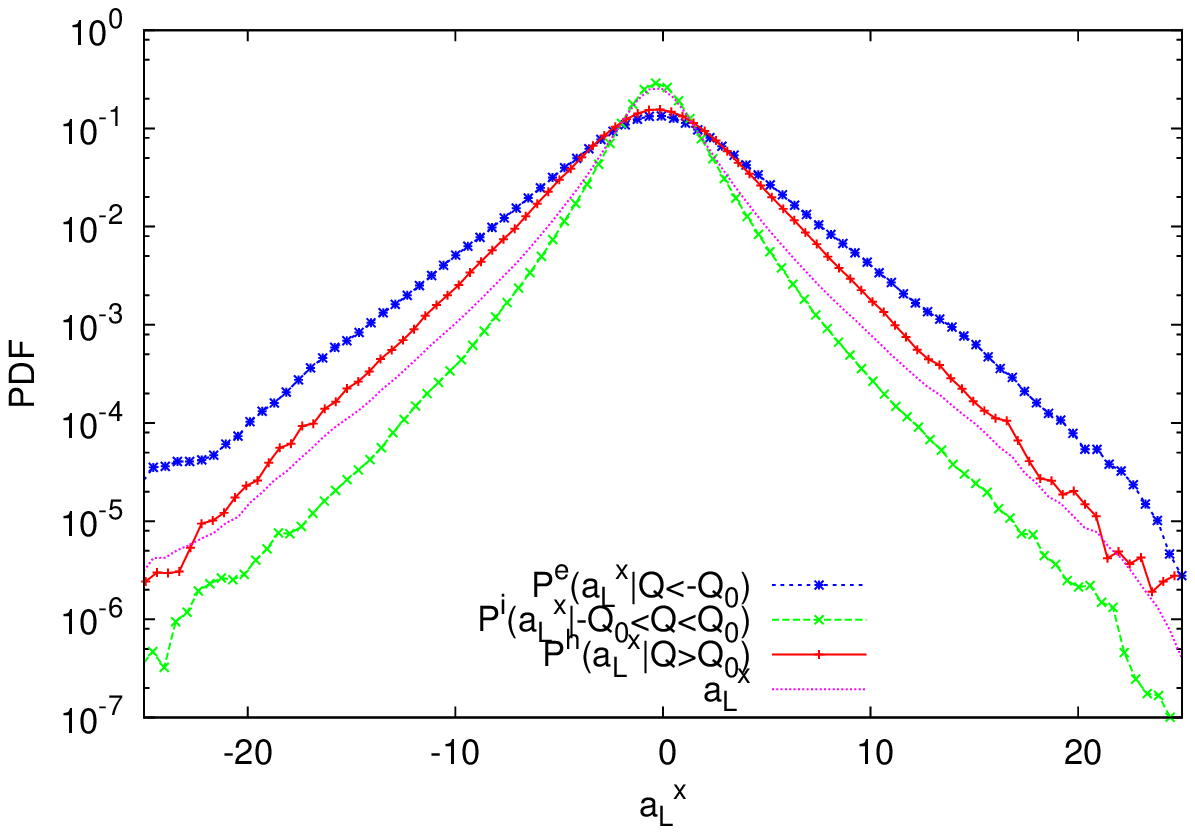}\\
   $c=4~cHW$ & $c=4~mHW$ \\   
   \includegraphics[scale=0.70]{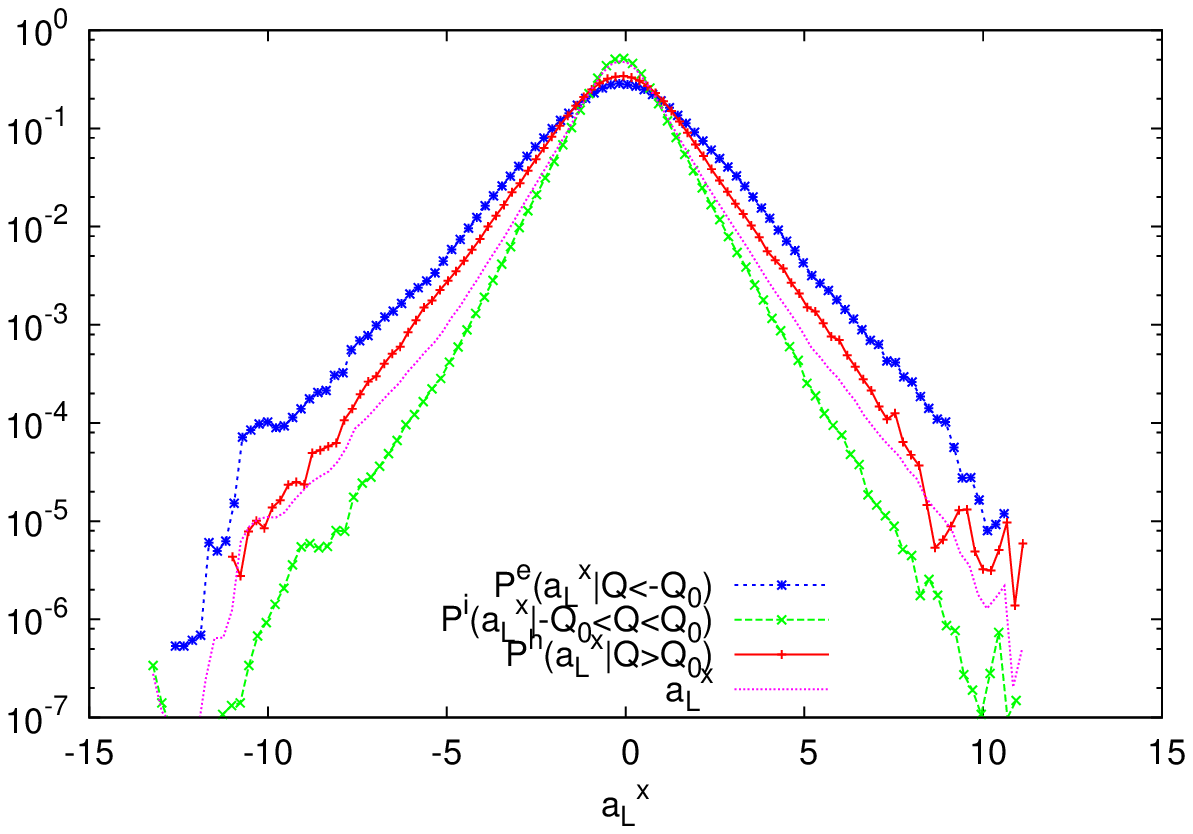}&
   \includegraphics[scale=0.70]{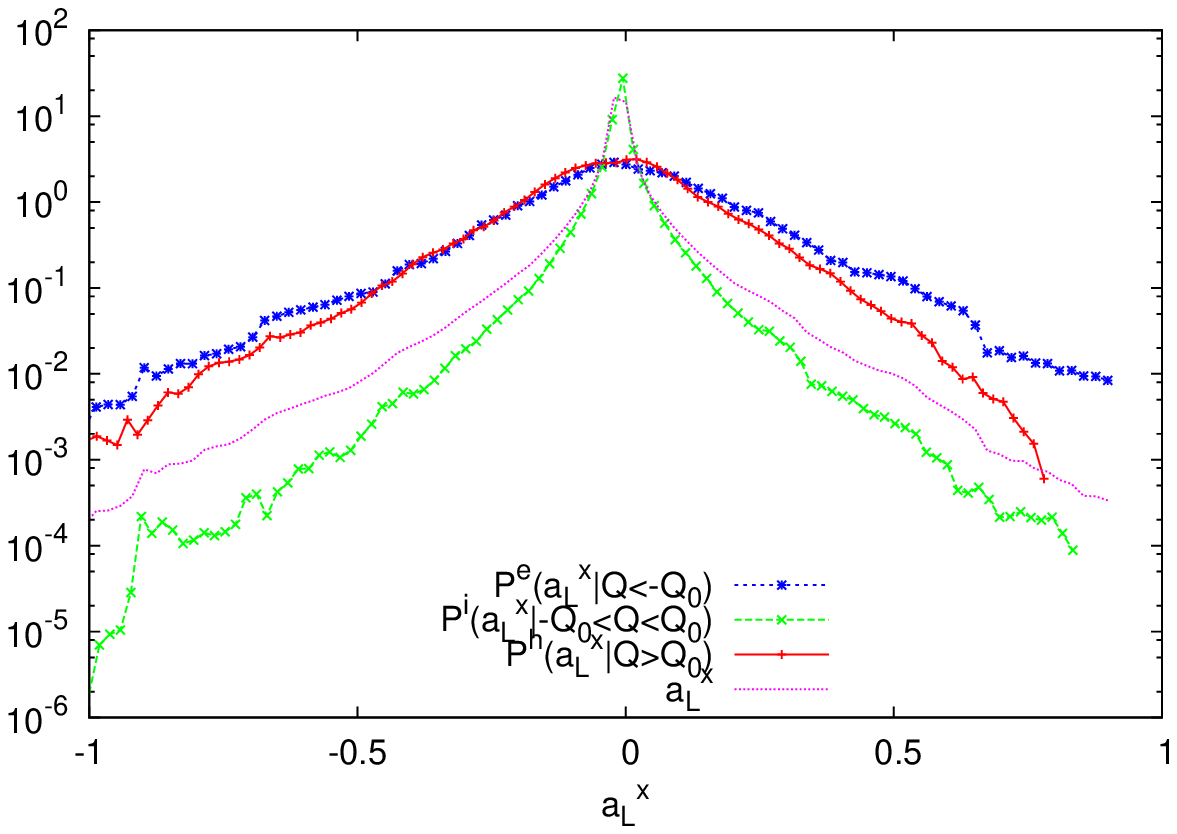}
   \end{tabular}
\caption{PDFs of Lagrangian conditional $x$-component acceleration with respect to the three-level Lagrangian Weiss value for {\it cHW} with $c=0.01$ (top, left), $c=0.7$ (top, right), $c=4$ (bottom, left) and {\it mHW} with $c=4$ (bottom, right).}
\label{Fig: PDF COND ACCX}
  \end{center}
\end{figure}

The PDFs of Lagrangian acceleration in $x$-direction conditioned with respect to the three-level Weiss value are given in Fig.~\ref{Fig: PDF COND ACCX}. The intermediate zones yield the largest contribution to the total PDF, however the contributions from elliptic and hyperbolic regions are not negligible. Similarly as for the Lagrangian velocities, the differences between the zones are less pronounced with increasing adiabaticity for classical HW and the discrepancies are more significant for modified HW. The same trend for the contribution of the different zones is observed for the $y$-component of the Lagrangian acceleration (not shown here).

%

\begin{figure}[!htb]
  \begin{center}
  \begin{tabular}{cc}
   $c=0.01~cHW$ & $c=0.7~cHW$ \\
   \includegraphics[scale=0.70]{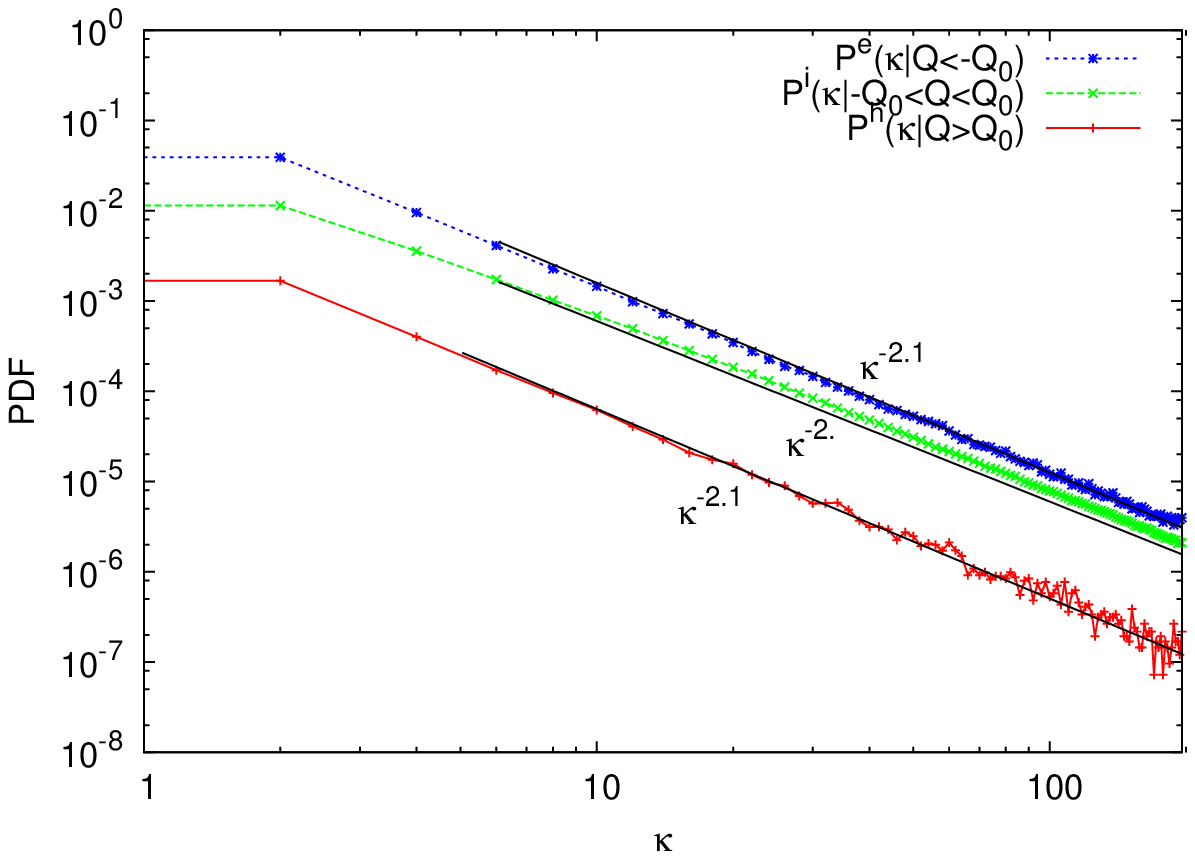}&
   \includegraphics[scale=0.70]{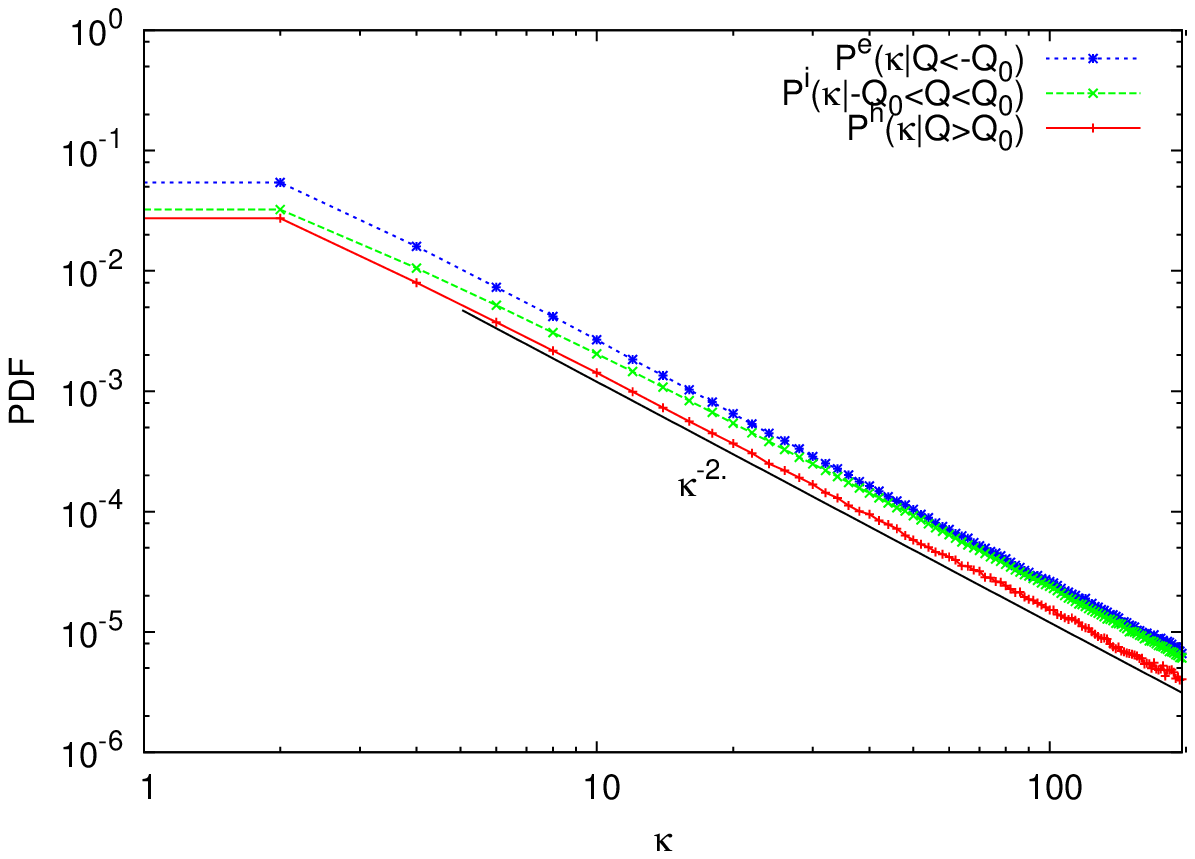}\\
   $c=4~cHW$ & $c=4~mHW$ \\   
   \includegraphics[scale=0.70]{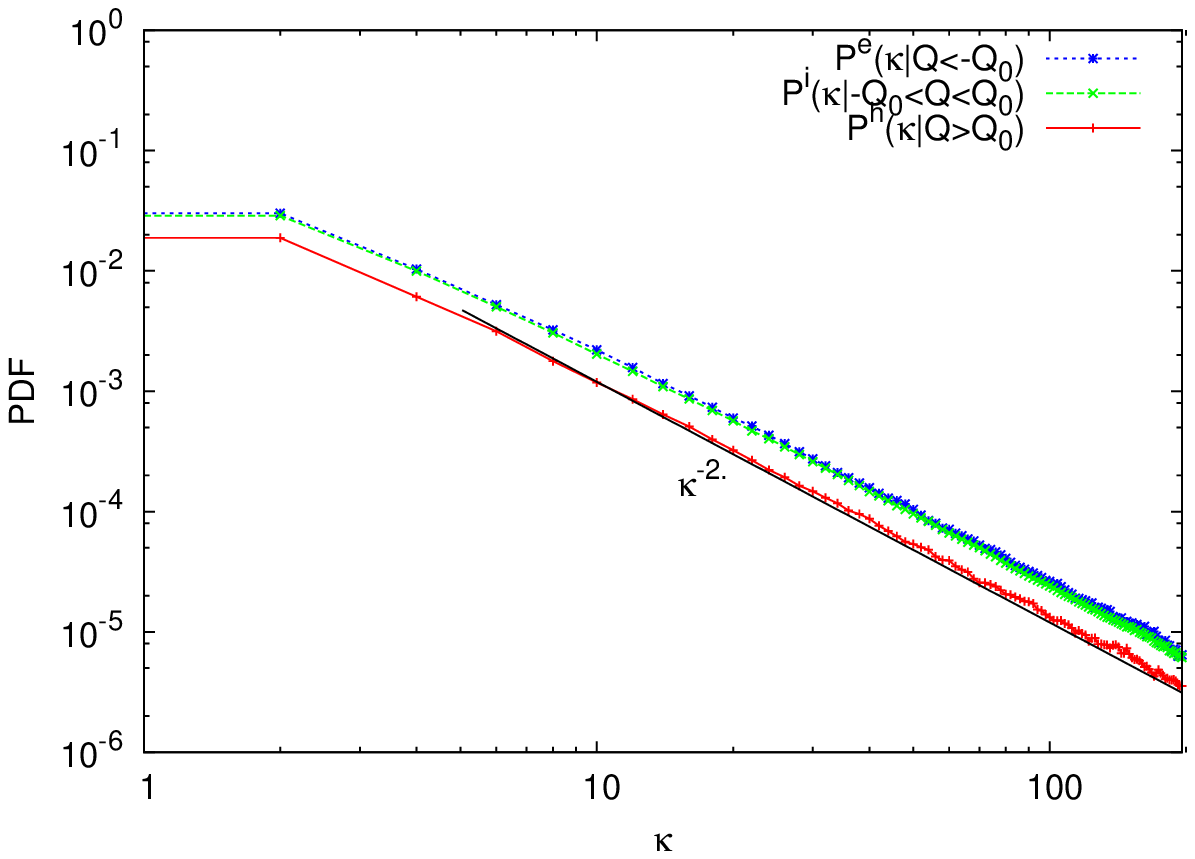}&
   \includegraphics[scale=0.70]{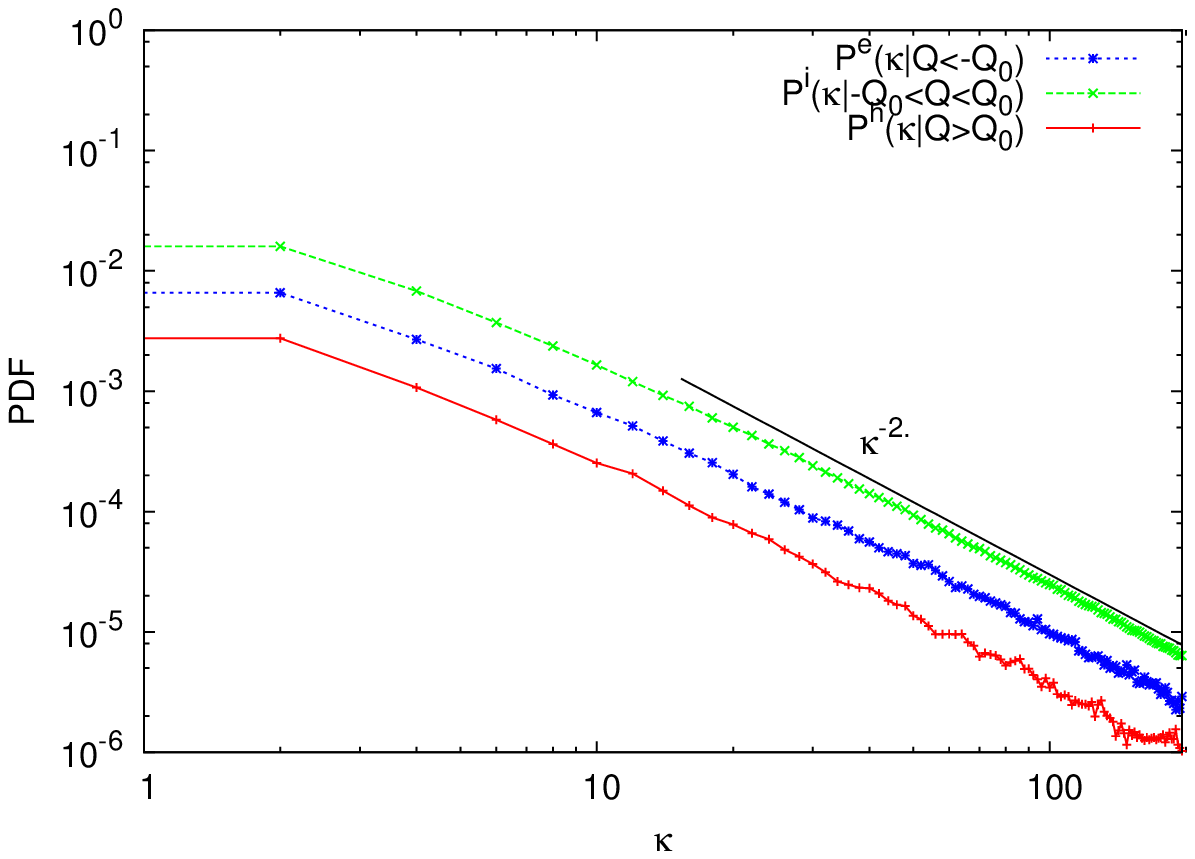}
   \end{tabular}
\caption{PDFs of Lagrangian conditional curvature $\kappa$ with respect to the three-level Lagrangian Weiss value for {\it cHW} with $c=0.01$ (top, left), $c=0.7$ (top, right), $c=4$ (bottom, left) and {\it mHW} with $c=4$ (bottom, right).}
\label{Fig: PDF COND CURV}
  \end{center}
\end{figure}

Figure~\ref{Fig: PDF COND CURV} shows the PDFs of curvature again conditioned with respect to the three-level Weiss value. We can see that they are very similar to the ones obtained for hydrodynamics shown in \cite{kadoch_etal_2011}. For small adiabaticity, the PDFs decay algebraically with an exponent somewhat stronger than $-2$ which demonstrates the presence of intermittency in these regimes.
For large adiabaticity, the decay exponent is precisely $-2$ which is exactly the exponent we predict from Gaussian PDFs of Lagrangian velocity in two-dimensions. This can be explained by the fact that $1/u^2$ yields an inverse chi-square distribution. As a consequence the decay exponent of the PDF of $1/u^2$ is $-2$ \cite{xu2007curvature}. The PDF of curvature is therefore an additional tool to quantify the presence of Lagrangian intermittency. From this latter point, we can conclude that the regimes become decreasingly intermittent for increasing adiabaticity since the deviation from the decay exponent $-2$ becomes less and less important. Furthermore, for the different regimes, we can claim that the intermediate regions are not intermittent from a Lagrangian point of view.\\
%

\begin{figure}[!htb]
  \begin{center}
  \begin{tabular}{cc}
   $c=0.7~cHW$ & $c=4~cHW$ \\
   \includegraphics[scale=0.70]{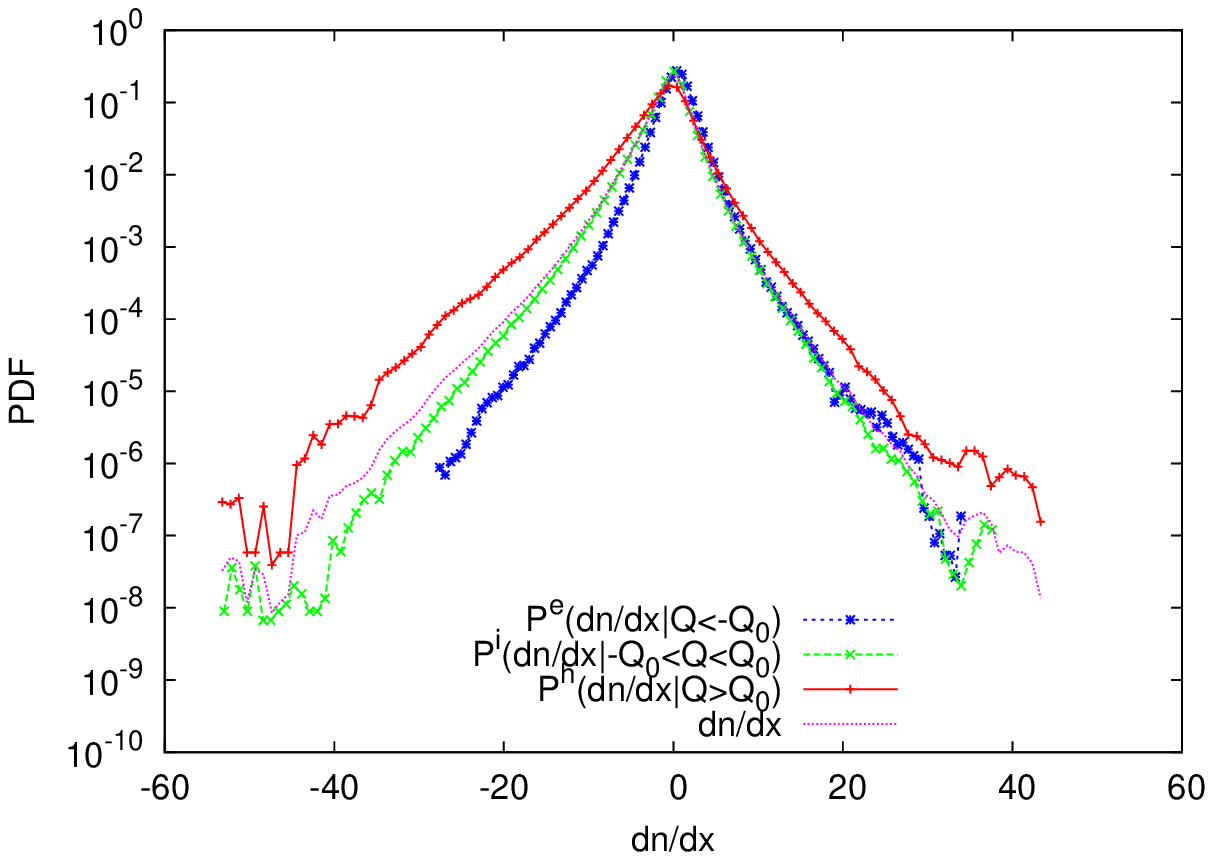}&
   \includegraphics[scale=0.70]{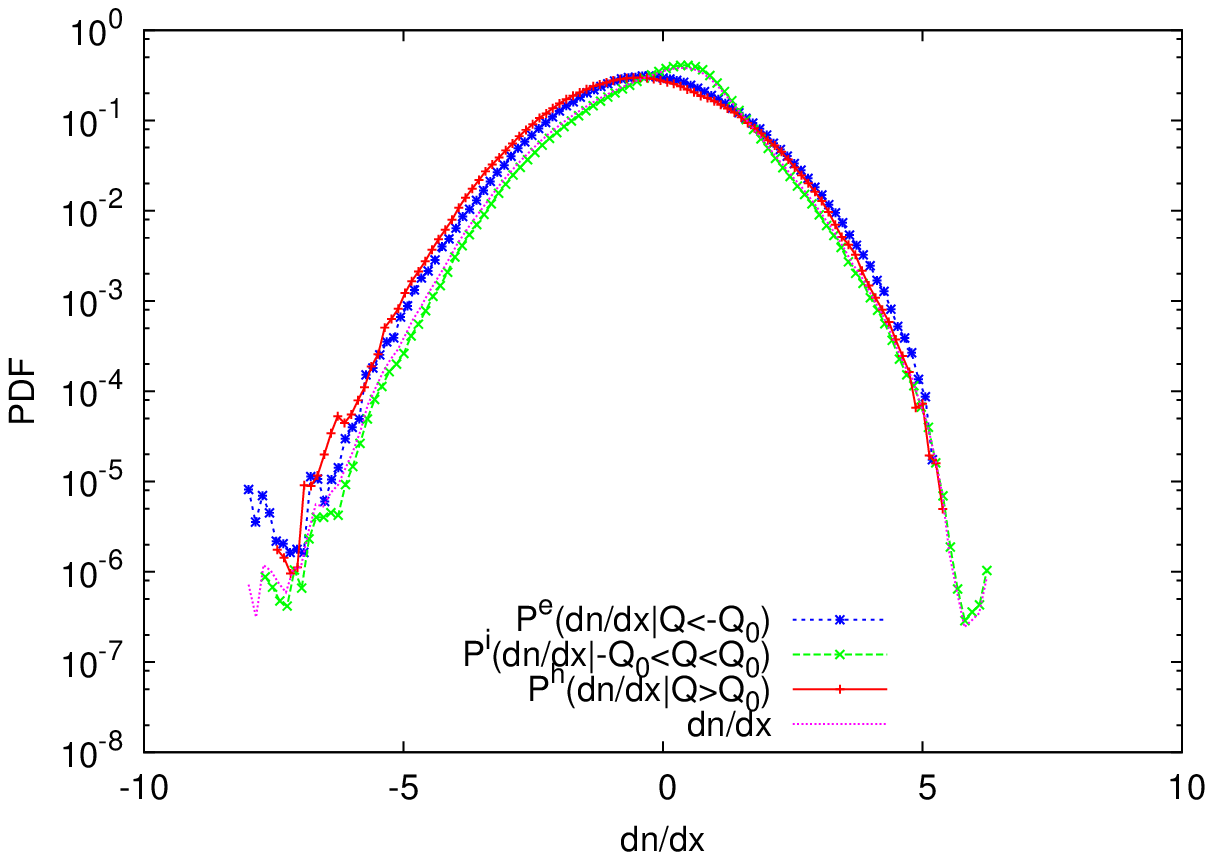}
   \end{tabular}
\caption{PDFs of Lagrangian conditional density gradient in the $x$-direction with respect to three level Weiss value for {\it cHW} with $c=0.7$ (left) and $c=4$ (right).}
\label{Fig: PDF COND DXN}
  \end{center}
\end{figure}

%
Motivated by \cite{Umansky_POF_2011}, where the authors studied the radial flux of plasma density and showed that it is similar to Bohm diffusion we analyze the density gradients in the different flow regions.
The PDFs of density gradients in the radial direction, $\partial n / \partial x$ , conditioned with respect to the three-level Weiss value are shown in Figures~\ref{Fig: PDF COND DXN} for two cases, $c=0.7$ and $4$ for {\it cHW}. 
The PDFs in the poloidal direction, $\partial n / \partial y$ are omitted as their shape is similar.
For the adiabatic regime ({\it cHW} with $c=0.7$) we find exponential tails for the total flow and the flow conditioned with the three level Weiss values. The deformation dominated regions yield the largest variance, while for intermediate and vortical regions similar values as in the PDF of the total flow can be observed.
In contrast in the geostrophic regime ({\it cHW} with $c=4$) all density gradient PDFs are Gaussian-like and thus have a parabola shape. Their variances are similar too.
For the hydrodynamic regime ({\it cHW} with $c=0.01$, not shown here) we observe heavy tails of the density gradients in both directions for negative Weiss values, {\it i.e.}, vorticity dominated regions. 
In the intermediate regions we find exponential tails, similar to the PDFs of the total field. 
For the deformation dominated regions the variance is strongly reduced.
Finally, for {\it mHW} (results not shown here) the zonal flows have a pronounced signature in the density gradient PDFs in the radial direction, while in the poloidal direction exponential tails can be seen.

\subsection{Mean curvature angle }

\begin{figure}[!htb]
 \begin{center}
 \begin{tabular}{cc}
   \includegraphics[scale=0.7]{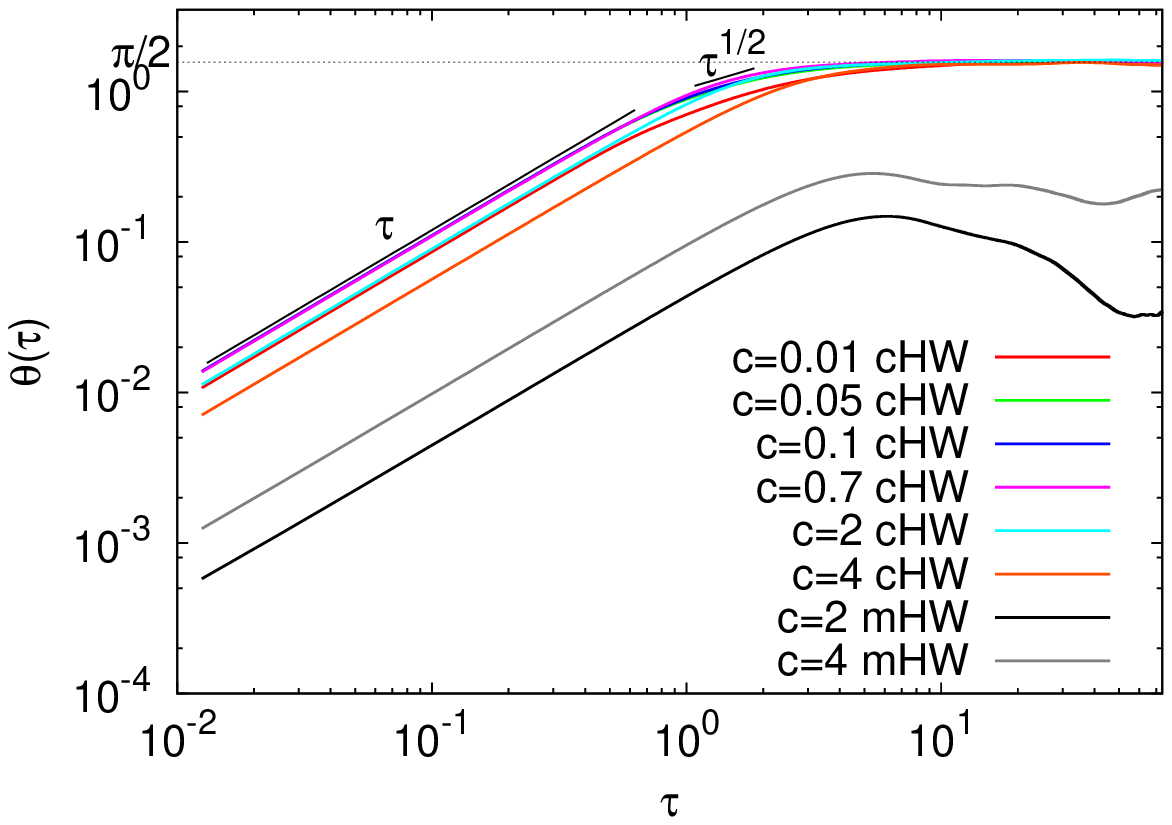}&
   \includegraphics[scale=0.7]{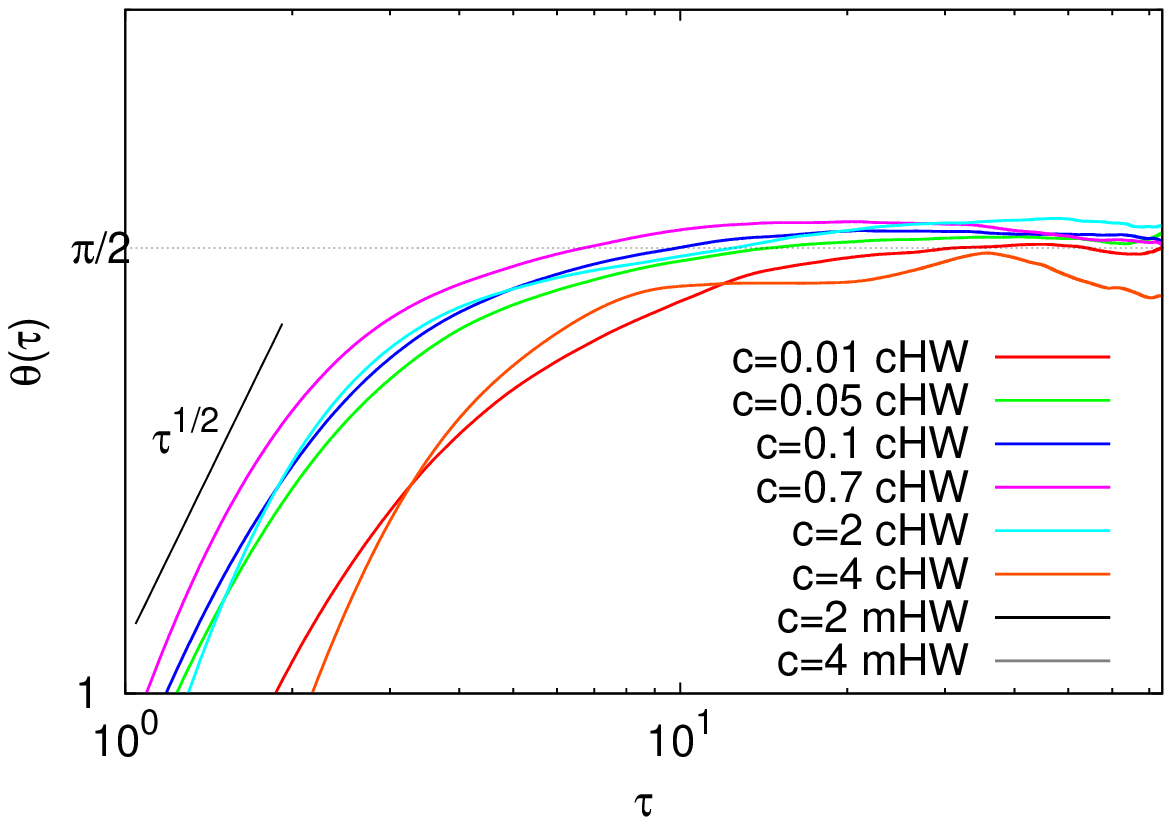}\\
  \end{tabular}
\vspace{-0.5cm}
\caption{Mean curvature angle $\theta(\tau)$ as a function of $\tau$ for the different flow regimes. A corresponding zoom for large increments $\tau$ is shown on the right.}
\label{Fig: MEAN ANGLE}
 \end{center}
\end{figure}

The directional change of tracer particles is analyzed by considering the curvature angle, as defined in eq.~(\ref{eq:cos}), and the coarse-grained curvature at different scales, introduced in section~\ref{subsec:lagrange}.
Figure \ref{Fig: MEAN ANGLE} (left) shows the mean curvature angle, $\theta(\tau)\equiv\left<|\Theta(t,\tau)|\right>$, where $\left< \cdot \right>$ denotes the ensemble and time average as function of the time increments for different adiabaticity values. The mean angles increase monotonically from 0 to $\pi/2$. This means that particles have for small $\tau$ the tendency going straight, similar to what was found in \cite{Bos_PRL_2015}, while for very large $\tau$ their motion becomes uncorrelated and they have the same probability traveling in any direction, which results in a mean angle of $\pi/2$, as we are taking the absolute value and do not distinguish between left and right turn.
For all regimes, {\it i.e.} {\it cHW} and {\it mHW}, we identify for small $\tau$ then a clear linear scaling behavior, followed by a $\tau^{1/2}$ transition, the so-called inertial scaling \cite{Bos_PRL_2015}, which is difficult to see, except in the hydrodynamic case.
In Fig.~\ref{Fig: MEAN ANGLE} (right), the zoom illustrates for large $\tau$ that for $c>0.01$ the mean angle is slightly larger than $\pi/2$. 
Concerning the modified case, where zonal flows are present, the multi-scale curvature is radically diminished at all time-lags, reflecting the reduced radial motion of the fluid particles and the anisotropy of the flow structure. We can also note that for {\it mHW} the behavior at large $\tau$ is different. Instead of presenting an asymptotic value for $c=2~mHW, ~4~mHW$, the mean angles decrease similarly to what is found for flows in porous media \cite{He_PRF_2018}. One possible explanation for the asymptotic behavior for the largest $\tau$ is that the value reflects the proportion of particles that change the flow domain.\\

\begin{figure}[!htb]
 \begin{center}
 \begin{tabular}{cc}
   \includegraphics[scale=0.7]{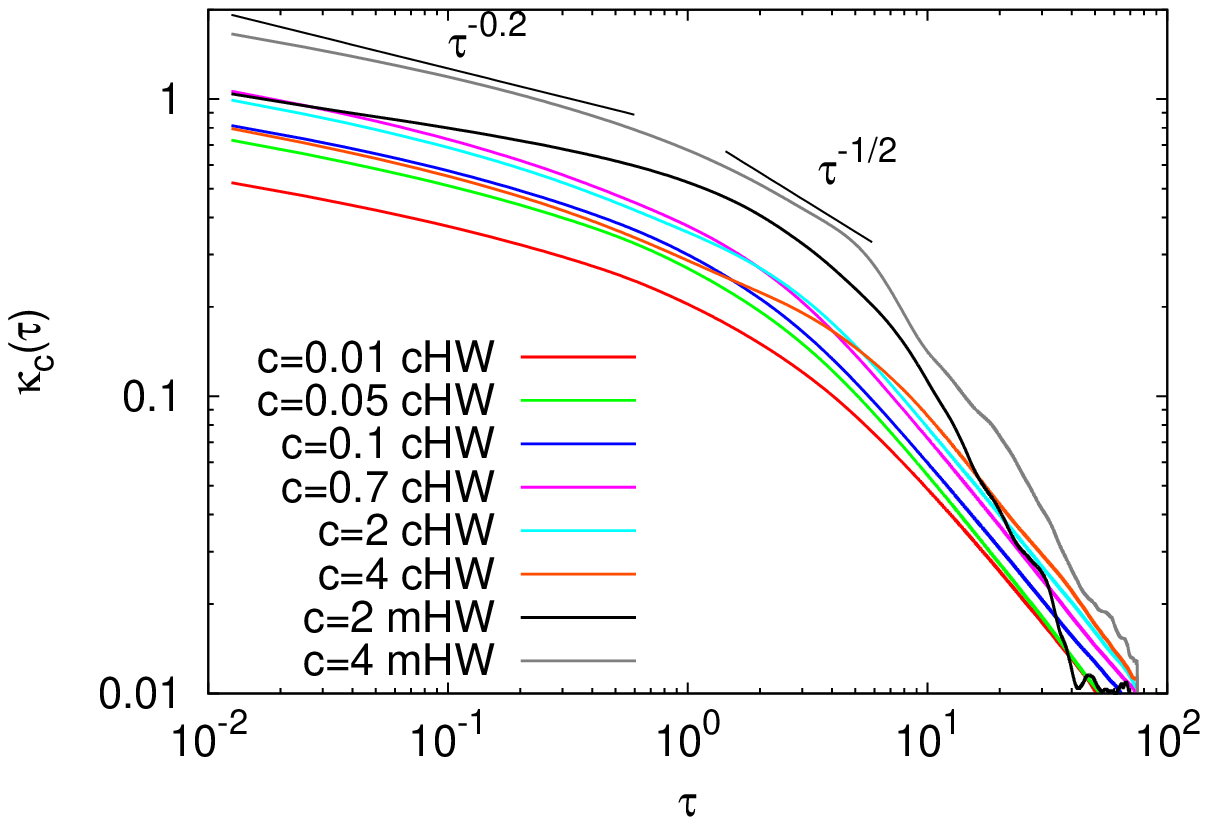}&
   \includegraphics[scale=0.7]{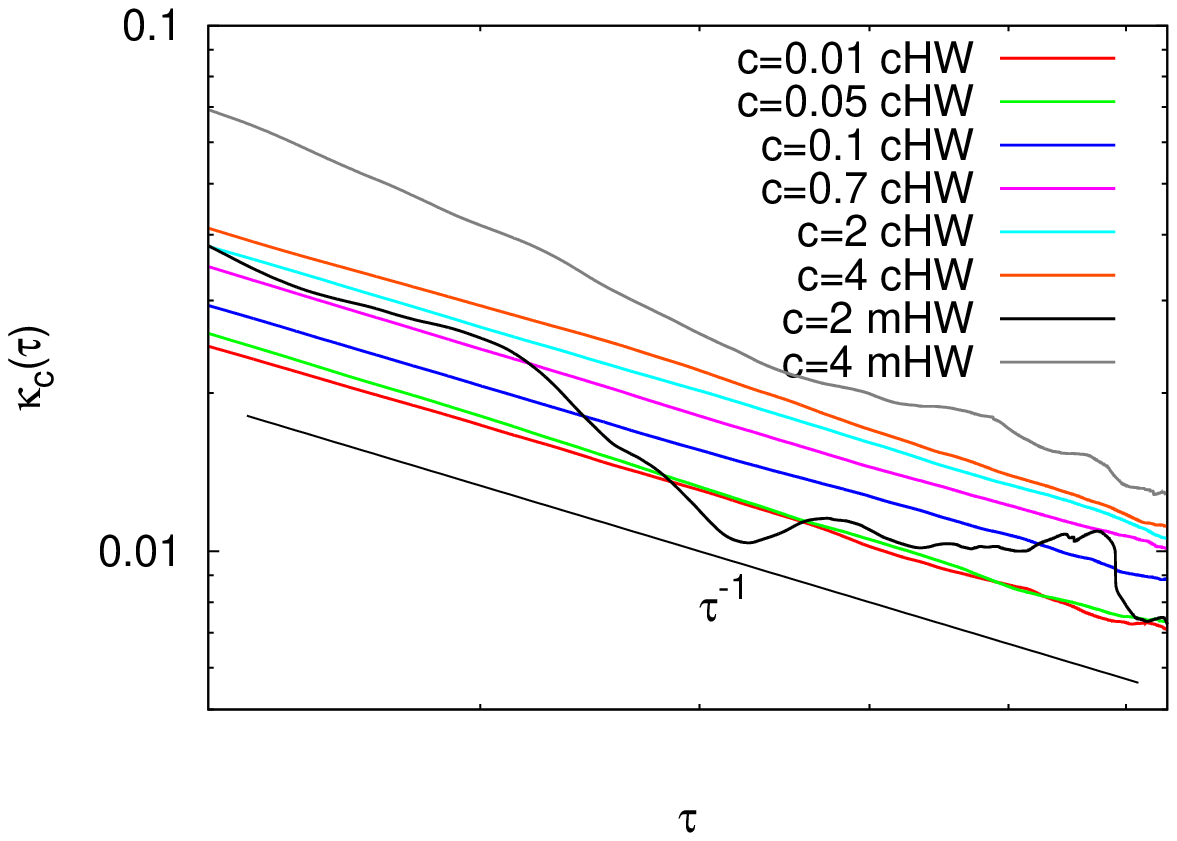}\\
  \end{tabular}
\vspace{-0.5cm}
\caption{Mean coarse-grained curvature $\kappa_c(\tau)$ as a function of $\tau$ for the different flow regimes. A corresponding zoom for large $\tau$ ($20 < \tau < 100$) is shown on the right.
}
\label{Fig: MEAN CURVATURE}
 \end{center}
\end{figure}

Figure \ref{Fig: MEAN CURVATURE} (left) shows the mean coarse-grained curvature angle $\kappa_c(\tau)$ with $\kappa_c(\tau)\equiv\left<|K(t,\tau)|\right>$, where $K(t,\tau) = \Theta(\tau) / (2 \tau || {\bm u} ||)$ and $\left< \cdot \right>$ denotes the ensemble and time average as function of the time increments for different adiabaticity values. 
We find that the mean curvatures decrease monotonically from values of $0.5-1.8$ to $6 \cdot 10^{-3}$, when $\tau$ increases. For all regimes and for $\tau < 1$, the coarse-grained curvature exhibits an algebraic scaling behavior which should tend to a constant $1/\left(2||\bm u(t)||\right)$ when $\tau \rightarrow 0$ because $\Theta(t,\tau\rightarrow 0)\sim \tau$. The observed powerlaw behaviors may testify the self-similarity of the trajectories.
For large $\tau$, the same algebraic scaling behavior $\propto \tau^{-1}$ is observed in all cases, as expected, since $\Theta(t,\tau\rightarrow \infty)\sim \pi/2$. In Fig.\ref{Fig: MEAN CURVATURE} (right), the zoom at large $\tau$ illustrates that the slope of the mean coarse-grained curvature angle is the same for different adiabaticity values $c$ in the case of {\it cHW}. In contrast for {\it mHW} with $c=2~mHW, ~4~mHW$ we find instead of a pronounced $\tau^{-1}$ scaling, that  $\kappa_c(\tau)$ fluctuates, which is probably due to the presence of zonal flows.

\section{Conclusions}
\label{sec:conclusions}

We performed extensive numerical studies for a wide parameter range of the classical Hasegawa--Wakatani model and its modified version governing the plasma flow and the observed electrostatic drift-wave turbulence in the tokamak edge. 
The modified Hasegawa--Wakatani model allows to assess the influence of zonal flows.
In the proposed Lagrangian approach different statistics were performed for ensembles of tracers along their trajectories.
Conditional averages using the Okubo--Weiss criterion were computed and the flow topology was decomposed into different regions, vortical or deformation dominated and intermediate regions.
The influence of the adiabaticity parameter $c$ which is related to the parallel dynamics of the plasma and the nonlinear cascades in the parallel direction has been investigated in a systematic way,
including the quasi adiabatic regime of relevance to edge plasma turbulence. 
Moreover, we performed simulations also for values beyond the ones relevant for describing plasma edge turbulence to explore a large parameter space.
We also studied the adiabatic limit ($c >> 1$) for which the model reduces to a Hasegawa--Mima type equation, {\it i.e.} the geostrophic flows and the limit $c << 1$ for which we recover a Navier--Stokes system, {\it i.e.} hydrodynamic flows.
Furthermore, motivated by the fact that in the classical system the turbulent flow remains isotropic for low values of the adiabaticity parameter and zonal flows are absent, we performed simulations for a modified Hasegawa--Wakatani system.

The Eulerian and Lagrangian statistics allowed us to characterize the unalike complex dynamics of the flows and the tracer transport, as mentioned above. 
Analyzing the co-spectrum of velocity and density fluctuations in the radial direction yields
insight into the contributions of the density flux at different length scales. The observed inertial scaling of $k^{-7/2}$ is found to be in agreement with predictions based on dimensional arguments.
Another important result is that the behavior of the residence time, observed in two-dimensional homogeneous isotropic turbulence \cite{kadoch_etal_2011}, is still valid for the different Hasegawa--Wakatani regimes. The influence of vortex trapping which explains the longest residence times in strong elliptic regions tends to disappear with increasing adiabaticity and the residence times from strong elliptic and hyperbolic regions become of the same order. Moreover, the presence of zonal flows that induces shear flows, for modified HW, implies a slightly larger contribution from strong hyperbolic regions compared to strong elliptical ones.
Furthermore, the conditional Lagrangian statistics with respect to three level Weiss values reveal that the intermediate zones are responsible for the total PDF and the differences between the different zones are reduced with increasing adiabaticity for classical HW. However, again due to the presence of shear flows, the contributions of intermediate zones are still of major importance for modified HW. Indeed, the strong elliptic and hyperbolic regions are located in the zone where shear flows appear.

An analysis of radial density gradients shows that PDFs have exponential tails in the cHW quasi-adiabatic regime.
These results are in agreement with Ref.~\cite{Umansky_POF_2011} where the authors showed that turbulence fluctuations have non-Gaussian characteristics, while the radial flux of plasma density is similar to Bohm diffusion.

Directional statistics considering
the angular change of the Lagrangian tracer particles reveal
a multiscale measure for the coarse grained curvature. We thus quantified the directional properties of the complex particle motion  at different time scales for the different flow regimes. 

In forthcoming work multiscale angular Lagrangian statistics can be considered to characterize the directional change of inertial particles and their clustering, {\it e.g.}, related to impurities in the plasma, at different time scales and show the significance of zonal flows for transport.

\subsection*{Acknowledgments}
DCN acknowledges support  from the Oak Ridge National Laboratory, managed by UT-Battelle, LLC, for the U.S. Department of Energy under contract DE-AC05-00OR22725. DCN also gratefully acknowledges the support and hospitality of the \'Ecole Centrale de Marseille for the visiting positions during the elaboration of this work. KS acknowledges partial support by the French Federation for Magnetic Fusion Studies (FR-FCM) and the Eurofusion consortium, funded by the Euratom research and training programme 2014-2018, 2019-2020 and 2021-2022 under grant agreement No 633053. The views and opinions expressed herein do not necessarily reflect those of the European Commission.
Centre de Calcul Intensif d'Aix-Marseille is acknowledged for granting access to its high performance computing resources. WJTB and KS thank Prof. Pat Diamond and the Organizers of the Festival de Th\'eorie in Aix-en-Provence, France for discussions which triggered some of the presented investigations on the modified Hasegawa--Wakatani system.

\bibliography{biblio}

\clearpage
\appendix
\section{flow dynamics}
\label{sec:appendix_flowdynamics}

\begin{figure}[!htb]
 \begin{center}
 \begin{tabular}{cc}
   \includegraphics[scale=0.7]{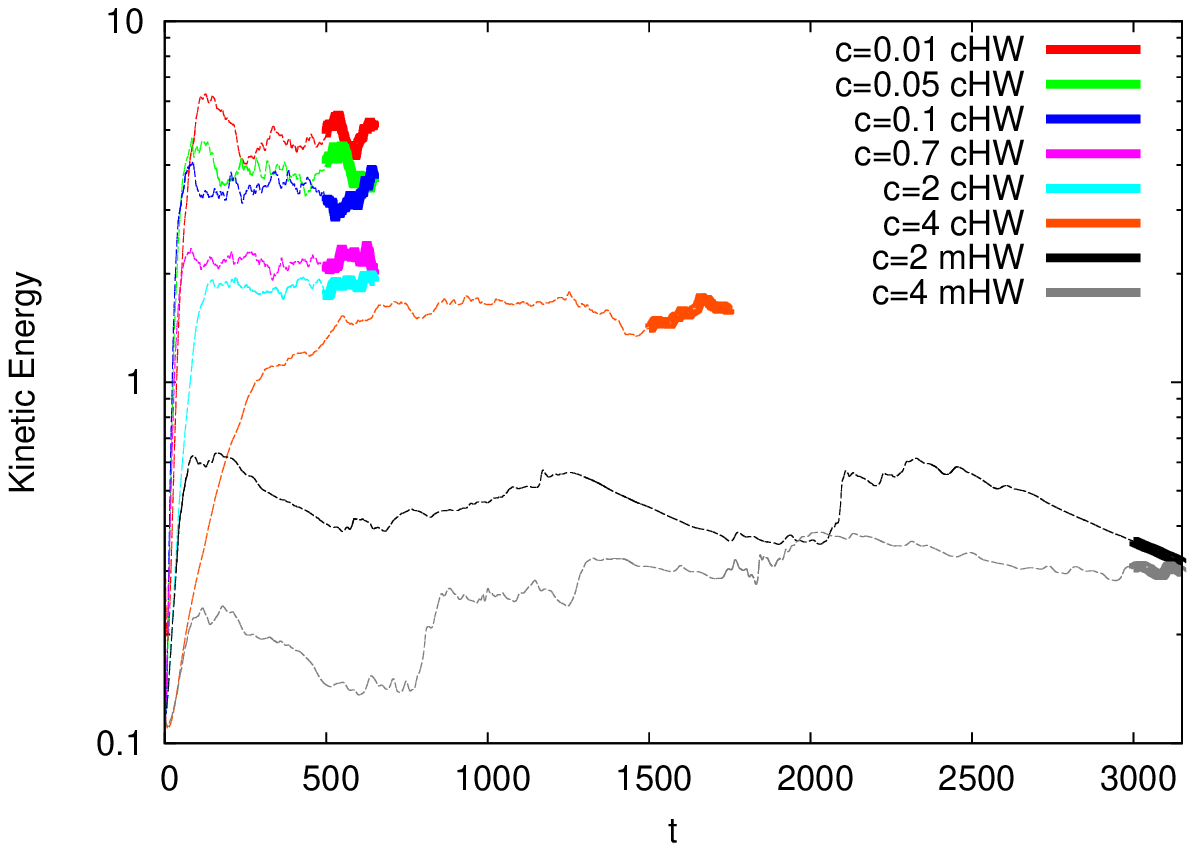}&
   \includegraphics[scale=0.7]{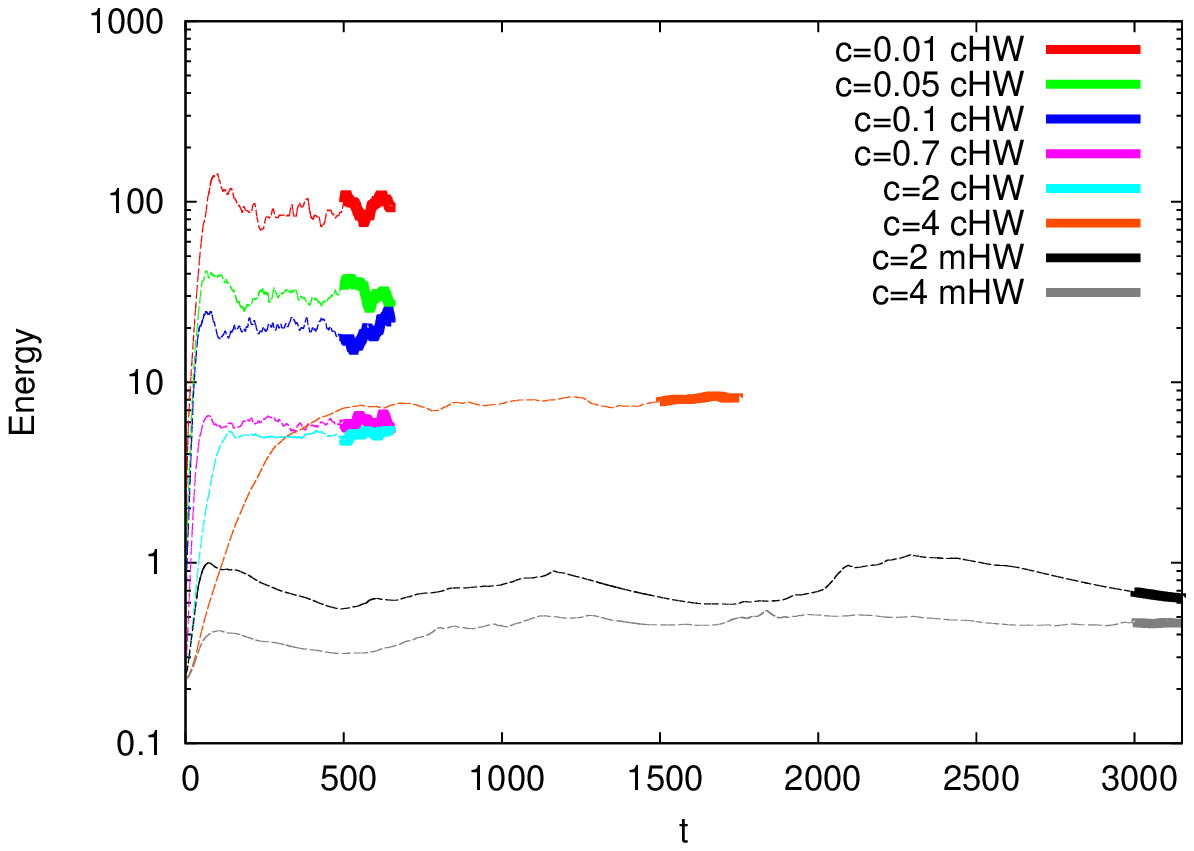}
  \end{tabular}
\caption{Time evolution of kinetic energy $E_{kin}=0.5 \langle u^2 \rangle$ (left) and total energy $E_{tot}=E_{kin} + 0.5 \langle n^2 \rangle$ (right). The thick lines indicate the time slab where the statistics of the particle trajectories are performed: From $t=500$  to $t=650$ for {\it cHW} with $c=0.01, 0.05, 0.1, 0.7$ and $2$. From $t=1000$ to $t=1150$ for {\it cHW} with $c=4$. From $t=3000$ to $t=3150$ for {\it mHW} with $c=2$ and $c=4$. 
}
\label{Fig: EVOL}
 \end{center}
\end{figure}

To get insight into the flow dynamics the time evolution of kinetic energy $E_{kin}=0.5 \langle u^2 \rangle$ and total energy $E_{tot}=E_{kin} + 0.5 \langle n^2 \rangle$ are shown in Fig.~\ref{Fig: EVOL}. After some transition time, exhibiting drift-wave instabilities, all configurations reach a statistically steady state, even if some fluctuations in the energy are still present. 
In the {\it cHW} case the fluctuations are more important for $c=4$ than for $c=2$.
The thick lines correspond to the time interval where the statistics of the particle trajectories have been performed.
%

\end{document}